\documentclass[preprint2]{aastex}

\usepackage{subfigure}
\usepackage{graphicx,dblfloatfix}
\usepackage{rotating}

\slugcomment{Not to appear in Nonlearned J., 45.}

\shorttitle{OM IN QUASARS: SPECTRAL VARIABILITY}
\shortauthors{Ram\'\i rez et al.}

\begin{document}

\title{Optical Microvariability in Quasars: Spectral Variability}

\author{A. Ram\'\i rez, D. Dultzin, AND J. A. de Diego}
\affil{Instituto de Astronom\'\i a de la Universidad Nacional Aut\'onoma de M\'exico, Apartado Postal 70-264, 04510 M\'exico, D.F., M\'exico}
\email{aramirez@astroscu.unam.mx,deborah@astroscu.unam.mx, jdo@astroscu.unam.mx}

\begin{abstract}

We present a method that we developed to discern where the optical microvariability (OM) in quasars originates: in the accretion disk (related to thermal processes) or in the jet (related to non-thermal processes). Analyzing nearly simultaneous observations in three different optical bands of continuum emission, we are able to determine the origin of several isolated OM events. In particular, our method indicates that from nine events reported by \cite{Ram09}, three of them are consistent with a thermal origin, three to non-thermal, and three cannot be discerned. The implications for the emission models of OM are briefly discussed.

\end{abstract}


\section{Introduction}

Optical microvariability (OM; variations with small amplitude in time scales from minutes to hours) in quasars can be a powerful tool to constrain the energy emission models of active galactic nuclei. Some studies indicate that OM does not depend on the radio properties of quasars (see \citealt{de98}, hereafter PI; \citealt{Stalin04}; \citealt{Gupta05}; \citealt{Ram09}, hereafter PII). However, the mechanism responsible for OM has not been characterized. Concerning this topic, variability studies in the optical bands acquired a particular relevance, because the optical/UV excess is identified with the emission from the accretion disk (e.g., \citealt{Krishan94}; \citealt{Lawrence05}; \citealt{Pereyra06}; \citealt{Li08}; \citealt{Wilh08}). Unfortunately, all the proposed scenarios are very complex when the emission from additional spectral components is considered.

 In photometric studies, the properties of the spectral energy distribution (SED) of quasars era characterized by the spectral index, which is defined as the slope of a curve in a plane $log(f_\nu)~versus~log(\nu)$, and calculated as magnitude differences at different bands (e.g., \citealt{Masa98}; \citealt{Treve02}). During variability events, changes in the shape of this SED can give us important information about the emission processes. For instance, evidence from variability studies strongly indicates that OM has non-thermal nature in both BL Lac objects and flat spectrum radio quasars (FSRQs; e.g., \citealt{Damicis02}; \citealt{Vagne03}; \citealt{Villata04b}; \citealt{Villata04a}; \citealt{Hu06a}, \citealt{Hu06b}, \citealt{Hu07}), but in the case of FSRQs the contribution of a thermal component must be considered in the spectrum of these objects (e.g., \citealt{Gu06}; \citealt{Hu06b}; \citealt{Ram04}). Concerning quasars, short-term variability (variations with amplitudes of tenths of a magnitude and time scales of weeks to months) might have thermal origin (e.g., \citealt{Treve02}). Nevertheless, all these studies usually employ two bands, or only the average of color values is used, losing thus information about {\it texture} of shortest variations (see \citealt{Giveon99}; \citealt{Webb00}; \citealt{Treve02};  \citealt{Vagne03}).

In order to discern where the OM in quasars originates, we assume that emission from the accretion disk must be related to thermal processes, while the emission from the jet must be related to non-thermal processes. During OM events, the spectral component responsible for the variation may display characteristic color changes. Then, we developed a method with quasi-simultaneous observations at three optical bands to analyze these color changes that accompanied microvariability events. This paper is organized as follows: in Section~\ref{datos}, we shortly comment on the data treatment; in Section~\ref{sed}, we define a spectral variability index and how to determine the OM origin; in Section~\ref{results}, we analyze the data; finally, we discuss the results in Section~\ref{discusion}; while a summary and conclusions are given in Section~\ref{summary}.

\section{Samples selection, observations, and data reduction}\label{datos}

Details of the selection criteria, observation strategy, and data reduction have been exposed in PI, PII, and \cite{de10}. These data consist of a sample of 22 core-dominated radio-loud quasars (CRLQ) and 22 radio quiet quasars (RQQs) observed in diverse epochs. Each object in the RQQ sample paired a CRLQ object in brightness and redshift minimizing selection effects when properties of microvariability between both samples were compared in PII. Here, we will only discuss the data corresponding to OM events reported in PII. This subsample is listed in Table \ref{log5}. Four telescopes were used, which are located in Mexico and Spain. $BVR$ filters of Johnson-Cousins series were used. The observational strategy consists of monitoring a CRLQ-RQQ couple during the same night in overlapping sequences; five images of each object in the $BVR$ sequence are taken ($\sim 1$ minute of exposure time per image, although it depends of the object brightness, the filters and the telescope used). Each object was observed at moderated air masses, always at least $30^o$ above the horizon. Standard stars were observed to obtain the flux level of the first data set of each objects in each night. These stars were observed each time that a sequence of each pair of objects was complete. This implies that standard stars were observed each hour, approximately. All these stars are taken from \citet{landolt92}. In addition, these standard stars are used to perform correction by atmospheric extinction.

\section{Quasar SED and SED variability.}\label{sed}

In order to discern where the OM originates, the analysis is divided into two main steps: first, a SED model is fitted to the {\it initial} data set, obtaining estimations of the values of the parameters of each spectral component; then, the observed spectral variations are modeled using these values.

The simplest description of the continuum emission for quasars is given by a thermal and non-thermal emission mixture (e.g., \citealt{Malkan82}; \citealt{Malkan86}). Thus, at any time $t$, the flux would be determined by the expression
\begin{equation}
\label{ec5.1} f_{\nu t} = f_{\nu n t} + f_{\nu T t},
\end{equation}
where the sub-index $t$ refers to the time when a particular data set was acquired; $f_{\nu t}$ refers to the total flux; $f_{\nu n t}$ to its non-thermal component; and $f_{\nu T t}$ to its thermal component (hereafter sub-indices $T$ and $n$ indicate thermal and non-thermal components, respectively). Usually, thermal and non-thermal components are associated with the accretion disk and with the relativistic jet, respectively. Hereafter, we will adopt this assumption.

At any time $t$, the contribution from each component to the total flux can be described as $a_{\nu t} \equiv {f_{\nu n t} / f_{\nu t}}$ for the non-thermal component, and $ b_{\nu t} \equiv {f_{\nu T t} / f_{\nu t}}$ for the thermal one (note that $b_{\nu t_0}= 1-a_{\nu t_0}$).

In a naive description, the non-thermal emission from the jet can be described by a simple power law, while the thermal emission from the disk can be modeled by a simple blackbody. Although a simple blackbody is an over-simplification for the accretion disk, this is the simplest model that fits our data (something similar has been made previously by \citealt{Malkan82}; \citealt{Malkan86}; among others).

When a given object is monitored in a particular night, we can determine an initial time, $t_0$, which corresponds to the first data set of observations. When a variation is detected we take the difference of fluxes between a given data set obtained at the time $t$, and that for the first data set; then, we normalize with respect to the initial total flux. We consider that the variation is provoked by only one of the spectral components in expression (\ref{ec5.1}), leading to the fact that in the difference of fluxes only the variable component survives, i.e.,
\begin{equation}
\label{ec5.2} \varpi_{\nu i t} \equiv {f_{\nu i t} - f_{\nu i t_0}
\over f_{\nu t_0}}.
\end{equation}
In this expression, $f_{\nu i t}$ ($i = n, T$) represents the variable component. Some advantages (which are easy to verify) of measuring the flux changes by this expression are: data transformation to standard photometric system is unnecessary; error propagation is diminished, and a larger accuracy is reached; correction by interstellar extinction and cosmological assumptions (as the ones required in order to give a value for $H_0$) are also avoided. It is important to stress that, when we fitted the model to the initial data set, it is necessary to correct for interstellar extinction to the first $BVR$ data set; while for the spectral variability description, this correction is not required.

On the other hand, we have to consider that the observed flux is related to the emission in the sources frame through the expression $f_{\nu} = (1+z) L(\nu [1+z])/ 4 \pi D^2 _L$ (see \citealt{Kembhavi99}). Here $f_{\nu}$ represents the observed flux at the observed frequency $\nu$; $L$ the luminosity of the source; $z$ the redshift; and $D_L$ the luminosity distance.

In addition, we can represent the brightness amplitude of each component by means of a pair of constants: $A_t$ for the non-thermal component and $\beta_t$ for the thermal one. The injection of new particles with similar energy, for the non-thermal case, or changes of size of the region of emission, for the thermal case, can provoke amplitude changes.

We initially fit the observed flux at the first data set, obtaining the initial values for the temperature of the disk, the spectral index of the non-thermal component, and the contribution of each component: $T_{t_0}$, $\alpha_{t_0}$, and $a_{\nu t_0}$ and $b_{\nu t_0}$, respectively. Taking the initial data, we can express for the non-thermal component the ratio of fluxes between any $t$ and $t_0$, i.e.,
\begin{equation}
\label{ec5.a}
{f_{\nu n t} \over f_{\nu n t_0}} = {A_{t} \over A_{t_0}} \nu^{\alpha_t}
\nu^{-\alpha_{t_0}} (1+z)^{\alpha - \alpha_{t_0}},
\end{equation}
from this expression we obtain 
\begin{equation}
\label{ec5.14}
f_{\nu n t} = {A_{t} \over A_{t_0}} \nu^{\alpha_t}
\nu^{-\alpha_{t_0}} (1+z)^{\alpha - \alpha_{t0}} a_{\nu t_0} f_{\nu}
10^{-0.4m_{\nu t_0}},
\end{equation}
where $m_{\nu t_0}$ refers to the magnitude in the standard system, corrected by interstellar extinction (corrections to these initial magnitudes were carried out using values reported by the NED); $f_{\nu}$ is a constant of reference flux, corresponding to a star of zero magnitude; and $\nu= \nu_B$, $\nu_V$, $\nu_R$, i.e., the effective frequency in the $B$, $V$, and $R$ bands. Something similar can be done for the thermal component, leading to
\begin{eqnarray}
\label{ec5.15} f_{\nu T t} &=& {\beta_{t} \over \beta_{t_0}} \nu^3
\nu^{-3} {e^{[({h\nu \over \kappa T_{t_0}}) (1+z)]} -1 \over e^{[({h\nu
\over \kappa T_t}) (1+z)]} -1} \times \nonumber \\ & & \times b_{\nu t_0}
f_{\nu} 10^{-0.4m_ {\nu t_0}}.
\end{eqnarray}

Concerning the contribution from each component, we can give only $a_{V t_0}$ or $b_{V t_0}$, since we can write $a_{\nu t_0}$ and $b_{\nu t_0}$ (where $\nu = \nu_B, \nu_R$) in terms of them, i.e.,
\begin{equation}
\label{ec5.9} a_{\nu t_0} = a_{V t_0} {f_{V} \over f_{\nu}}
10^{-0.4(m_{V t_0} - m_{\nu t_0})} {f_{\nu n t_0} \over f_{V n
t_0}},
\end{equation}
while
\begin{equation}
\label{ec5.10} b_{\nu t_0} = b_{V t_0} {f_{V} \over f_{\nu}}
10^{-0.4(m_{V t_0} - m_{\nu t_0})} {f_{\nu T t_0} \over f_{V T
t_0}}.
\end{equation}

\begin{figure}[htp]
\begin{center}
\epsscale{1} \plotone{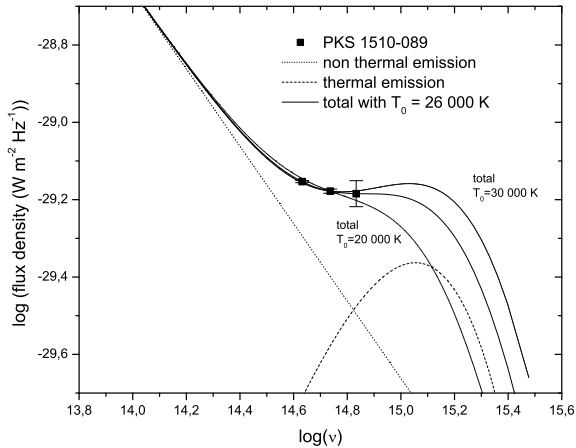} \caption{\sl \footnotesize  Combination of two components can fit the first data set for PKS~1510-089 to obtain $T_{t_0}$, $\alpha_{t_0}$, and $a_{V t_0}$. Dotted line corresponds to the non-thermal component, while dashed line corresponds to the thermal component. Spectral index of the power law for the non-thermal component has a value of $-1$, while the temperature of the disk would be in $26,000$ K. Continuum thick line represents the mixture of these components. However, these data can be fitted by different mixtures. Continuum thin lines represent the combination of a power law, with index $-1$ and black bodies of $20,000$ K and $30,000$ K.}
\label{pks15dos}
\end{center}
\end{figure}

Thus, evaluating expressions (\ref{ec5.14}) and (\ref{ec5.15}) at $t=t_0$ and replacing them in Equation (\ref{ec5.1}), we can model the emission at each frequency for the first data set to get values for $T_{t_0}$, $\alpha_{t_0}$, and $a_{V t_0}$ (see Figure \ref{pks15dos}). We will test these values when we will describe the spectral variations, generating curves on a plane defined by $\varpi_{V, R}~vs.~ \varpi_B$ (see Figure 2). Comparing data and models on this plane, we obtain a {\it spectral variability diagnosis plane}.

\subsection{Emission models for quasars on diagnosis plane for spectral variability}\label{varpi}

Using expression (\ref{ec5.14}), i.e., a power law for the non-thermal component, expression (\ref{ec5.2}) can be written as
\begin{equation}
\label{ec5.16} \varpi_ {\nu n t} = a_{\nu t_0}(n_{nt}(\nu +
z\nu)^{\alpha_t-\alpha{t_0}} -1),
\end{equation}
while using expression (\ref{ec5.15}), i.e., a blackbody as the thermal component, expression (\ref{ec5.2}) can be written as
\begin{equation}
\label{ec5.17} \varpi_ {\nu T t} = b_{\nu t_0} (n_{Tt} [{\nu \over \nu_V}]^3~
{e^{(h\nu_V/kT{t_0}) (1+z)} -1 \over e^{(h\nu / kT_t) (1+z)}-1} -1),
\end{equation}
where $n_{Tt} = {\beta_t / \beta_{t_0}}$ and $n_{nt} = {A_t / A_{t_0}}$. In this way, changes in amplitude are easily obtained by varying the values of these ratios from unity, at $t_0$, to their final values at $t$.

\begin{figure}[htp]
\begin{center}
\epsscale{0.9} \plotone{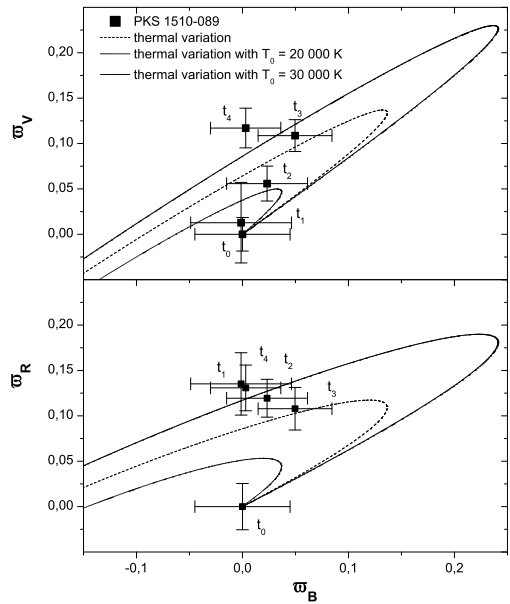} \caption{\sl \footnotesize Each curve corresponds to a similar thermal change for the fits of Figure \ref{pks15dos}. The curves correspond to different values of $T_{t_0}$.} \label{pks15vardos}
\end{center}
\end{figure}

\begin{figure}[htp]
\begin{center} 
\epsscale{0.9} \plotone{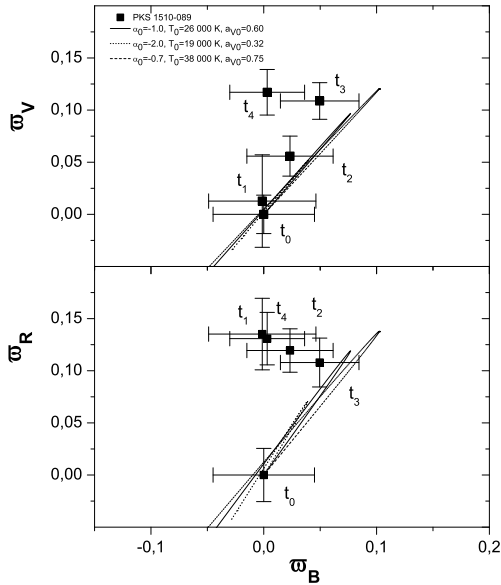} \caption{\sl \footnotesize Curves show simultaneous variations of $\alpha_t$ and $n_{n t}$ in the non-thermal component for the data of PKS~1510-089. Similar changes for $\alpha_{t_0}$ of $-0.7$, $-1.0$ and $-2.0$ are simulated. Temperatures for thermal component are $38,000$ K, $26,000$ K and $19,000$ K, respectively. Contributions of non-thermal component are $0.75$, $0.6$ and $0.32$, respectively. The choice of $\alpha_{t_0} = -1$ does not seem to be a decisive factor in the variability source discernment.} \label{pks15alfas}
\end{center}
\end{figure}

\begin{figure}[htp]
\begin{center}
\epsscale{0.8} \plotone{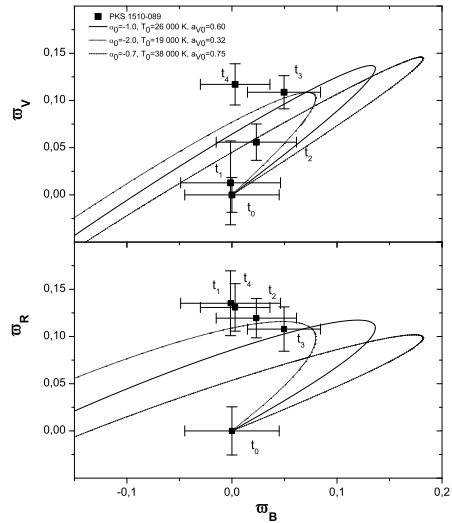} \caption{\sl \footnotesize  Three characterization curves for simultaneous variations of $T_t$ and $\beta_t$ are shown for data of PKS~1510-089. The three curves correspond to similar changes in the thermal component for $\alpha_{t_0} = -0.7$, $-1.0$, and $-2.0$. Initial temperatures are $38,000$ K, $26,000$ K and $19,000$ K, respectively. It is feasible to explain the variability of this quasar with the assumption of changes in the thermal component. The choice of $\alpha_{t_0} = -1$ does not seem to be a decisive factor in the variability source discernment.} \label{pks15tes}
\end{center}
\end{figure}

\begin{figure}[htp]
\begin{center}
\epsscale{0.8} \plotone{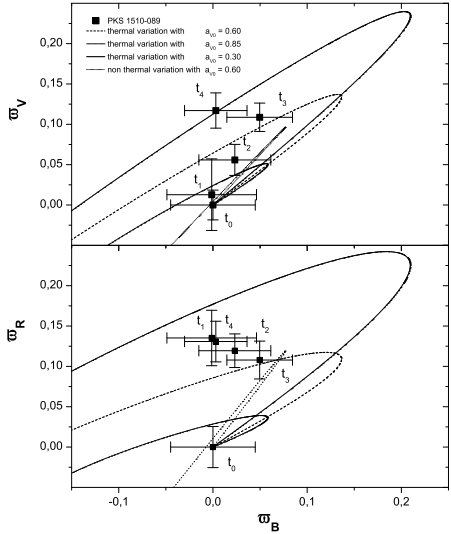} \caption{\sl \footnotesize  Variation of the thermal component. Curves represent similar changes of $T_{t}$ and $n_T$, but with $a_{V t_0} = 0.60, 0.85, 0.30$ ($b_{V t_0} = 1 - a_{V t_0}$). A variation from the non-thermal component cannot generate behaviors like those observed. On the other hand, regardless the exact values for $T_{t_0}$, $\alpha_{t_0}$, and $a_{V t_0}$, the observed variation is consistent with changes of the thermal component.} \label{pks15a0}
\end{center}
\end{figure}

\begin{figure}[htp]
\begin{center}
\epsscale{0.8} \plotone{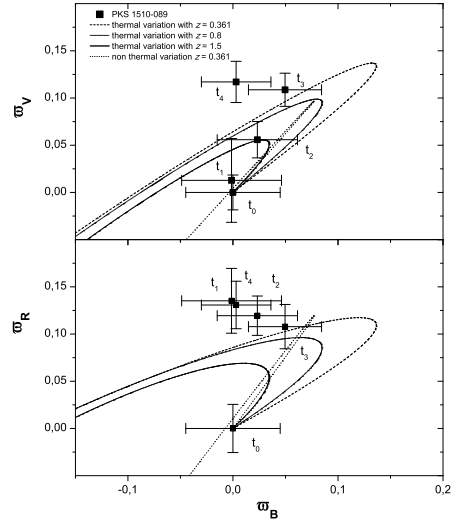} \caption{\sl \footnotesize Variation of the thermal component. Similar changes in $T$ and $n_T$, but at different redshifts. $z=0.361$ corresponds to the redshift of the quasar. It is interesting to notice that with the same conditions (same $T_{t_0}$, $\alpha_{t_0}$ and $a_{V t_0}$), objects with different redshifts will have a different appearance from the same variation. Additionally, a curve representing a change of $\alpha_t$ and $n_{nt}$ is shown.} \label{pks15z}
\end{center}
\end{figure}

Once we get $T_{t_0}$, $\alpha_{t_0}$, and $a_{\nu t_0}$, changes of $T_t$ and/or $n_{Tt}$, or $\alpha_t$ and/or $n_{nt}$, in expression (\ref{ec5.2}), generate the trajectories in the $\varpi_{V, R}: \varpi_B$ plane (see Figures \ref{pks15vardos}-\ref{pks15z}). Simultaneous changes in $T_t$ and $n_{T_t}$, or $\alpha_t$ and $n_{n_t}$ can generate diagnosis trajectories with a returning behavior (see Figure \ref{pks15vardos}). A similar case can happen in an expansion of the thermal component with a decrease in its temperature.

Unfortunately, for the data presented in PII it is not possible to establish a unique set of initial parameters, due to the lack of data in more bands. Figure \ref{pks15dos} shows this problem. In this figure, we present three reasonable fits for disk temperatures of $26,000$ K, $20,000$ K and $30,000$ K for the first data set of PKS~1510-089. For this reason, we must select the most appropriate values for $\alpha_{t_0}$, $T_{t_0}$, and $a_{\nu t_0}$, based on the typical values found in the literature for similar objects. Some investigations suggest that the typical maximum disk temperature, close to the central object, should be between $20,000$ K and $30,000$ K, although temperatures outside of this range are not discarded (e.g., \citealt{Malkan82}; \citealt{Malkan86}; \citealt{Czerny87}; \citealt{Pereyra06}). The spectral index of the non-thermal component might be around $-1$ (e.g., \citealt{Malkan82}; \citealt{Elvis86}; \citealt{Brown89a}; \citealt{Brown89b}). On the other hand, for blazars, the non-thermal component contributes with most of the flux (e.g., \citealt{Malkan82}), and at low brightness levels, it is expected that both components contribute the same (e.g., \citealt{Gu06}).

Nevertheless, regardless the exact value for these parameters at $t=t_0$, we can discern the OM origin analyzing the spectral variations. The procedure consists of taking several $T_{t_0}$-$\alpha_{t_0}$-$a_{V t_0}$ sets, trying to reproduce the spectral variations with each of them. First, the simplest component (a power law) is fitted to the initial data, obtaining $\alpha_{t_0}$. From this fit, we look for an explanation to the spectral variation, i.e., performing changes of amplitude, of the slope, or both, in the diagnosis plane. If the explanation is satisfactory, for simplicity we consider that the OM event is consistent with changes in this unique component. Otherwise, an extra (thermal) component must be added to fit the initial data.

When this second (thermal) component is added to fit the data, two families of characterization curves are generated: one for thermal variations (as those showed in Figure \ref{pks15vardos}) and one for non-thermal variations (as those showed in Figure \ref{pks15alfas}). Contrasting the data with each family of curves, we can determine whether a spectral variation is consistent with a change in the thermal component (i.e., changes of $T_t$ and/or $n_{T t}$) or in the non-thermal one (i.e., changes of $\alpha_t$ and/or $n_{n t}$).

After comparing the data with all the curves, we find that we can set the initial spectral index of the non-thermal component, $\alpha_{t_0}$, at $\sim -1$, {\bf without affecting our conclusions. So, when we include the thermal component, we have only $T_{t_0}$ and $a_{V t_0}$ as free parameters.} Then, we take this fit as representative of each family of curves, as shown in Figures 7. For this reason, the estimations for the values of $T_{t_0}$, $\alpha_{t_0}$, and $a_{V t_0}$ must be considered more as a suggestion than as the actual conditions for both disk and jet.

We have explored how these assumptions can affect the discernment of the OM origin. As it can be appreciated in Figures \ref{pks15vardos}-\ref{pks15z}, depending on their origin, the curves show a different behavior. With different assumptions of values of $T_{t_0}$, $\alpha_{t_0}$, and $a_{V t_0}$, the trajectories for non-thermal variations are very similar. For this reason, we believe that the choice of $\alpha_{t_0} = -1$ has no influence on the description of the spectral variation. Expression (\ref{ec5.16}) shows, in fact, that $a_{V t_0}$, $\alpha_t -\alpha_{t_0}$, and $n_{n t}$ have more influence on the behavior of the trajectories than $\alpha_{t_0}$. Additionally, it is easy to see from expression (\ref{ec5.16}) that considering another power law as the second component, instead of a thermal one, does not allow us to reproduce the observed variability.

To illustrate this procedure, we have taken the case of PKS~1510-089. First, a unique non-thermal component has been fitted to the initial data. Since this fit does not describe accurately the variation, it is necessary to add the thermal emission from the disk (Figure \ref{pks15dos}). We then fixed $\alpha_{t_0}$ at $-1$ and calculated the corresponding value for $T_{t_0}$ for different choices of $a_{V t_0}$. Then, we analyzed the variations of the thermal component (see Figure \ref{pks15vardos}). We have used $T_{t_0} \sim 26,000$ K (although we also explored the cases with $20,000$ K and $30,000$ K, solid thin lines in Figure \ref{pks15dos}) and a non-thermal component contribution of $\sim 60\%$.

Afterward, $\alpha_{t_0}$ was fixed at different values, and the corresponding values for $T_{t_0}$ and $a_{V t_0}$ were calculated. Figures \ref{pks15alfas} and \ref{pks15tes} show spectral variability curves for changes of the non-thermal and thermal component, respectively. Curves correspond to $\alpha_{t_0} = -1, -0.7, -2$. As it can be observed in Figure \ref{pks15alfas} for a variation of the non-thermal component, the range of values used for the parameters has poor influence on the behavior of the spectral variations. On the other hand, the data have a behavior that is closer to the one expected from thermal changes (Figure \ref{pks15tes}).

Following a different approach, $a_{V t_0}$ was fixed at different values ($a_{0V} =0.60, 0.85, 0.30$), and the corresponding values for $T_{t_0}$ and $\alpha_{t_0}$ were calculated. Figure \ref{pks15a0} shows different curves for similar changes in $T_t$ and $n_T$, and for clarity in the Figure, we have included only a curve for changes in $\alpha_t$ and $n_n$, since the other curves are similar (as those shown in Figure \ref{pks15a0}). Again, the spectral variation has a behavior that corresponds more to what is expected from thermal variations.

Finally, we have explored the effect of redshift on variations (Figure \ref{pks15z}). At the same values of $T_{t_0}$, $\alpha_{t_0}$, and $a_{V t_0}$, the same variation produced at different redshifts ($z=0 .0361, 0.8$, $1.5$) causes different observable magnitude changes; but with greater effect on thermal changes than on non-thermal ones.

From this exploration, we find that variations of the non-thermal component cannot explain the data; while these are consistent with thermal variations. In such a case, the temperature of the thermal component varies at least $8~000$ K for the best fit.

In summary, in the case of PKS~1510-089, a change of the thermal component is the most plausible explanation for the observed variation, regardless of the exact values for $T_{t_0}$, $\alpha_{t_0}$, and $a_{Vt_0}$. A similar procedure has been carried out for each object reported as variable in PII. In the next section we present the results for each object.

\subsection{Temporal resolution of the spectral variability}\label{retem}

A problem that has to be solved when dealing with color studies refers to the time delay between optical bands (e.g., \citealt{GP96}; \citealt{Villata00}; \citealt{Papadakis03}; \citealt{Hu06a}; \citealt{Hu06b}; \citealt{Papadakis07}; \citealt{Stalin09}). For spectral OM studies, it is particularly important to avoid any possibility of a time lag, because a shorter lag than the image acquisition time between bands could provoke a false spectral variation. Thus, we must observe an object at different bands in sequences that reasonably guarantee an interval duration less than the physical lags. For the blazar 0716+714, \citet{Villata00} found that a complete sequence between the $V$ and $I$ bands should not exceed $\sim 10$ minutes. \citet{Papadakis03} calculated a $0.4$ hrs time lag between $B$ and $I$ bands in most cases. On the other hand, \citet{Hu06b} established a time lag of $11$ minutes between the {\it BATC} {\it e} and {\it m} bands. Nevertheless there is little knowledge about the minimum temporary scales for microvariability. So, it is necessary a criterion for maximum acquisition time that help us to verify that, during a variation, each $BVR$ sequence was make at the same level of flux. Usually, time lag could be determined by the discrete correlation function (e.g., \citealt{Edelson88}; \citealt{Hufnagel92}; \citealt{Villata00}; \citealt{Villata04a}), but for our observational strategy, we developed a time-lag criterion based on the change rate of each OM event. This criterion permits to be relatively sure that the data sets were acquired when each object was at the same flux level.

We introduce a {\it simultaneity} criterion that relates the rate of brightness change with the time of acquisition. A {\it simultaneity} index is defined by the expression
\begin{equation}
\label{ec5.18} S_{\nu t} \equiv {{\Delta m \over \Delta t_{obs}}
\Delta t_{cap} \over 2.5 e_{\nu t}},
\end{equation}
where $\Delta m$ refers to the maximum variation amplitude; $\Delta t_{obs}$ refers to the time lasted between the observations to obtain $\Delta m$; $\Delta t_{cap}$ is the required time to take a complete sequence of $BVR$ images; and $e_{\nu t}$ is the standard error in each group of five images. Then, it will be considered that a $BVR$ set of observations is simultaneous when a magnitude difference larger than $2.5 e$ is avoided, i.e., when the measured error is smaller than a marginal variation. Thus, a set of images taken at the time $t$ is simultaneous when each $S_{\nu t} < 1$. As this criterion is applied on each sequence of observations for an object in each night, we have an average $S_{\nu}$ for the total group of observations. We will show this average in the discussion for each object when none of the $S_{\nu t}$ is greater than $1$. In other case, each $S_{\nu t}$ will be showed separately.

Although this criterion requires that the variations have constant change rates, in the next analysis we will see that it is possible to make this assumption, fulfilling this condition.




\section{Results}\label{results}

\subsection{Model fits and initial data}\label{fits}

The results presented in this section show that it is feasible to distinguish qualitatively between thermal and non-thermal origin of some microvariation events. In those cases where both components can explain the spectral variation, for simplicity we suppose that the unique non-thermal component is responsible for the variation. This ambiguity may be eliminated by increasing
the monitoring time and observing more optical bands.



\begin{sidewaystable}
\centering
{\footnotesize \caption[Estadistica]{\bf Log of observations, and parameters values from fitting the model with one and two components}
\medskip
\begin{tabular}{lccccccccccc}
\hline \hline
\smallskip
Object &        &  Observation &            & Monitoring & Magnitude & Magnitude &                &                 & $T_{t_0}$  & $a_{V t_0}$ & Variation      \\
Name   &  $z$   &  Data        & Telescope  & Time       &   ($V$)     &  Error    & $\alpha_{t_0}$ & $\delta \alpha$ &  (K)       & (\%)        & Origin         \\
\hline

3C~281      & 0.599 & 2000 Mar 01  & EOCA  & 02:40  & 17.40  & 0.03 & -0.03 & 0.01 & 24 000 & 46  & Thermal      \\
3C~281      &  ''   & 2000 Mar 04  & EOCA  & 03:45  & 17.40  & 0.02 &  0.13 & 0.03 & 26 900 & 46  & Non-thermal  \\
PKS~1510-089& 0.361 & 2000 Mar 20  & JKT   & 04:22  & 16.86  & 0.02 & -0.23 & 0.03 & 26 000 & 60  & Thermal      \\
PKS~0003+15 & 0.45  & 2001 Oct 11  & Mx1   & 03:08  & 15.36  & 0.02 &  0.08 & 0.02 & 19 200 & 36  & Thermal (VB  \\
MC3~1750+175& 0.507 & 1999 Jun 20  & JKT   & 02:59  & 16.18  & 0.02 &  0.64 & 0.02 & 33 700 & 26  & Thermal (VB) \\
CSO~21      & 1.19  & 2001 Dec 19  & Mx1   & 02:32  & 17.02  & 0.03 & -0.67 & 0.11 & 29 500 & 78  & Non-thermal  \\
1628.5+3808 & 1.461 & 2001 Jun 13  & Mx1   & 02:23  & 16.78  & 0.02 & -1.75 & 0.02 & 10 000 & 65  & Thermal      \\
US~3472     & 0.532 & 2001 Dec 20  & Mx2   & 03:16  & 16.25  & 0.01 &  0.48 & 0.01 & 27 900 & 29  & Thermal (?)  \\
Mrk~830     & 0.21  & 1999 Agou 17* & JKT   & 02:11  & 17.55  & 0.03 & -3.06 & 0.49 & ------ & 100 & Non-thermal  \\
Mrk~830     &  ''   & 1999 Agou 18* & JKT   & 02:01  & 17.56  & 0.03 & -2.51 & 0.35 & ------ & 100 & Non-thermal  \\
Mrk~830     &  ''   & 2001 Jun 14  & Mx1   & 01:08  & 17.66  & 0.01 & -1.81 & 0.04 & ------ & 100 & Non-thermal  \\
\hline \label{log5}
\end{tabular}
}
\begin{list}{}
{\setlength{\rightmargin}{0.5truecm}
\setlength{\leftmargin}{0.5truecm} \setlength{\parsep}{0.3ex}
\footnotesize  }
\item[]{\footnotesize Log of observations for the microvariability events reported in PII (Columns 1-5); values for the fit to the first data set; and possible origin of the variation. Column 1 indicate the object name; Column 2, the redshift; the date of the observations are given in Column 3; the used telescopes are given in Column 4; and Column 5 shows the duration for each monitoring. Telescopes are: EOCA, Estaci\'on Observacional de Calar Alto (Spain); JKT, Jacobus Keptein Telescope (Spain); Mx1, 2.01m telescope of OAN (Mexico); Mx2, 1.5m telescope of OAN (Mexico). Table shows initial values for $\alpha_t$ when only a power law is used to fits the first data set (Columns 6-9). The columns indicate the observed $V$ magnitude (Column 6); error in this magnitude (Column 7); and the index of the power law and its error (Columns 8 and 9, respectively). The values of the two components fit to the first data set ($\alpha_{t_0} = -1$ in all the cases), and the possible origin of the variation are shown in Columns 9-12: blackbody initial temperature (Column 10); initial contribution to total flux of non-thermal component (Column 11; contribution in the others bands are related to the contribution in the $V$ band, see the text); and the variation origin, thermal or non-thermal (Column 12).NOTE: *~Variations are only marginal.}
\end{list}
\end{sidewaystable}

Figures \ref{fig81}- \ref{fig811} show the procedure described in the last section for the case of PKS~1510-089 for all the objects.  The figures show the fits for a unique spectral component for each object (panel a). The fits for the thermal and non-thermal mixed components are shown in panel b. Panel (c) shows a plane $\varpi_{V, R}: \varpi_B$ for the case of the fit showed in panel (a). Finally, panel (d) shows a $\varpi_{V, R}: \varpi_B$ plane for the fit showed in panel (b). In all cases, a $\chi^2$ test indicates that the fits are accurate. In panels (c) and (d), the error bars have been obtained by error propagation in the calculation of $\varpi_{\nu}$. Additionally, time labels are shown, corresponding $t_0$ to initial data. In all cases where two components have been considered, a value of $\alpha_{t_0}=-1$ has been set for the initial slope of the non-thermal component. Contribution to total flux refers to the $V$ band. In the next, values for $a_{V t_0}$ will be given in percentage.

Table \ref{log5} shows the log of the observations, and the results of fitting the model with one and two components. Column 1 indicates the name of the object; Column 2 redshift; Column 3 the date of observations. The telescope used is given in Column 4; while Column 5 shows the duration for each monitoring. The parameters for a single component are given in Columns 6-9: the magnitude in $V$ and its error (Columns 6 and 7, respectively); index $\alpha_{t_0}$ (Column 8) and its error, $\delta \alpha_{t_0}$ (Column 9). The parameters for two components are given in Columns 10-12: $T_{t_0}$ (Column 10); the initial contribution of the non-thermal component, $a_{V t_0}$ (Column 11); and the possible cause (thermal or non-thermal) of the variation (Column 9).

\subsection{Radio loud quasars}\label{RLQ}

\begin{figure*}[ht]
\begin{center}
\figurenum{7-1}
\subfigure[]{\scalebox{0.17}{\includegraphics{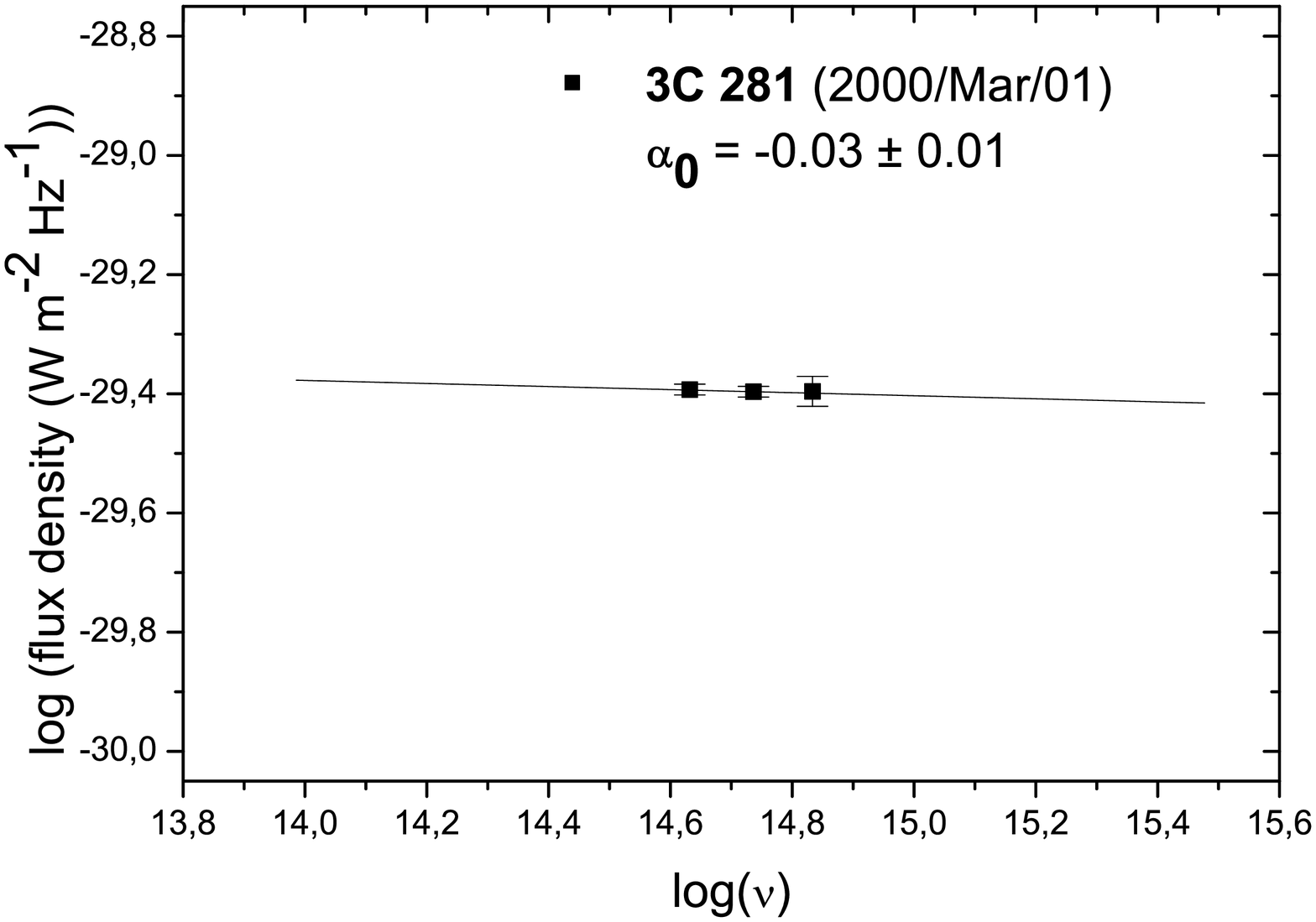}}\figurenum{7-1a}\label{fig81a}}
\subfigure[]{\scalebox{0.17}{\includegraphics{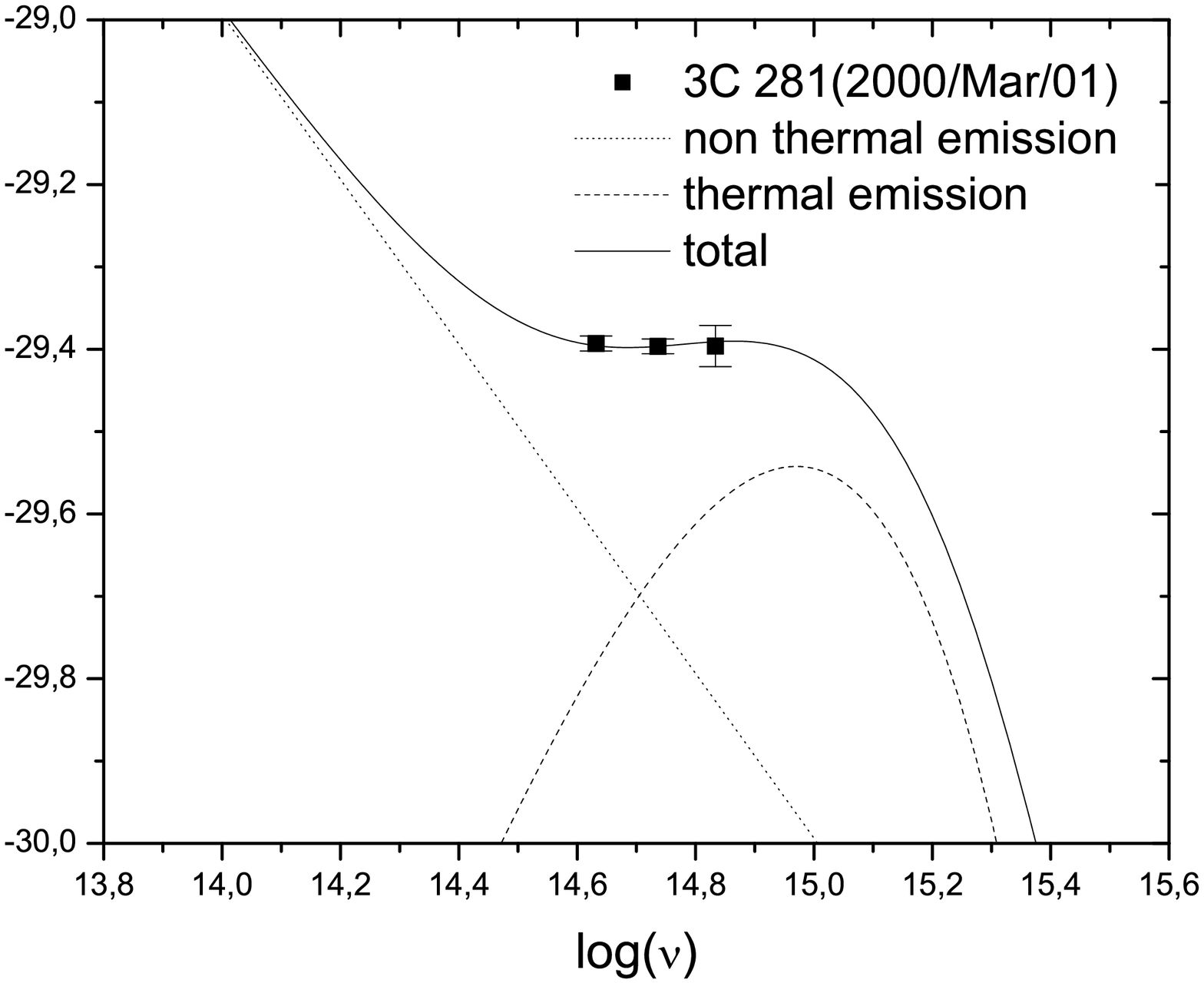}}\figurenum{7-1b}\label{fig81b}}
\subfigure[]{\scalebox{0.28}{\includegraphics{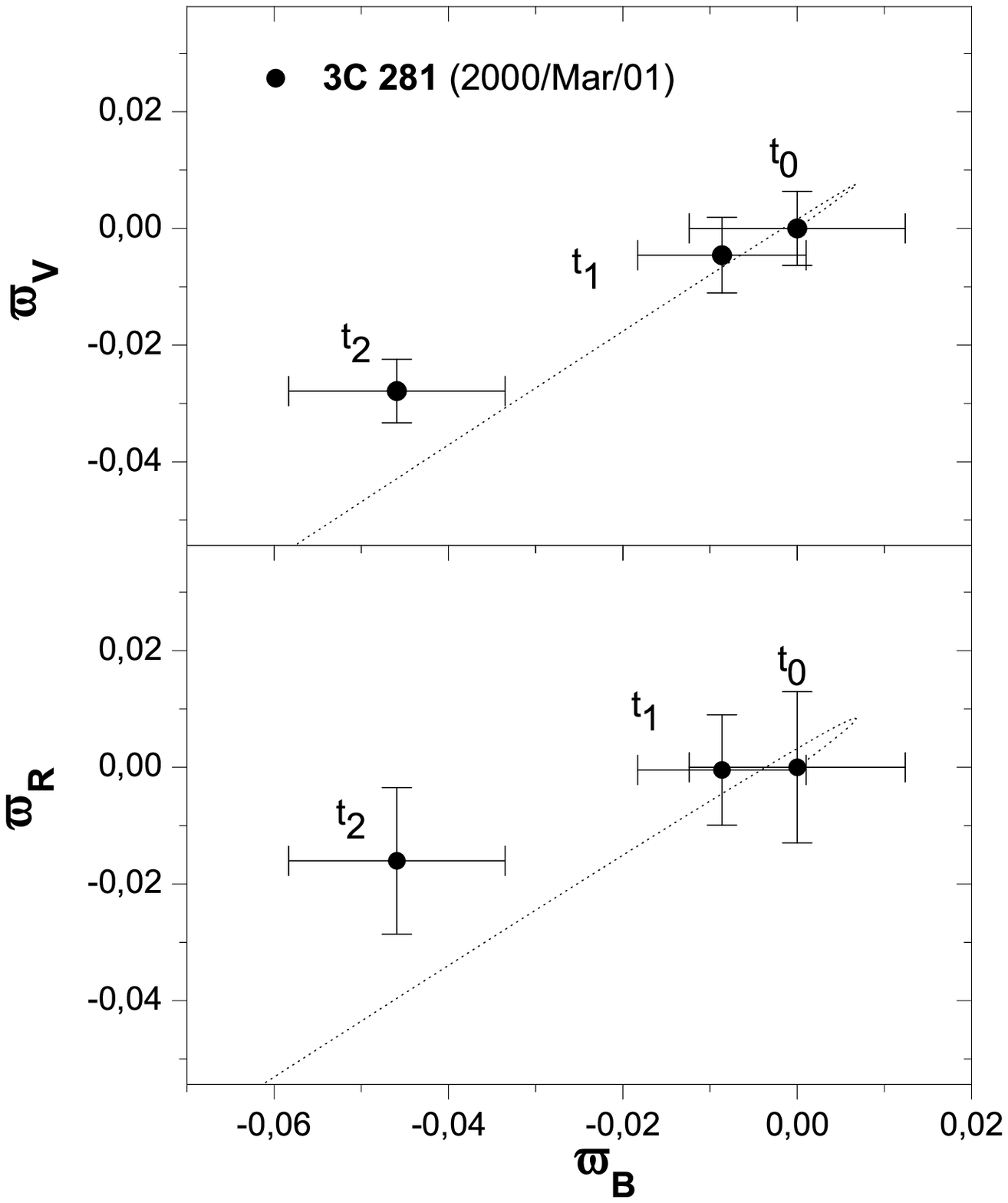}}\figurenum{7-1c}\label{fig81c}}
\subfigure[]{\scalebox{0.28}{\includegraphics{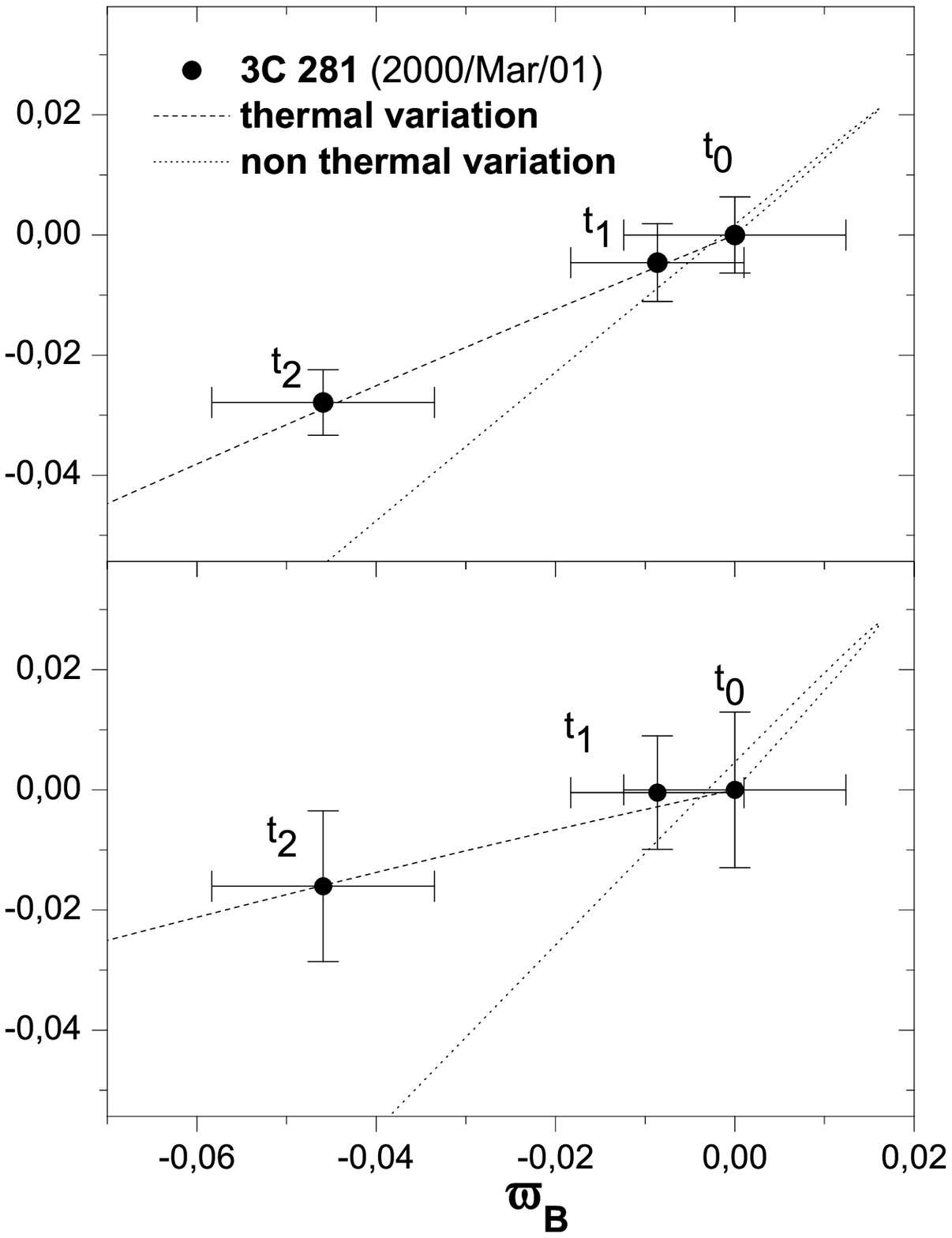}}\figurenum{7-1d}\label{fig81d}}
\caption{\sl \footnotesize   Fits to the first data sets and spectral variability: (a) first set of data is fitted by a power law; (b) first set of data is fitted by a power law and a blackbody mixture; continuum line represents the sum of thermal emission of accretion disk (dashed line), and non-thermal emission of the jet (dotted line). (c) the color change is fitted from the fit shown in panel (a); (d) the color change is fitted from the fit shown in panel (b); dashed lines represent thermal variations, while dotted lines represent non-thermal variations. Errors were obtained from error propagation when $\varpi_ {\nu}$ is calculated from data. $t_i$ ($i=0 ,1,2...$; $t_0$ corresponds to the beginning of each monitoring) indicates the temporary sequence for the data.} \label{fig81}
\end{center}
\end{figure*}

{\bf 3C~281.} Two OM events were detected in 2000 on March 1 and 4 (see PII).

{\it 2000 March 1.-} The object brightness was $m_V = 17.40 \pm 0.03$ (a little lower than the brightness reported by \citealt{Sandage65}: $m_V = 17.02$) at the begin of the monitoring. A single power law with $\alpha_{t_0} = -0.03 \pm 0.01$ fits the initial data (Figure \ref{fig81a}). However, variations of this component cannot explain the OM event (see Figure \ref{fig81c}). Then, we added in the model a thermal component with temperature $T_{t_0} \sim 24,000$ K (Figure~\ref {fig81b}). With this new fit, the non-thermal component contributes with $\sim 46\%$ to the total emission. Still non-thermal changes cannot explain the variation. On the contrary thermal changes can generate a behavior as that observed (Figure~\ref{fig81d}). In such a case, temperature decrease of at least $\sim 1,300$ K and an increase of the amplitude of $7\%$ ($n_{T_{t_2}} \sim 1.07$) are required. Scrutiny of the light curves (see Figure 1(a) in PII) indicates that variation developed after the second set of observations. A change of $1~300$ K during 2.7 hr implies a change of $700$ K in $1.44$ hr (i.e., between the first and second data sets). Simultaneously, the amplitude of the thermal emission would have an increase of $4\%$ ($n_{T_{t_1}} \sim 1.04$). Such change might have been present during the observations, and not being detected (see Figure \ref{fig81d}). The simultaneity criterion indicates no brightness changes during acquisition time ($S_B=0 .43$, $S_V=0 .67$, and $S_R=0 .17$). If indeed the OM event showed constant rate, our simultaneity criterion shows that the spectral change is not due to observational artifices.

\begin{figure*}[ht]
\begin{center}
\figurenum{7-2}
\subfigure[]{\scalebox{0.17}{\includegraphics{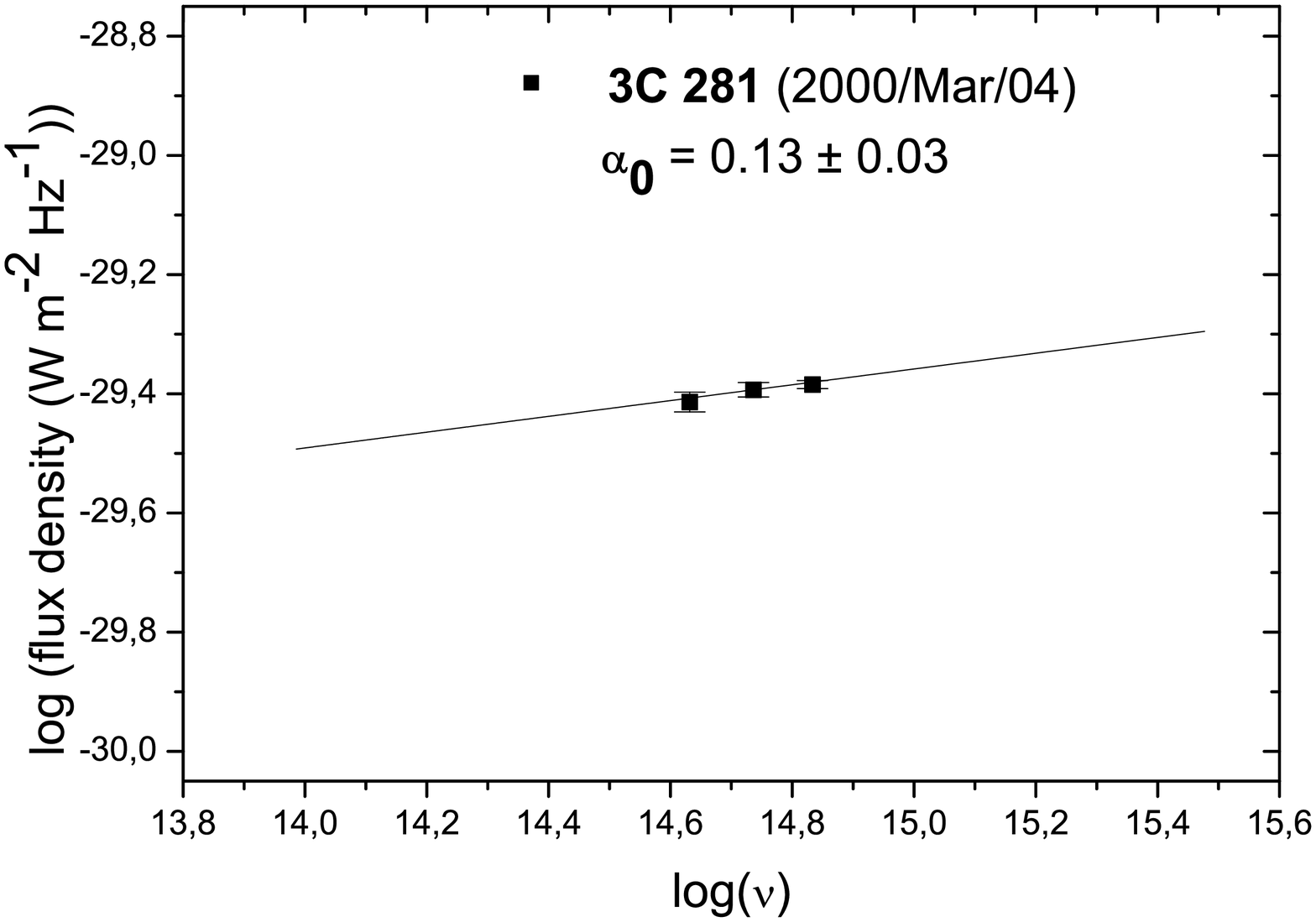}}\figurenum{7-2a}\label{fig82a}}
\subfigure[]{\scalebox{0.17}{\includegraphics{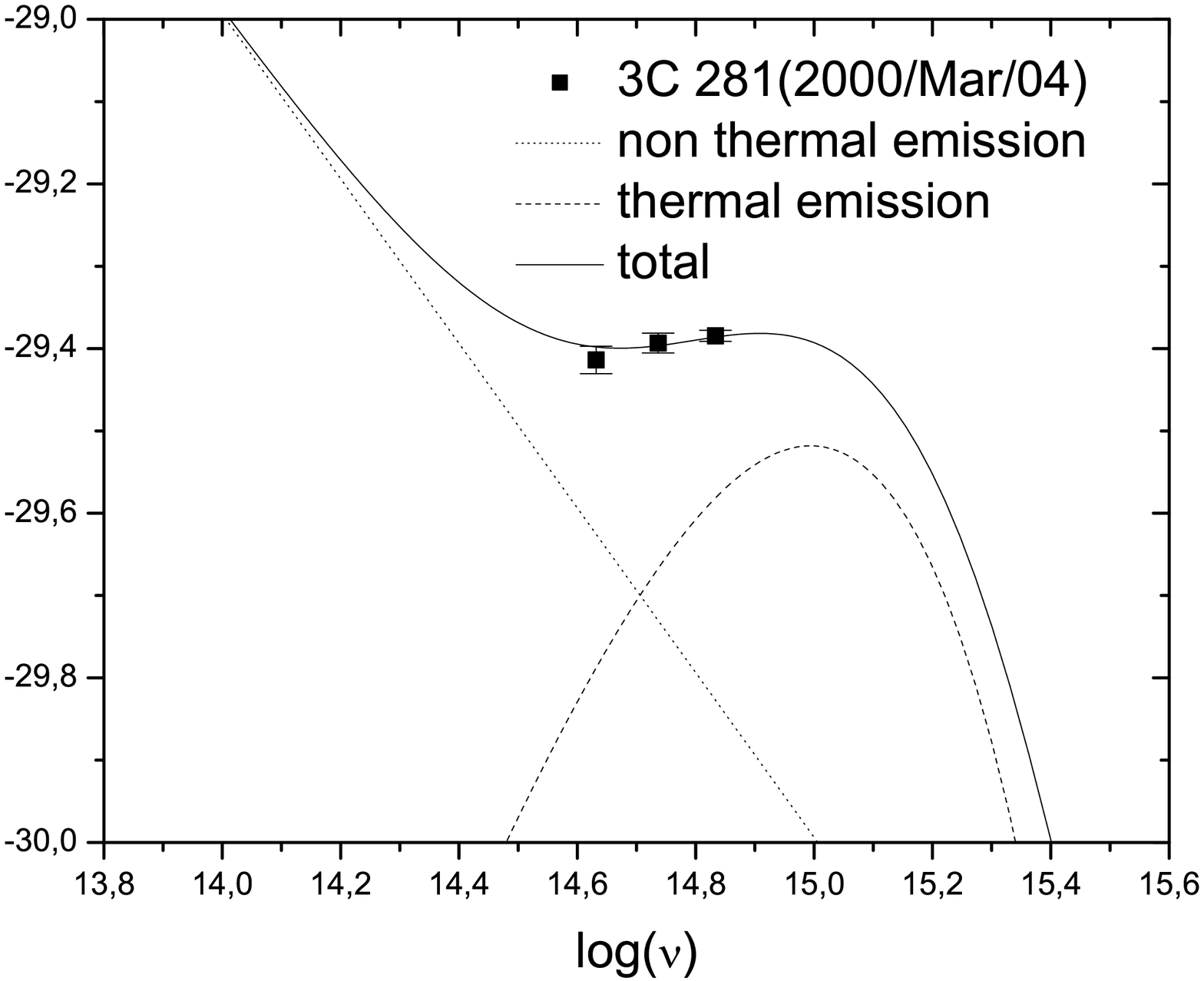}}\figurenum{7-2b}\label{fig82b}}
\subfigure[]{\scalebox{0.28}{\includegraphics{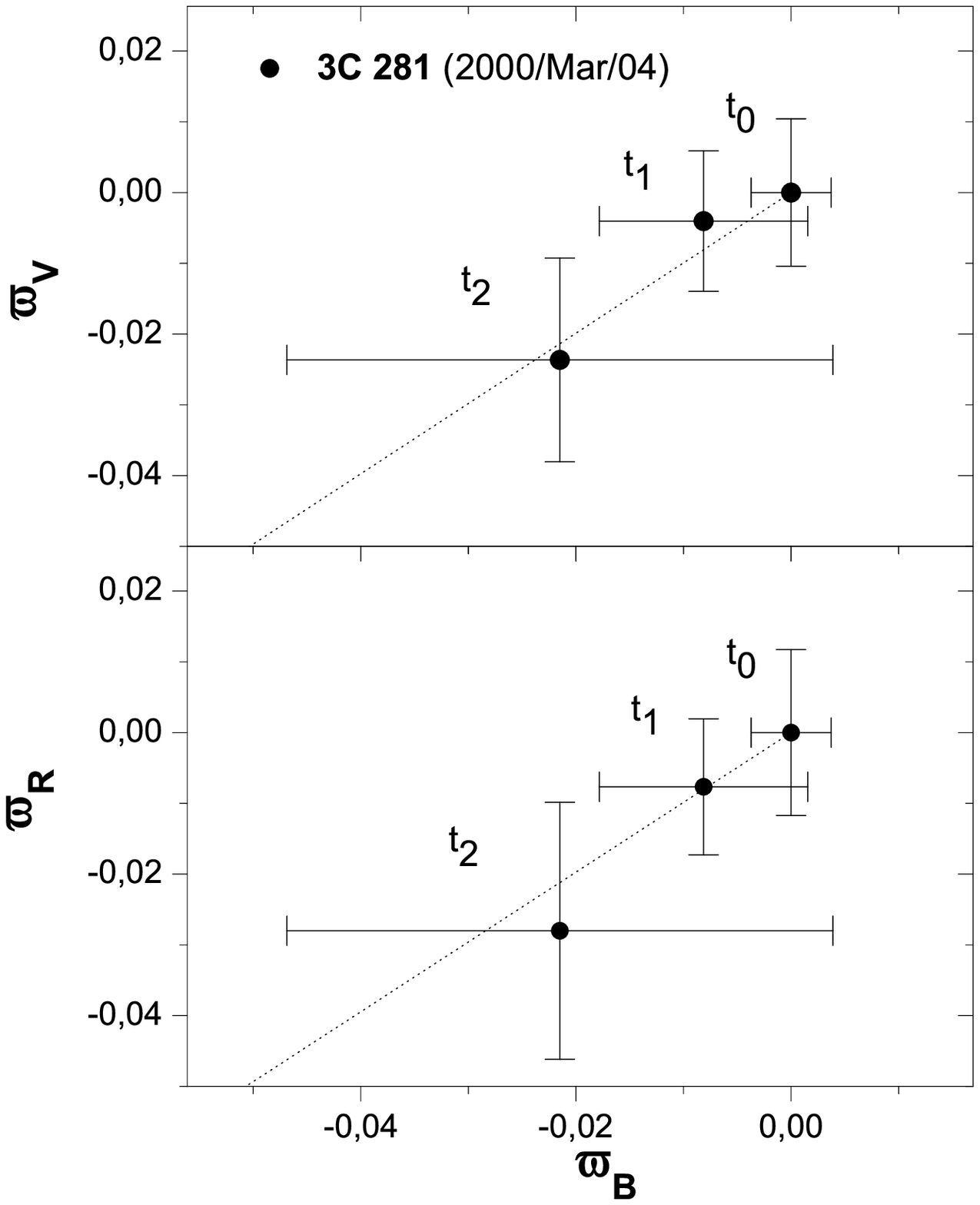}}\figurenum{7-2c}\label{fig82c}}
\subfigure[]{\scalebox{0.28}{\includegraphics{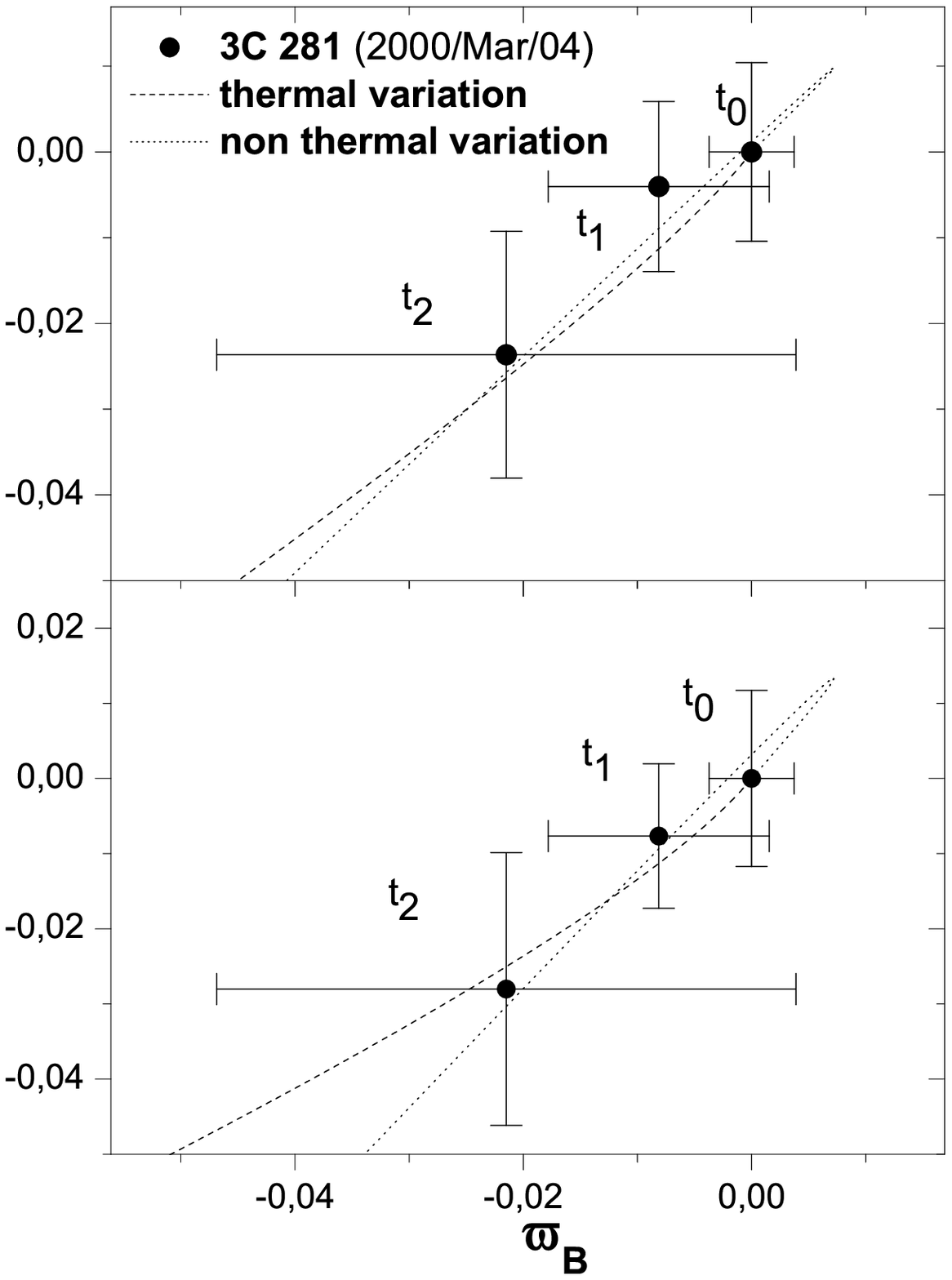}}\figurenum{7-2d}\label{fig82d}}
\caption{\sl \footnotesize Fits and variability: continuation.}
\label{fig82}
\end{center}
\end{figure*}

{\it 2000 March 4.} We measure $m_V = 17.40 \pm 0.02$ at the beginning of the monitoring. A power law can fit these data (see Figure \ref{fig82}), but with a rather low value for the spectral index $\alpha_{t_0} = 0.13 \pm 0.03$, with respect to that of March 1. It is possible to explain the variation without appealing a second component. In such a case, a change of the spectral index of $\Delta \alpha \sim 0.031$ is required (Figure~\ref{fig82c}). However, according to the results of March 1, and the fact that $\alpha_{t_0} > 0$, the emission of this object should contain a thermal component to explain the color variation for that night. Since a similar brightness was detected in both nights, we included this second component. The thermal component has $T_{t_0} \sim 26,900$ K, while $a_{V t_0} = 46\%$ and $\alpha_{t_0} = -1$ (Figure \ref{fig82b}). Fitting the variations with two components, the solution is degenerated in the sense that both factors, thermal and non-thermal, can account for the observed variability (Figure~\ref{fig82d}); although the best fit corresponds still to the non-thermal component. Thus, an index decrease of $\Delta \alpha \sim 0.02$ in this component, while its amplitude increased by $n_{n_{t2}} \sim 1.08$, can generate the observed OM. Such non-thermal change implies a rate of variation for $\alpha$ of $\sim 5.3 \times 10^{-3}~hr^{-1}$. Therefore, it is possible to hint a non detectable variation for the second data set, with $\Delta \alpha \sim 0.013$ (see Figure \ref{fig82d}). For a thermal variation case, it is necessary a decrease in $T$ of $\sim 1~800$ K, while $n_{T_{t_2}} \sim 1.09$. The simultaneity criterion indicates that during the acquisition of each set of images the brightness was the same ($S_B=0 .14$, $S_V=0 .48$, and $S_R=0 .39$).

\begin{figure*}[ht]
\begin{center}
\figurenum{7-3}
\subfigure[]{\scalebox{0.17}{\includegraphics{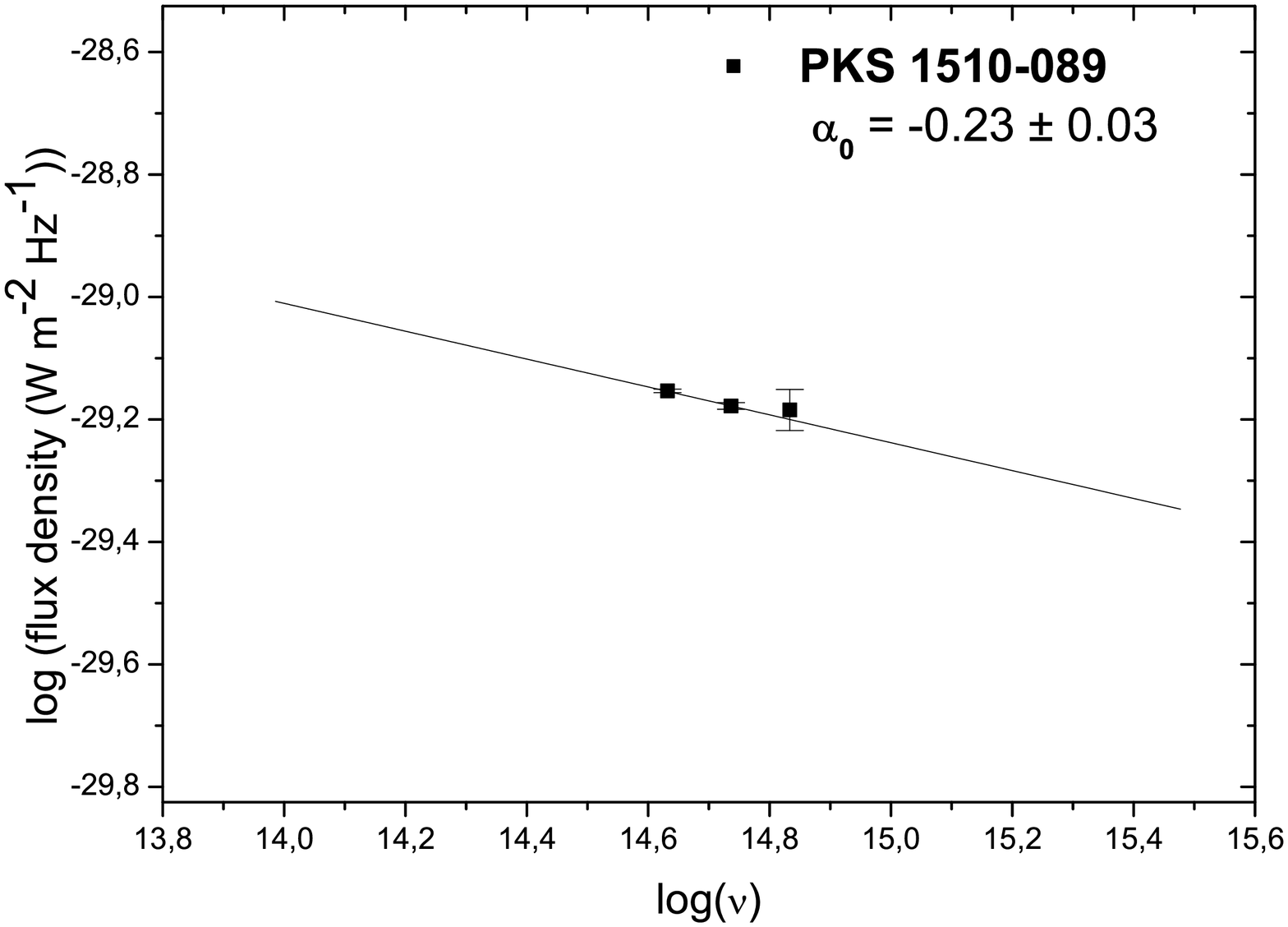}}\figurenum{7-3a}\label{fig83a}}
\subfigure[]{\scalebox{0.17}{\includegraphics{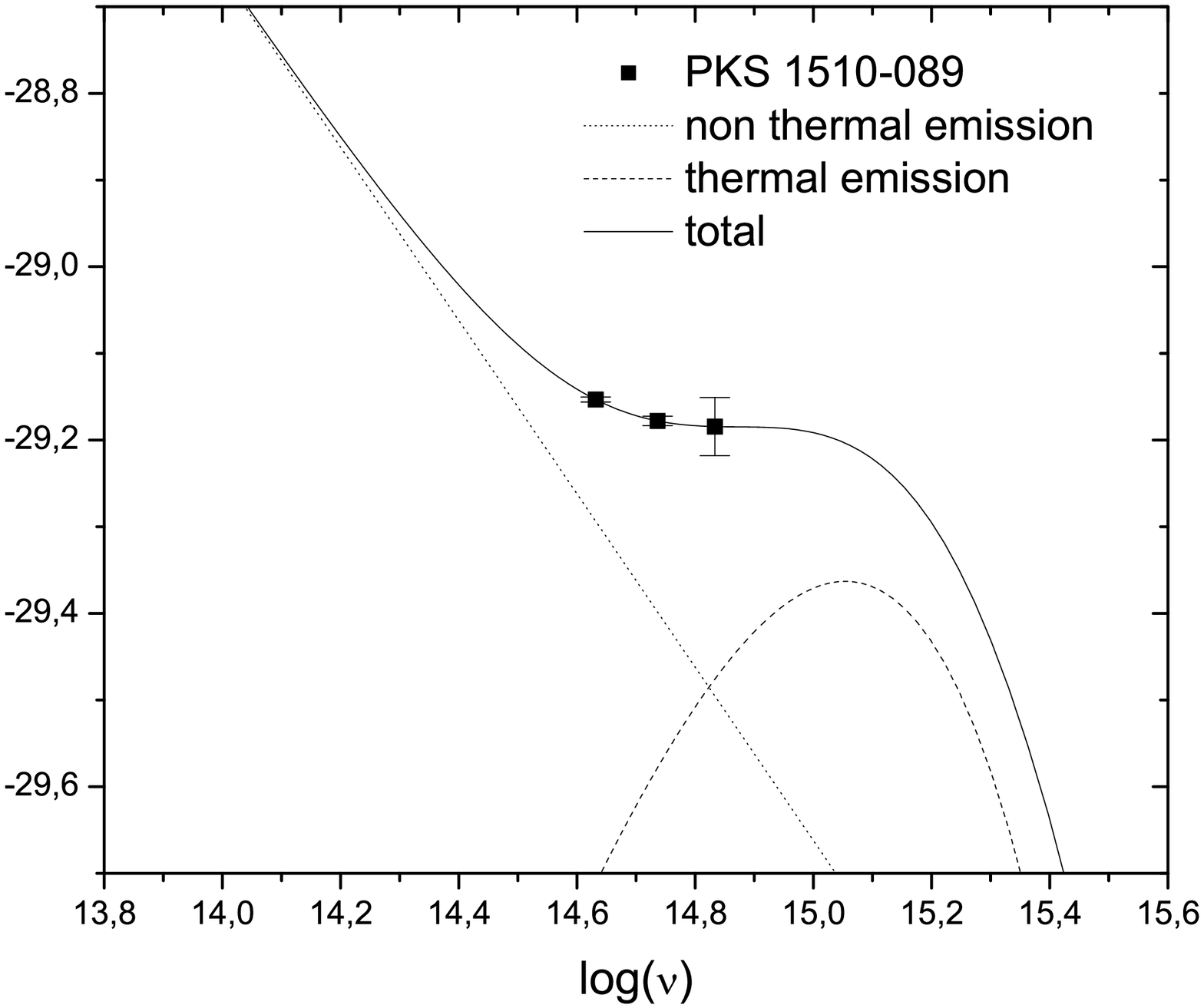}}\figurenum{7-3b}\label{fig83b}}
\subfigure[]{\scalebox{0.28}{\includegraphics{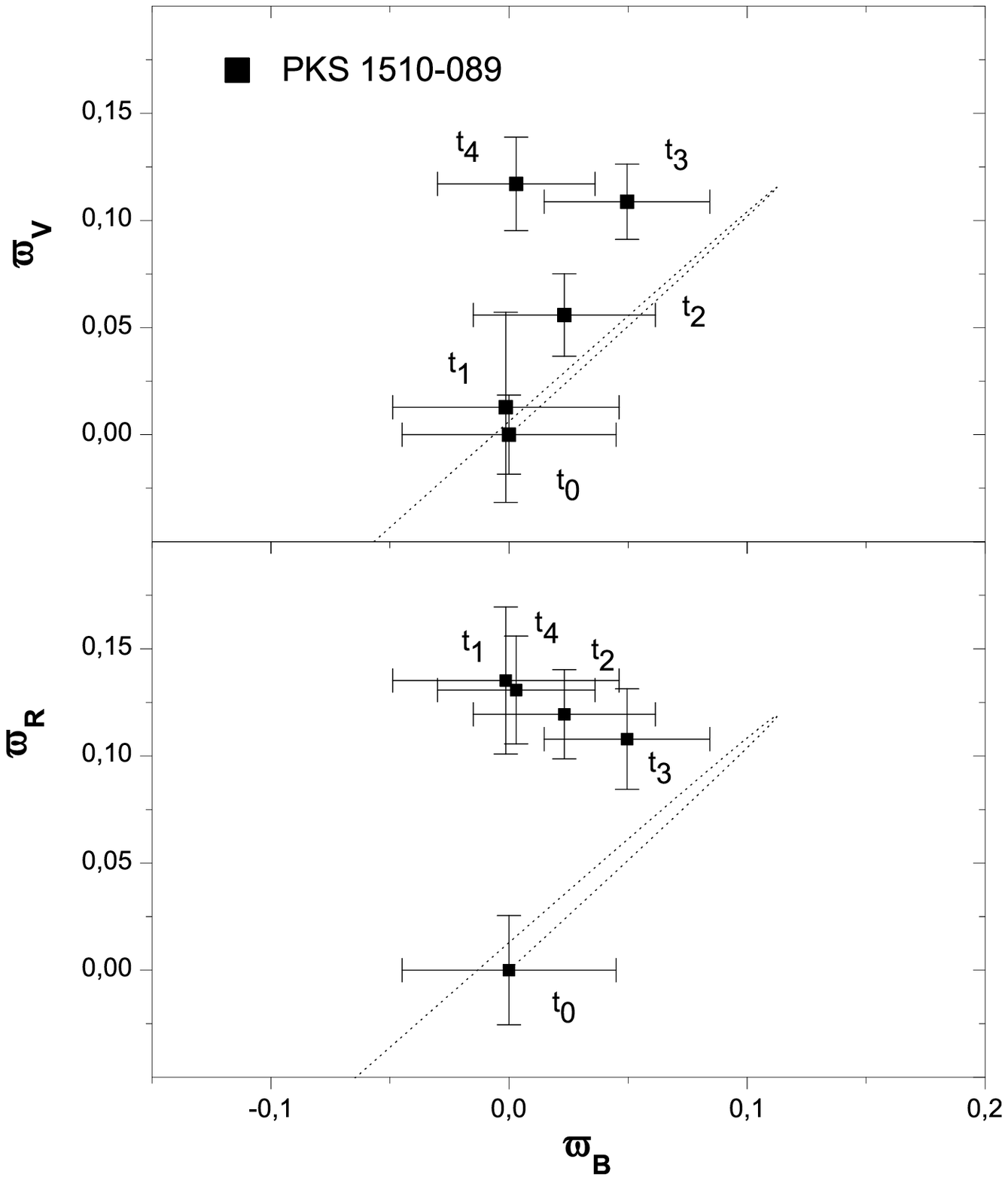}}\figurenum{7-3c}\label{fig83c}}
\subfigure[]{\scalebox{0.28}{\includegraphics{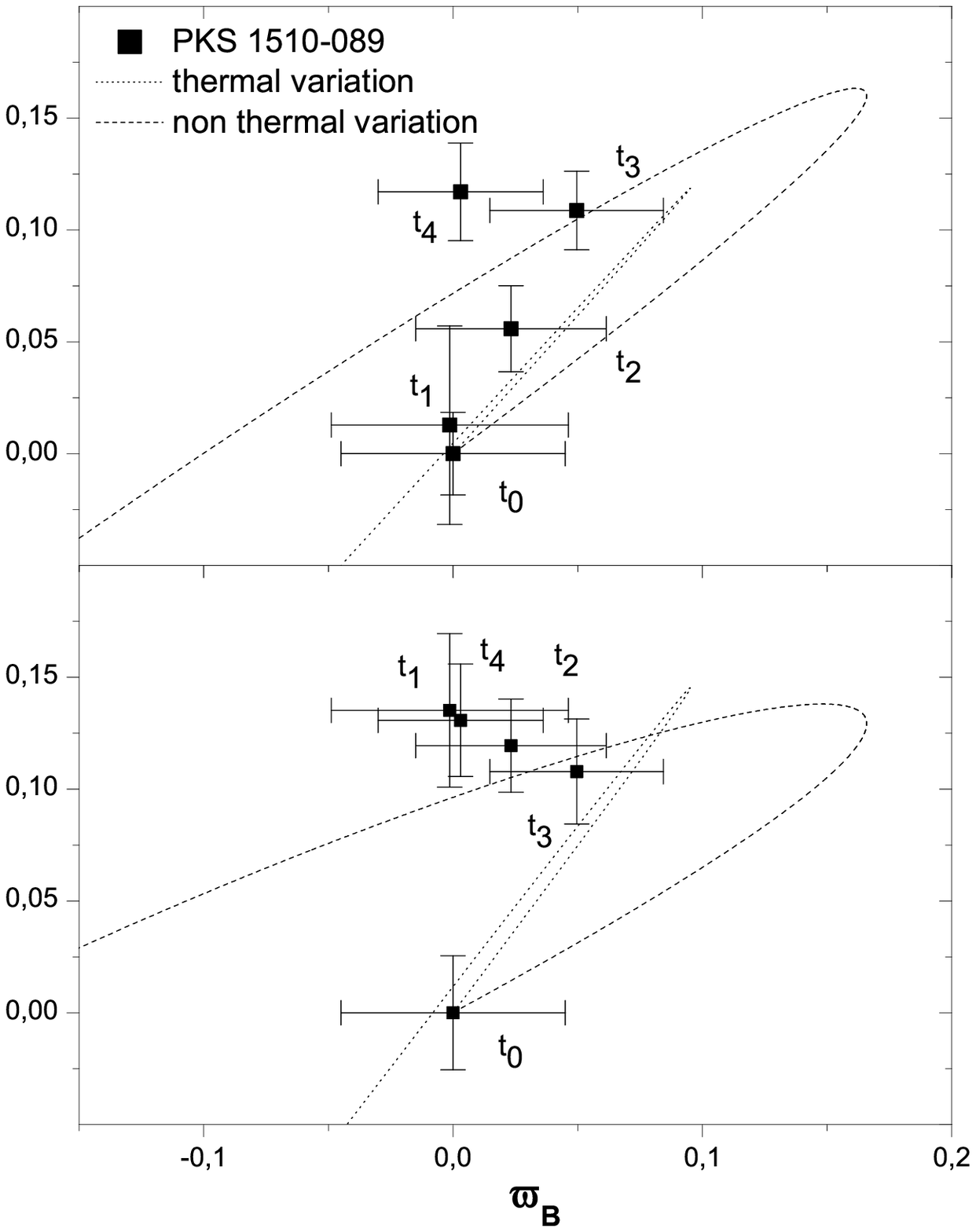}}\figurenum{7-3d}\label{fig83d}}
\caption{\sl \footnotesize Fits and variability: continuation.}
\label{fig83}
\end{center}
\end{figure*}

{\bf PKS~1510-089.} The reported magnitude for this quasar is $m_V=16.3$ (\citealt{Xie01}). On 2000 March 20, we measured $m_V = 16.86 \pm 0.02$ at the first data set. A power law with $\alpha_{t_0} = -0.23 \pm 0.03$ fits these data (Figures \ref{fig83a}). However, the spectral variation cannot be explained appropriately only with a single non-thermal component (Figure~\ref{fig83c}). Thus, a thermal component with $T_{t_0} \sim 26,000$ K and $b_ {V t_0} \sim 40\%$ is added (Figure \ref{fig83b}). With this configuration only thermal variations can reproduce the observed behavior (Figure \ref{fig83d}). Nevertheless, the amplitude of the thermal component should increase by a rather high factor ($n_{T_{t_5}} > 2.5$), while the temperature should fall $8~000$ K. In such a case, the rate of variation has been $\Delta T \sim 1~800$ K $hr^{-1}$ and $\Delta n_T \sim 0.57 h^{-1}$. The {\it return} observed in Figure \ref{fig83c} can be caused by a variation which is dominated first by changes in $n_T$ and then by changes in temperature. This behavior might have been evident if observations had been prolonged a few hours. With a constant rate of change (as it seems to have happened, see the V-band light curve in Figure 2(c) of PII), the simultaneity criterion indicates that observations were carried out satisfying the simultaneity criterion ($S_B=0.09$, $S_V=0 .22$, and $S_R=0 .86$).

\begin{figure*}[ht]
\begin{center}
\figurenum{7-4}
\subfigure[]{\scalebox{0.17}{\includegraphics{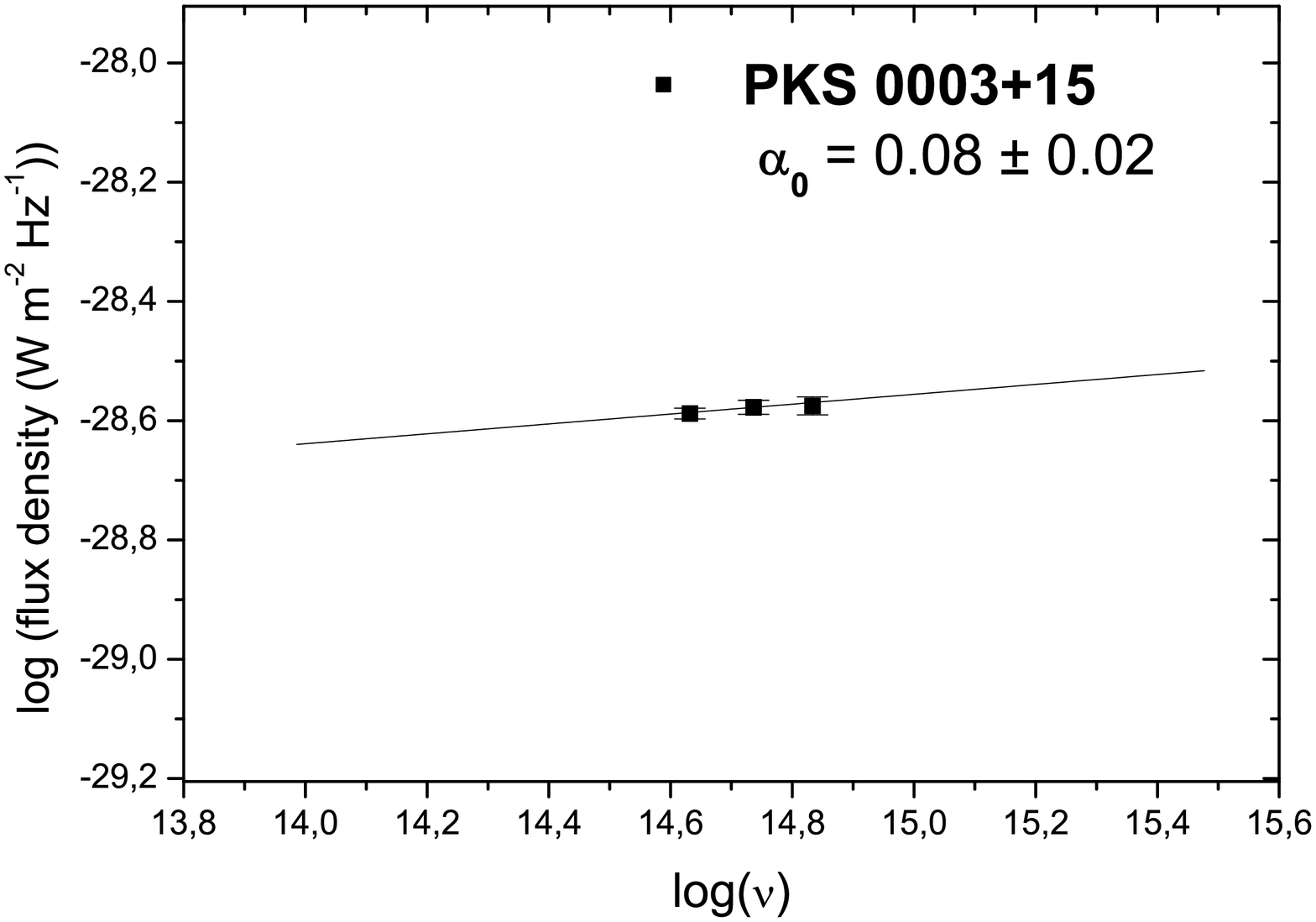}}\figurenum{7-4a}\label{fig84a}}
\subfigure[]{\scalebox{0.17}{\includegraphics{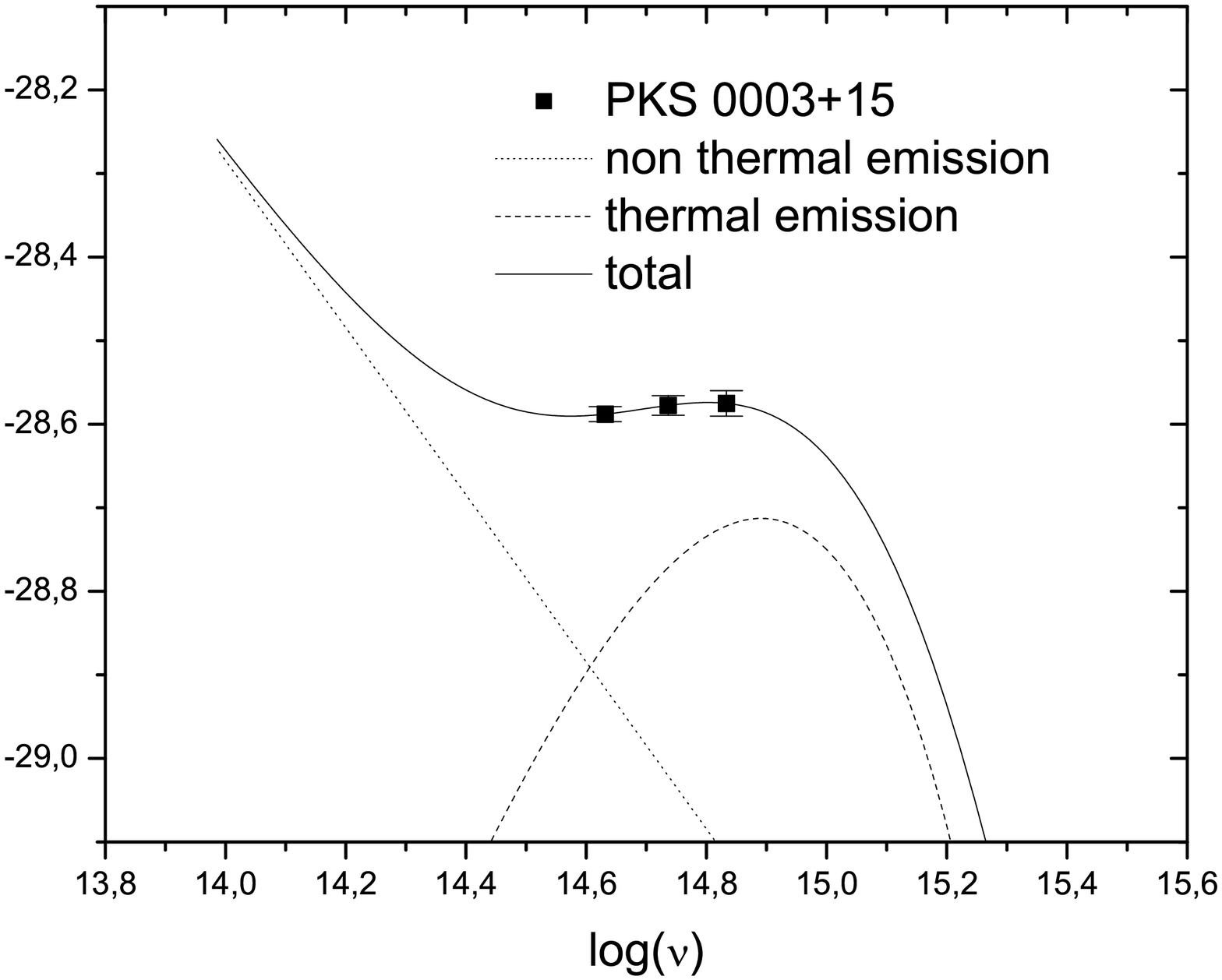}}\figurenum{7-4b}\label{fig84b}}
\subfigure[]{\scalebox{0.28}{\includegraphics{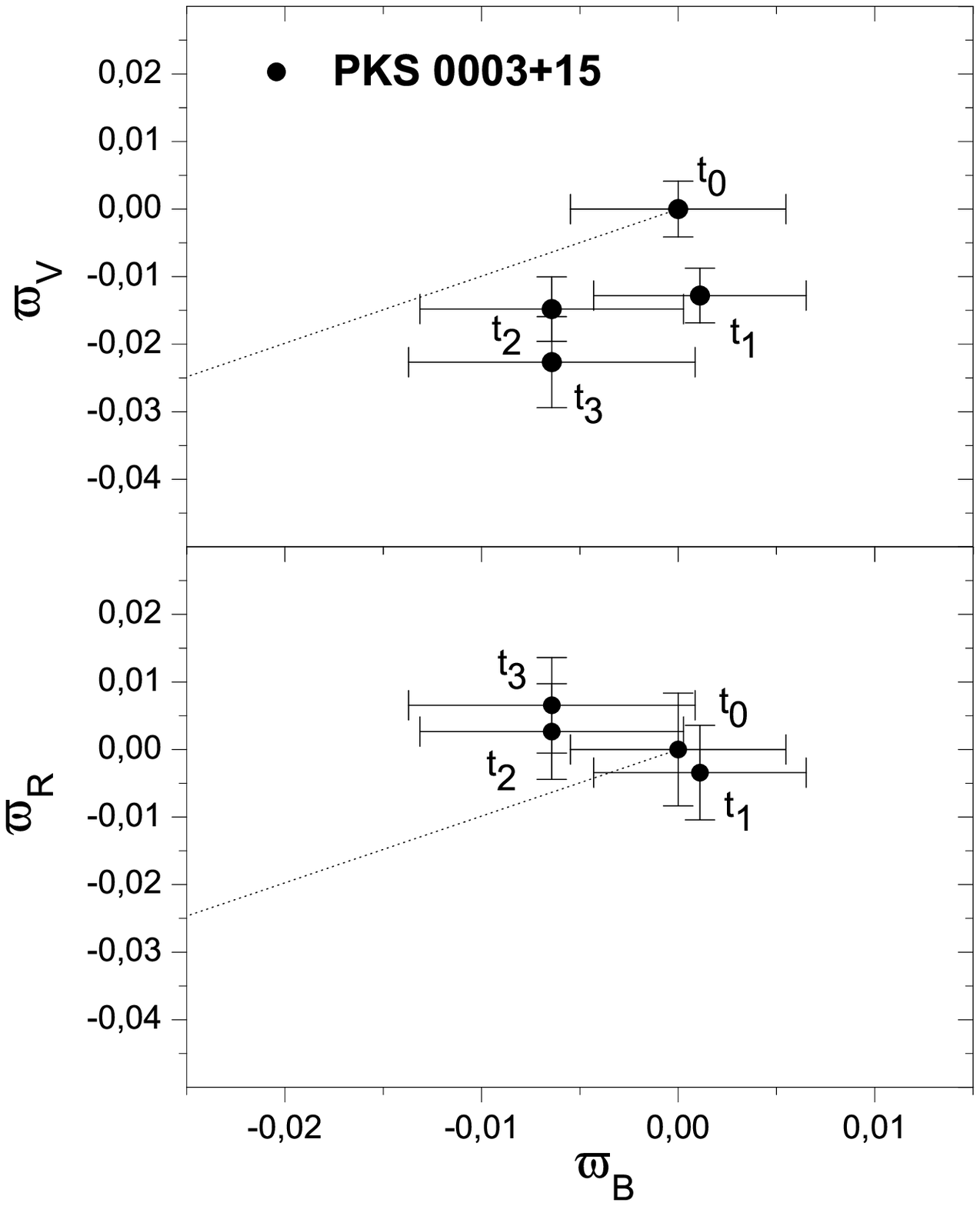}}\figurenum{7-4c}\label{fig84c}}
\subfigure[]{\scalebox{0.28}{\includegraphics{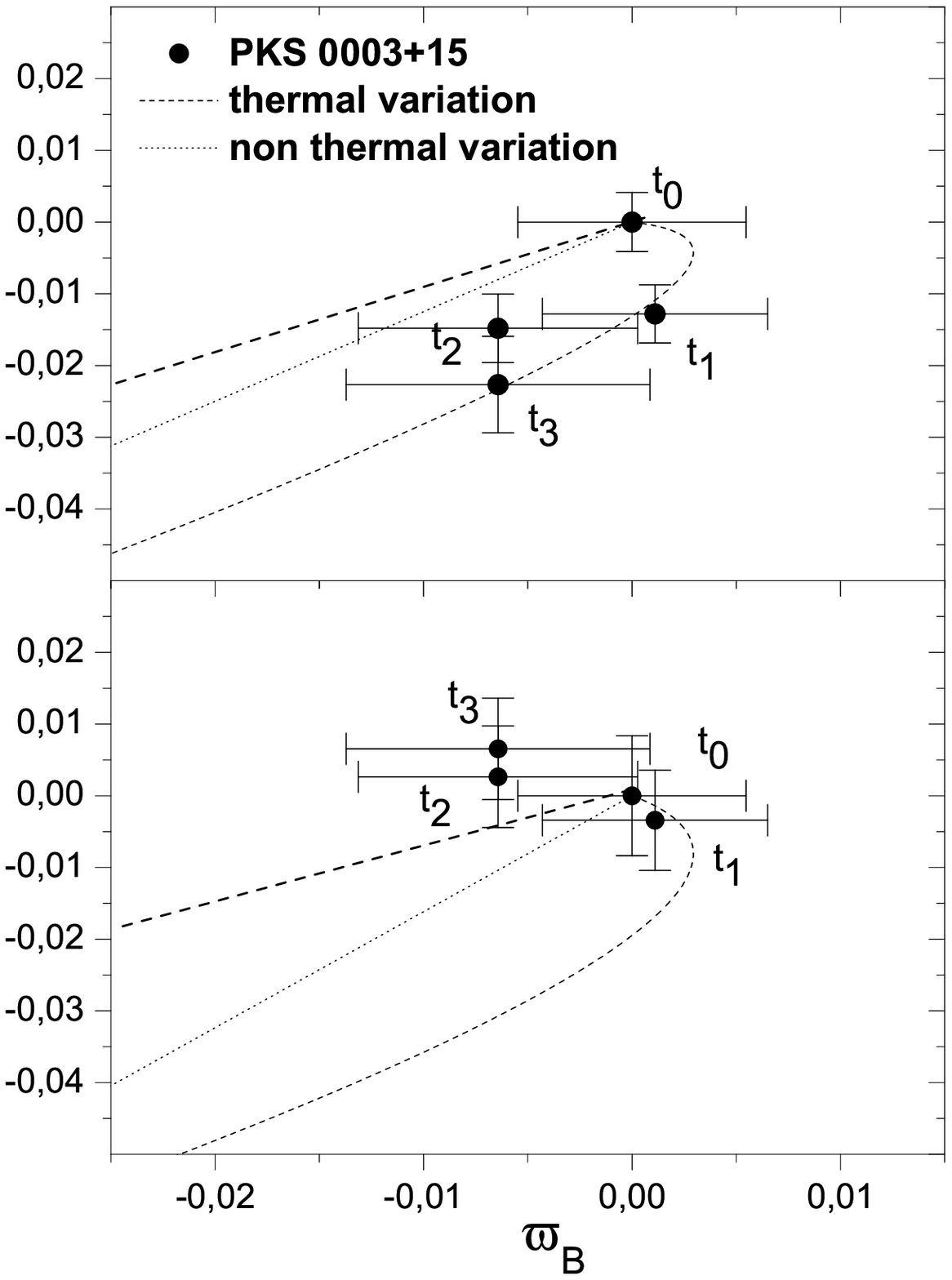}}\figurenum{7-4d}\label{fig84d}}
\caption{\sl \footnotesize Fits and variability: continuation.}
\label{fig84}
\end{center}
\end{figure*}

{\bf PKS~0003+15.} At the beginning of the observations we measured $m_V = 15.36 \pm 0.02$, brighter than $m_V=15.96$ reported by \citet{Barvainis05}. A single non-thermal component fits the first data set with $\alpha_{t_0} = 0.08 \pm 0.02$ (Figure \ref{fig84a}). However, this fit cannot describes satisfactory the color variation (Figure \ref{fig84c}), and a second component must be added. The new fit has $a_{V t_0} \sim 36\%$ and $T_{t_0} \sim 19,200$ K (Figures \ref{fig84b}). Both thermal and non-thermal changes may account for the showed variations. The thermal variation is represented by the thick line in Figure \ref{fig84d}. With a thermal variation, as the $BV$ data suggest (superior panel of Figure \ref{fig84d}), temperature could have increased in $\sim 1~200$ K, and amplitude decreased by $15\%$ ($n_{T_{t_3}} \sim 0.85$). This implies a rate of change of $380$ K $hr^{-1}$ in $T$ and $0.05 h^{-1}$ for $n_T$. The simultaneity criterion indicates that this spectral behavior is reliable ($S_B=0.02$, $S_V=0.19$, and $S_R=0 .01$). However, this change would produce a detectable variation in the $R$ band, which is not detected. Perhaps these data have not the required quality, or our {\it simplified model} is insufficient for explaining this behavior.

\subsection{Radio quiet quasars}\label{RQQ}

\begin{figure*}[ht]
\begin{center}
\figurenum{7-5}
\subfigure[]{\scalebox{0.17}{\includegraphics{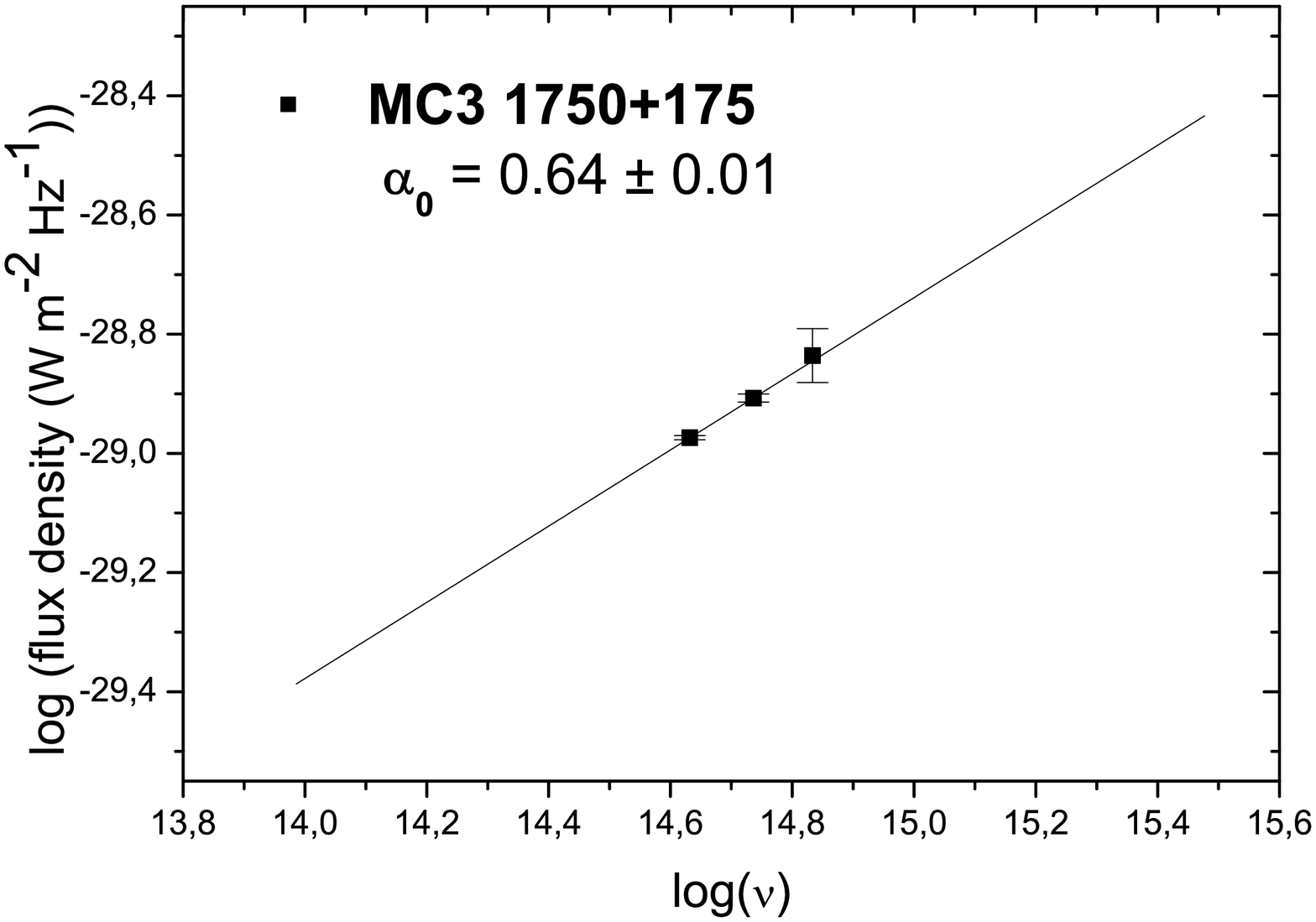}}\figurenum{7-5a}\label{fig85a}}
\subfigure[]{\scalebox{0.17}{\includegraphics{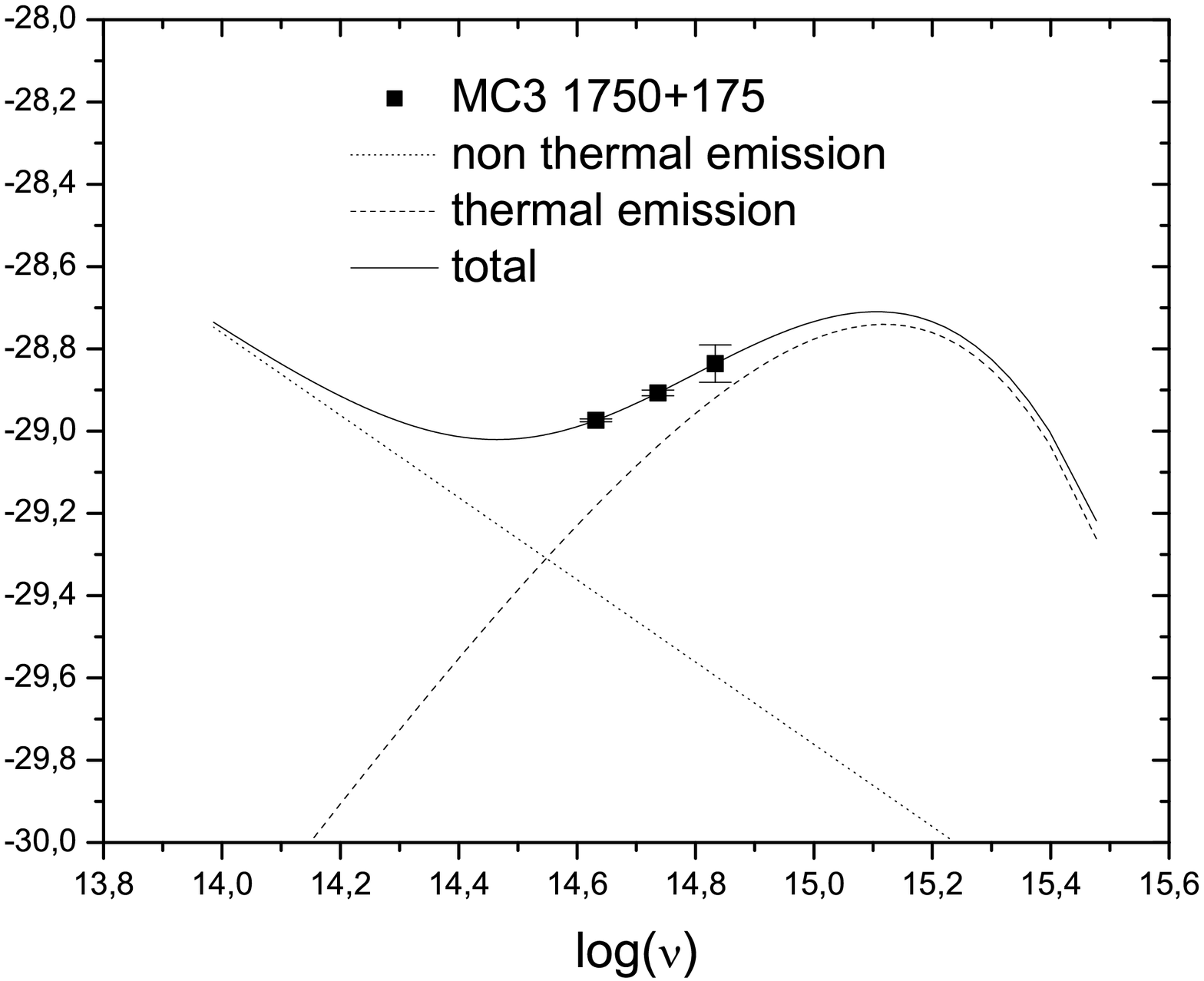}}\figurenum{7-5b}\label{fig85b}}
\subfigure[]{\scalebox{0.28}{\includegraphics{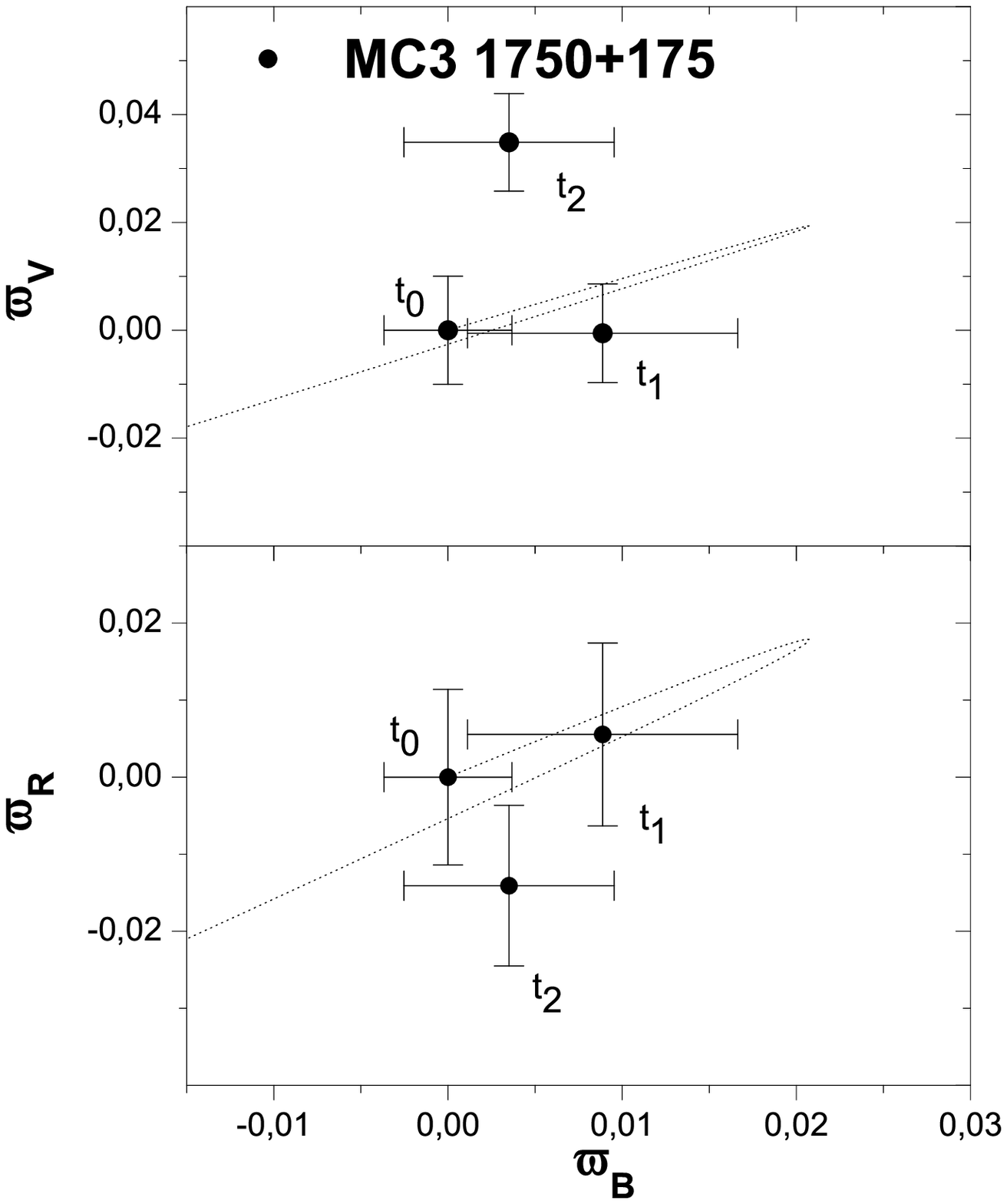}}\figurenum{7-5c}\label{fig85c}}
\subfigure[]{\scalebox{0.28}{\includegraphics{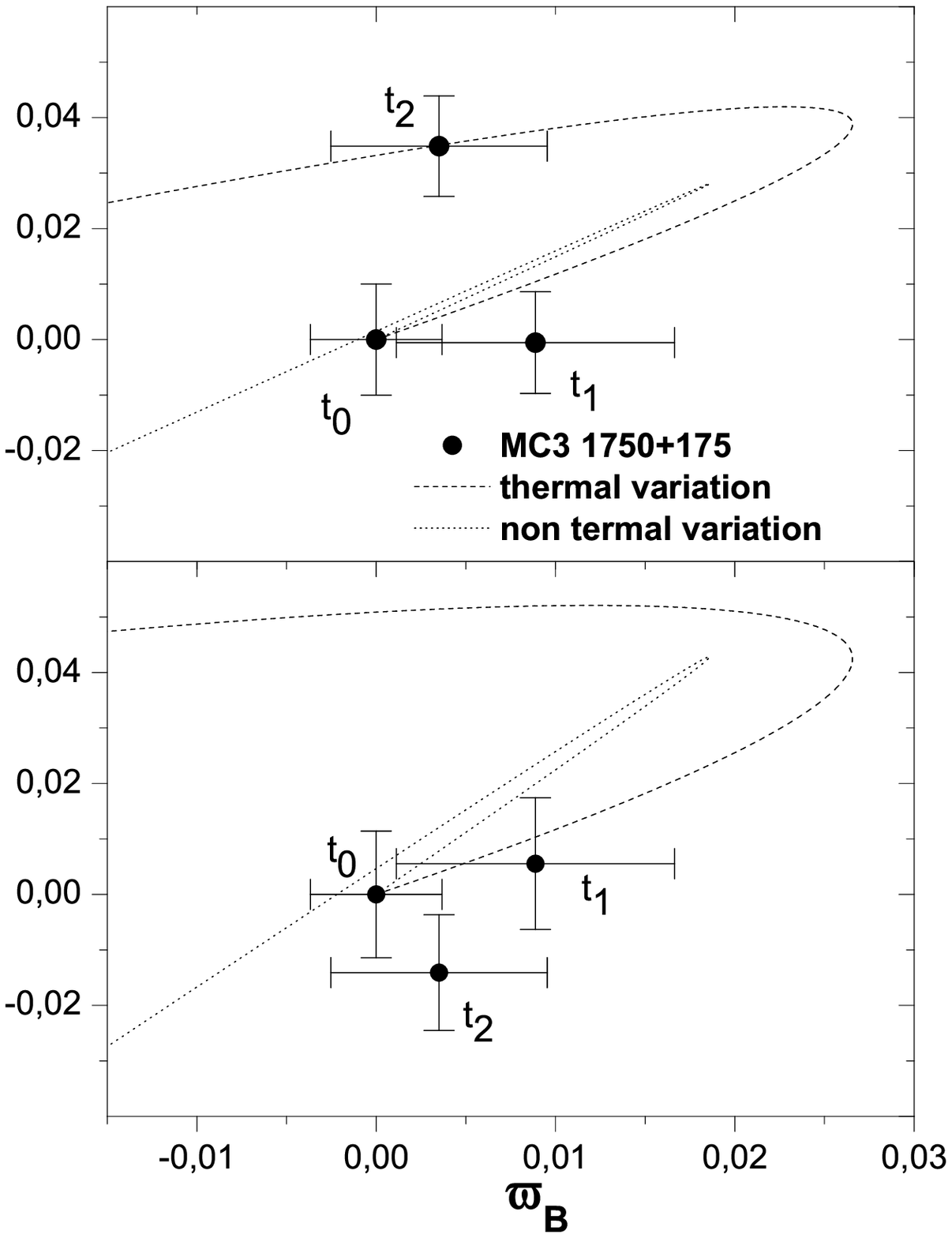}}\figurenum{7-5d}\label{fig85d}}
\caption{\sl \footnotesize Fits and variability: continuation.}
\label{fig85}
\end{center}
\end{figure*}

{\bf MC3~1750+175.} For this object, we measure $m_V=16.18 \pm 0.02$ at the beginning of the observations, i.e., brighter than the value reported by \citet{Veron01} ($m_V=16.60$). A power law with $\alpha_{t_0} = 0.64 \pm 0.02$ fits this first data set (Figure \ref{fig85a}). However, a second component is necessary to reproduce the observed spectral variation (Figure \ref{fig85c}). A mix of thermal and non-thermal components fits the data with $T_{t_0} \sim 33,700$ K and $a_{V t_0} \sim 27\%$ (Figure \ref{fig85b}). Non-thermal changes cannot explain this variation. Nevertheless, changes in thermal component only can explain the $BV$ data set (Figure \ref{fig85d}). As it happened in the case of PKS~0003+15, a variation should be observed in the $R$ band. A variation in the thermal component requires a temperature fall of $\sim 5~600$ K, while the amplitude should increase by $44\%$ ($n_{T_{t_2}} \sim 1.44$) (as it is suggested from the data in the superior panel of Figure \ref{fig85d}). In such a case, the simultaneity criterion indicates the reliability of the spectral evolution ($S_B=0.01$, $S_V=0.21$, and $S_R=0.10$); however, this variation is something much more complicated to describe with our {\it toy model}, as it is easy to see in the Figure \ref{fig85d}.

\begin{figure*}[ht]
\begin{center}
\figurenum{7-6}
\subfigure[]{\scalebox{0.17}{\includegraphics{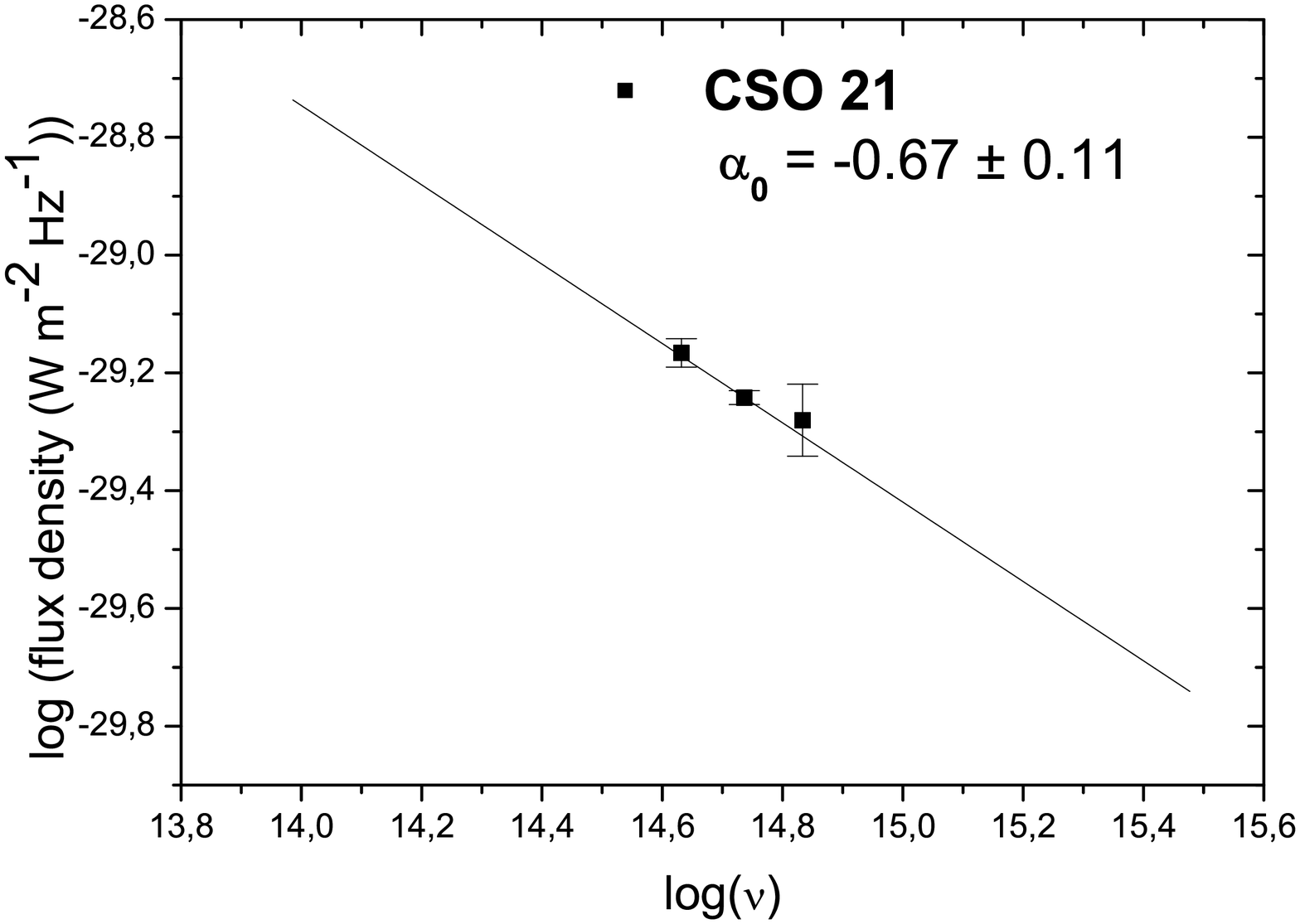}}\figurenum{7-6a}\label{fig86a}}
\subfigure[]{\scalebox{0.17}{\includegraphics{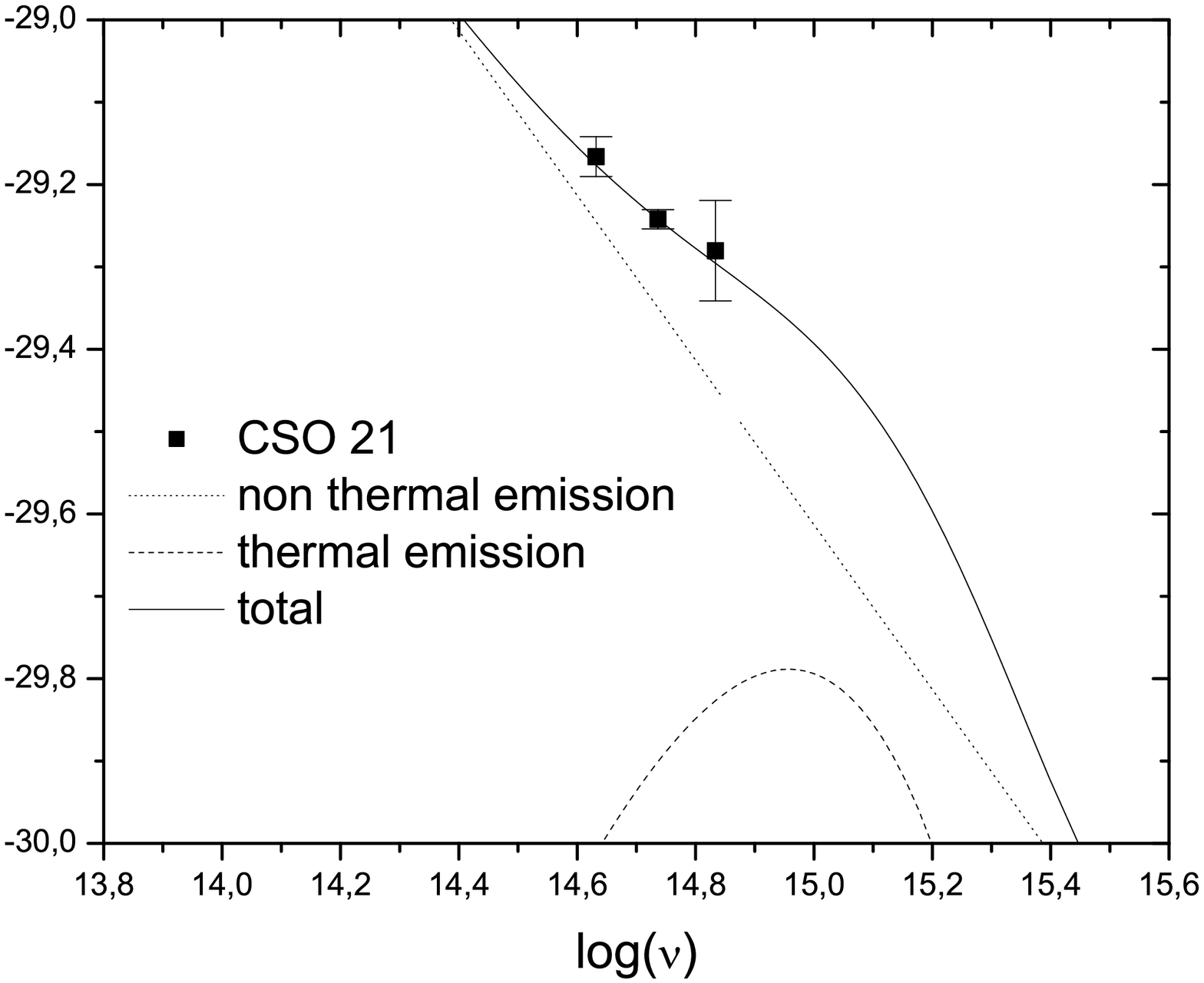}}\figurenum{7-6b}\label{fig86b}}
\subfigure[]{\scalebox{0.28}{\includegraphics{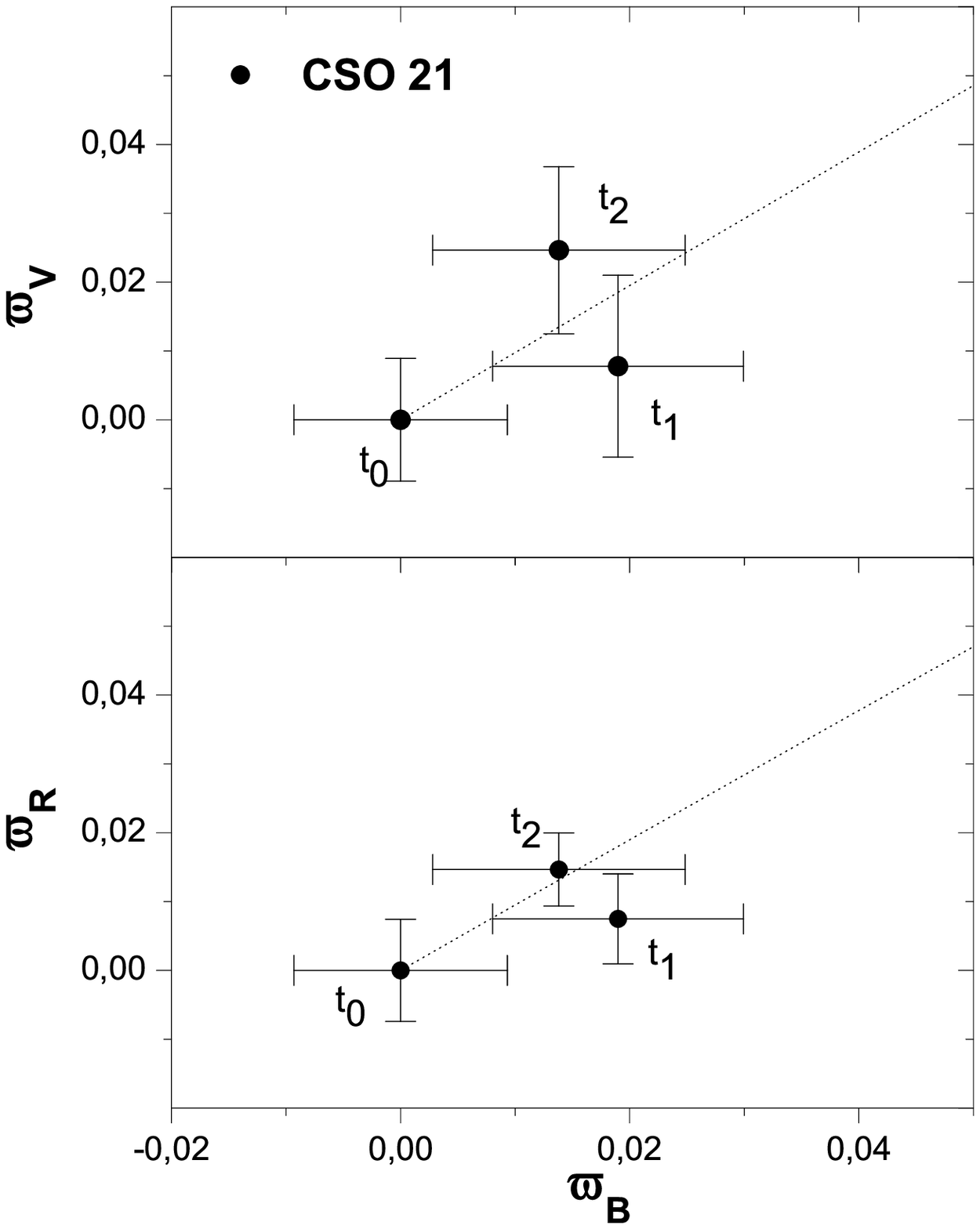}}\figurenum{7-6c}\label{fig86c}}
\subfigure[]{\scalebox{0.28}{\includegraphics{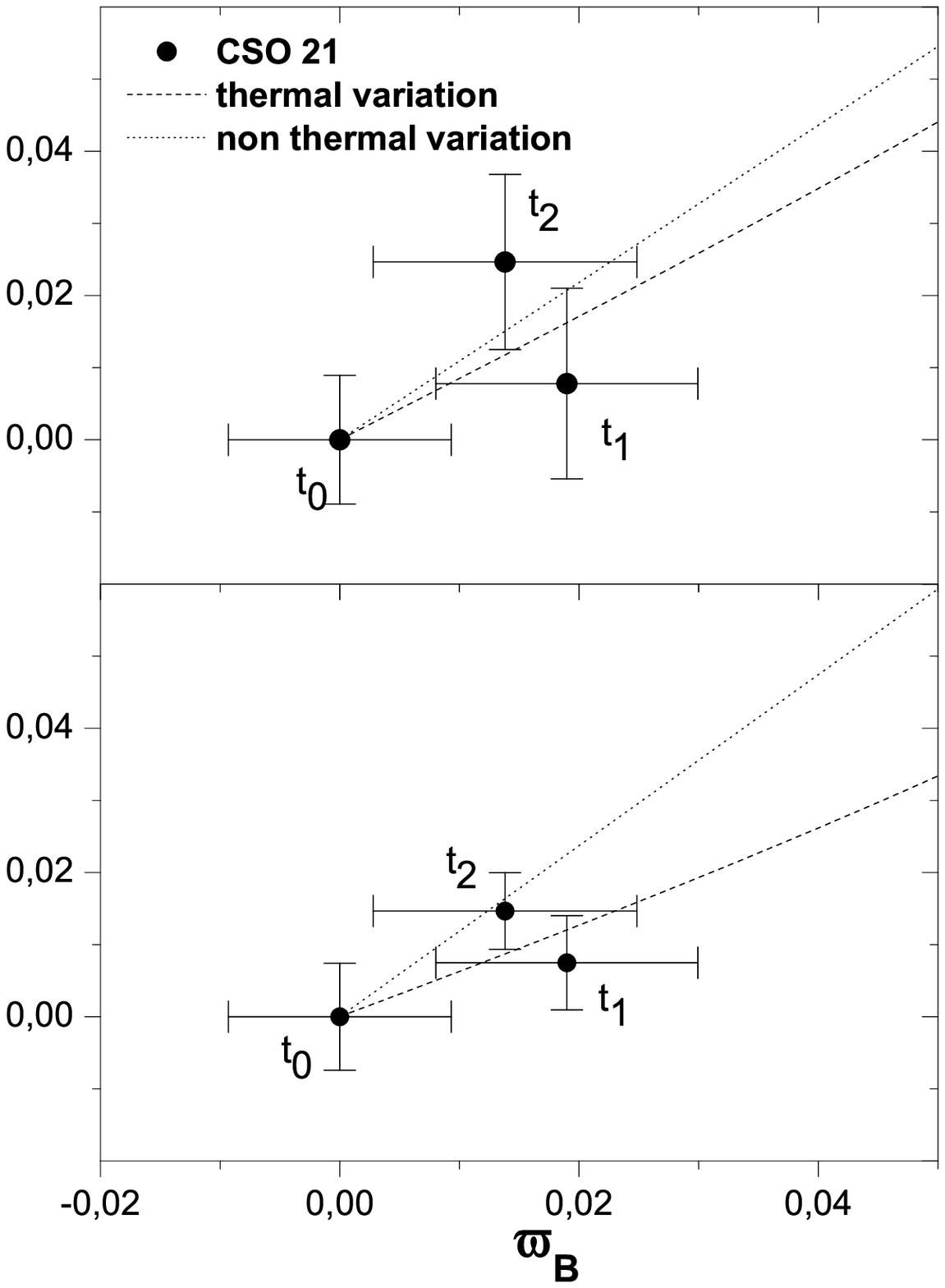}}\figurenum{7-6d}\label{fig86d}}
\caption{\sl \footnotesize Fits and variability: continuation.}
\label{fig86}
\end{center}
\end{figure*}

{\bf CSO~21.} For this object, we measure a brightness of $m_V=17.02 \pm 0.03$ at the beginning of the observations, which is similar to the value reported by \citet{Veron01}, $m_V=17.0$. This first set of data is fitted with $\alpha_{V t_0} = -0.67 \pm 0.11$ (Figure \ref{fig86a}). The variation detected can be explained by an increase of $0.002$ of the spectral index and a decrease in amplitude of $\sim 4\%$ (Figure \ref{fig86c}). It is possible that the OM detected for the $V$ and $R$ bands is also present in the $B$ band, but the variation in this band was not detected due to observational errors. The simultaneity criterion indicates that the level of brightness had not changed during observations of a $BVR$ sequence ($S_B=0.07$, $S_V=0.35$, and $S_R=0.11$). For simplicity, we first suppose that this OM event results from a variation in a non-thermal unique component. Nevertheless, the possibility of a second component is revised. In such a case, the temperature of the thermal component would be $T_{t_0} \sim 30,000$ K, and $a_{V t_0} \sim 78\%$ (Figure \ref{fig86b}). Including this thermal component the variations may be generated by any component (Figure \ref{fig86d}). In the case of a thermal change, the temperature should fall $400$ K, while amplitude would increase by $10\%$ ($n_{T_{t_2}} \sim 1.10$). For the case of variations of the non-thermal component, the spectral index stays constant, and the amplitude increases to $n_{n_{t_2}} \sim 1.02$.

\begin{figure*}[ht]
\begin{center}
\figurenum{7-7}
\subfigure[]{\scalebox{0.17}{\includegraphics{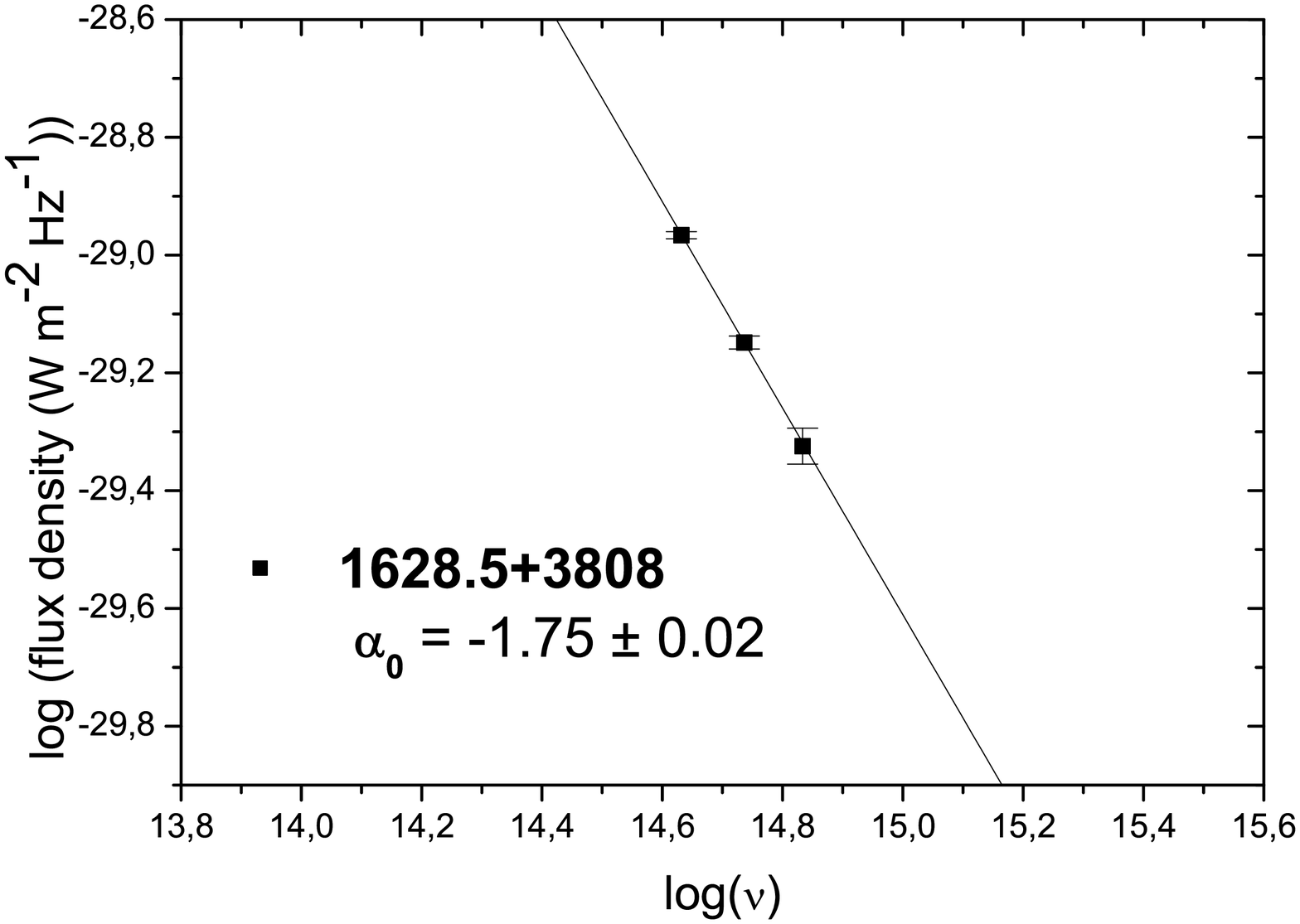}}\figurenum{7-7a}\label{fig87a}}
\subfigure[]{\scalebox{0.17}{\includegraphics{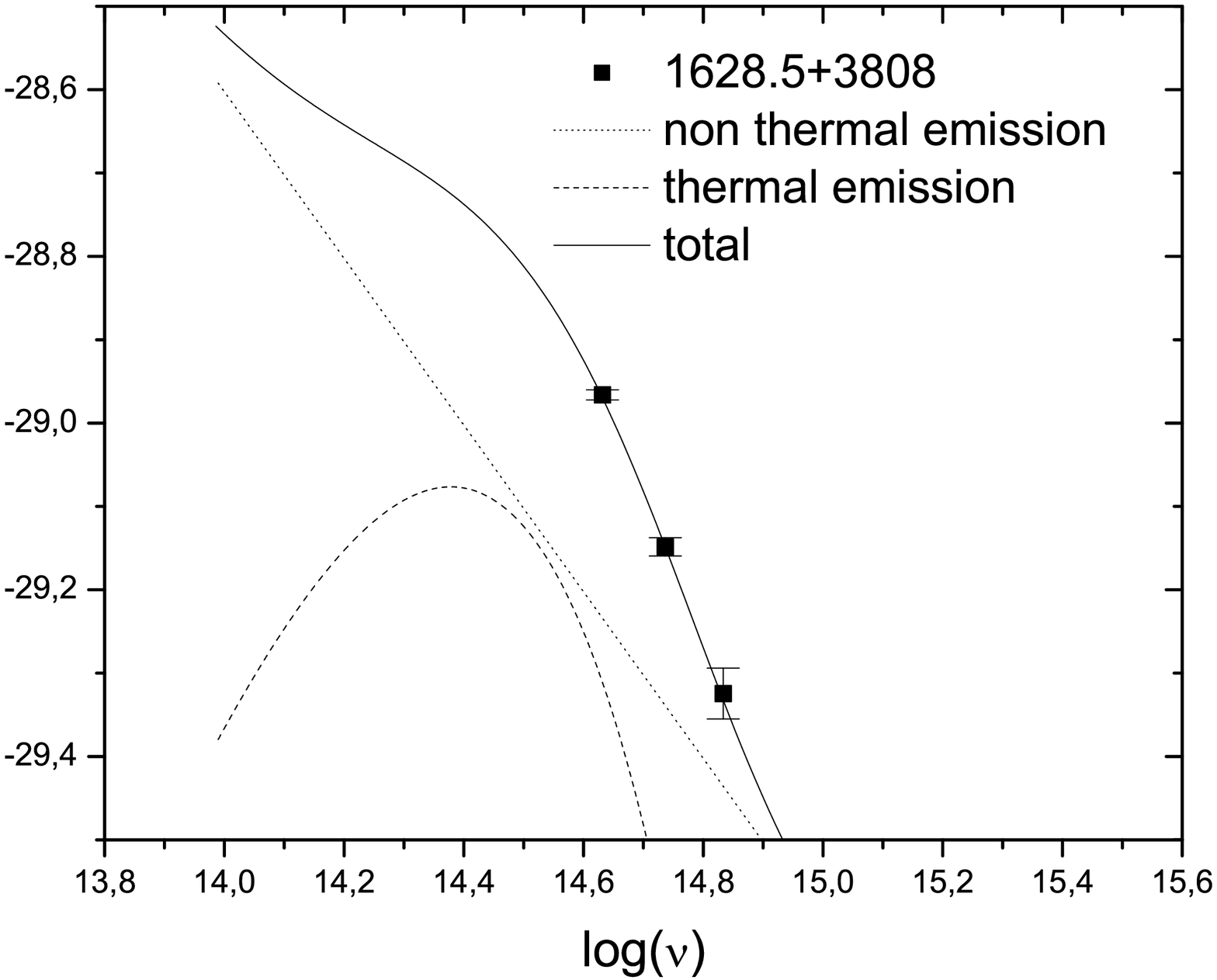}}\figurenum{7-7b}\label{fig87b}}
\subfigure[]{\scalebox{0.28}{\includegraphics{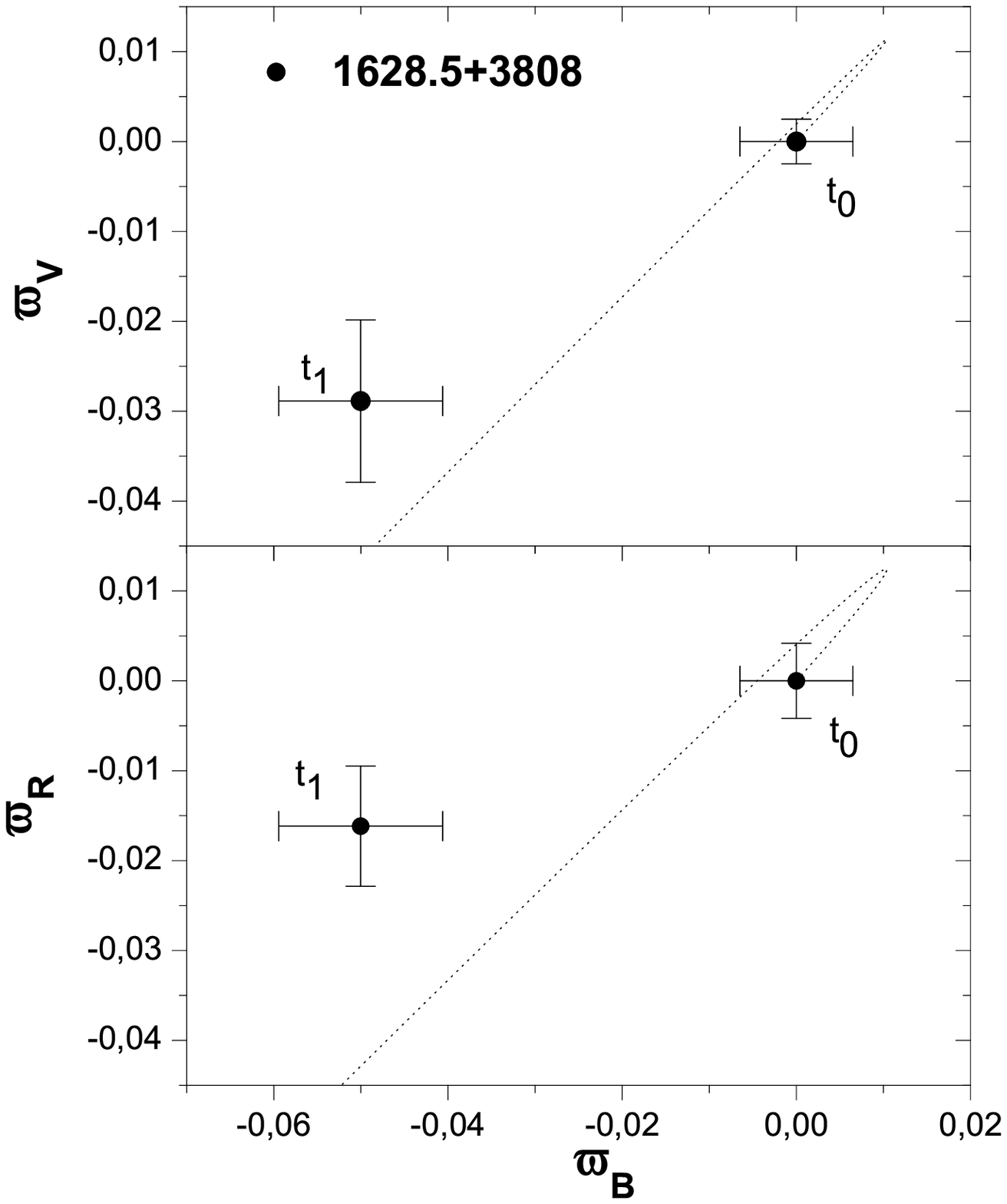}}\figurenum{7-7c}\label{fig87c}}
\subfigure[]{\scalebox{0.28}{\includegraphics{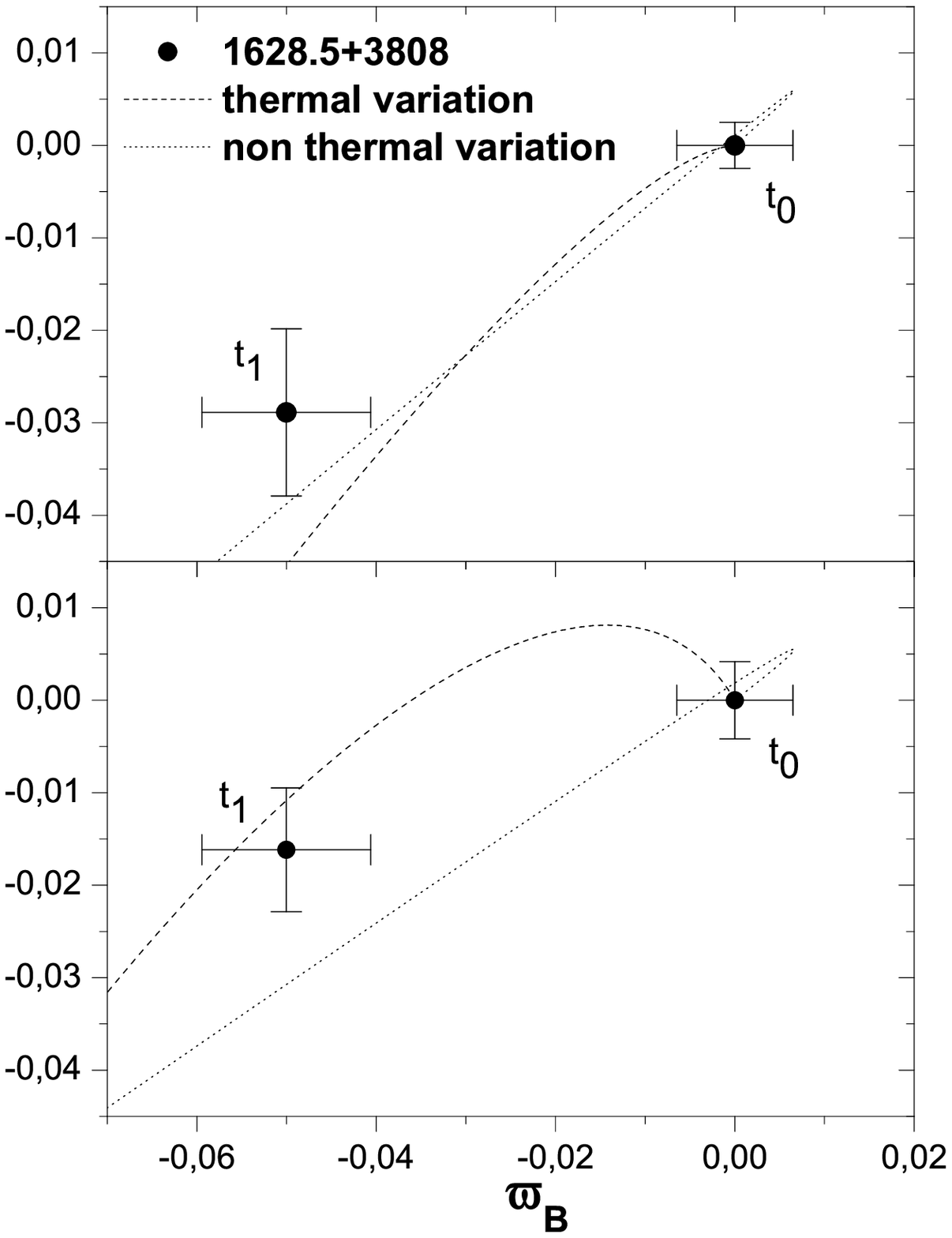}}\figurenum{7-7d}\label{fig87d}}
\caption{\sl \footnotesize Fits and variability: continuation.}
\label{fig87}
\end{center}
\end{figure*}

{\bf 1628.5+3808.} The magnitude of this quasar was $m_V=16.78 \pm 0.02$ at the beginning of the observations, i.e., similar to the reported brightness $m_V=16.80$ by \citet{Veron01}. A single non-thermal component fits these data, with $\alpha_{t_0} = -1.75 \pm 0.02$ (Figure \ref{fig87a}). However, the spectral variation cannot be reproduced by changes in this component (Figures \ref{fig87c}). Then, a thermal component is added to model the data with $T_{t_0} \sim 10,000$ K and $a_{V t_0} \sim 65\%$ (Figure~\ref{fig87b}). A variation of non-thermal origin can explain marginally the spectral evolution. On the contrary, a thermal variation can reproduce more accurately the observed behavior (Figure~\ref{fig87d}). In such a case, $T$ would have fallen by $\sim 760$ K, while $n_{T_{t1}} \sim 1.48$. According to the simultaneity criterion, the spectral variation should be real ($S_B=0.16$, $S_V=0.12$, and $S_R=0.14$).

\begin{figure*}[ht]
\begin{center}
\figurenum{7-8}
\subfigure[]{\scalebox{0.17}{\includegraphics{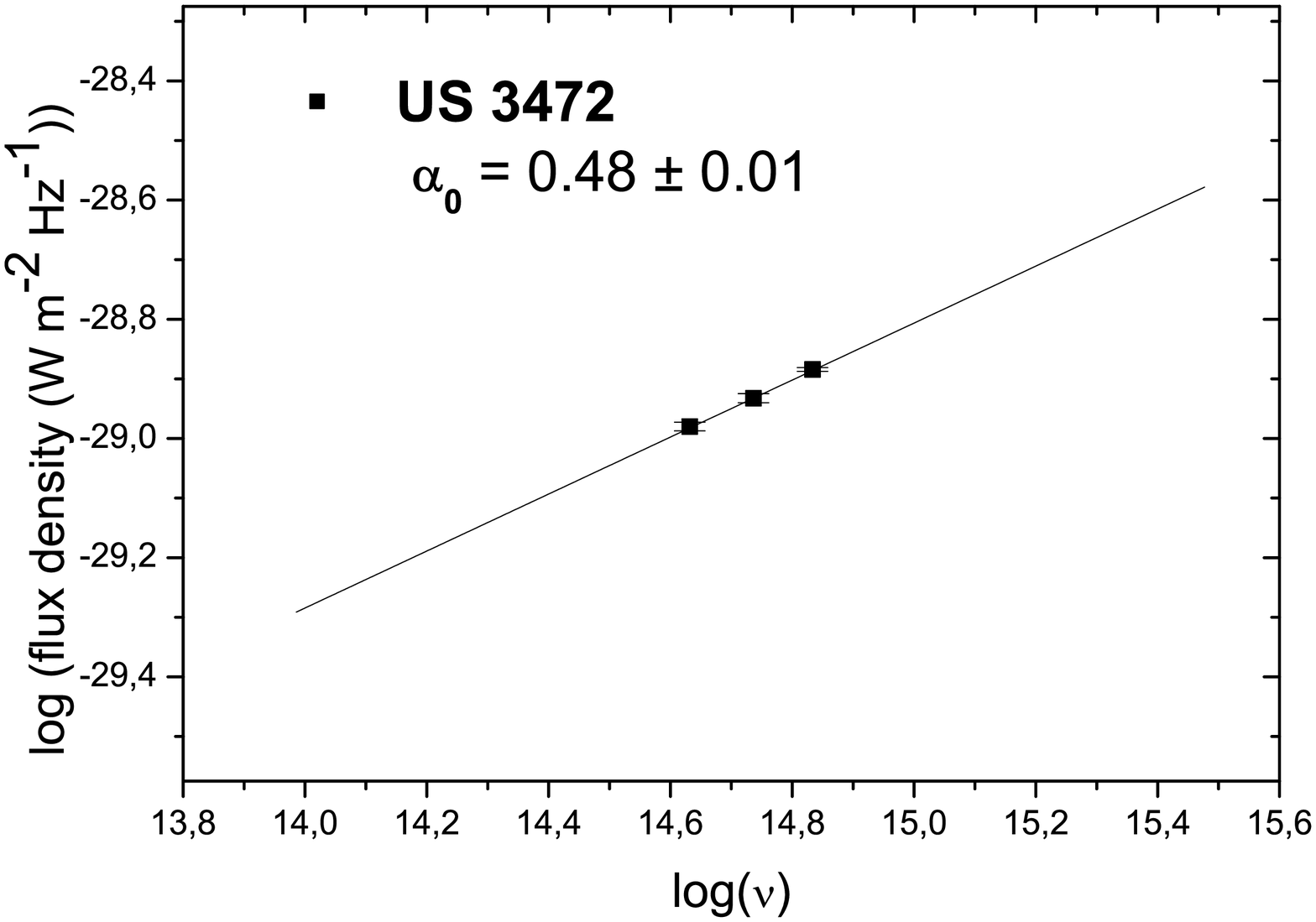}}\figurenum{7-8a}\label{fig88a}}
\subfigure[]{\scalebox{0.17}{\includegraphics{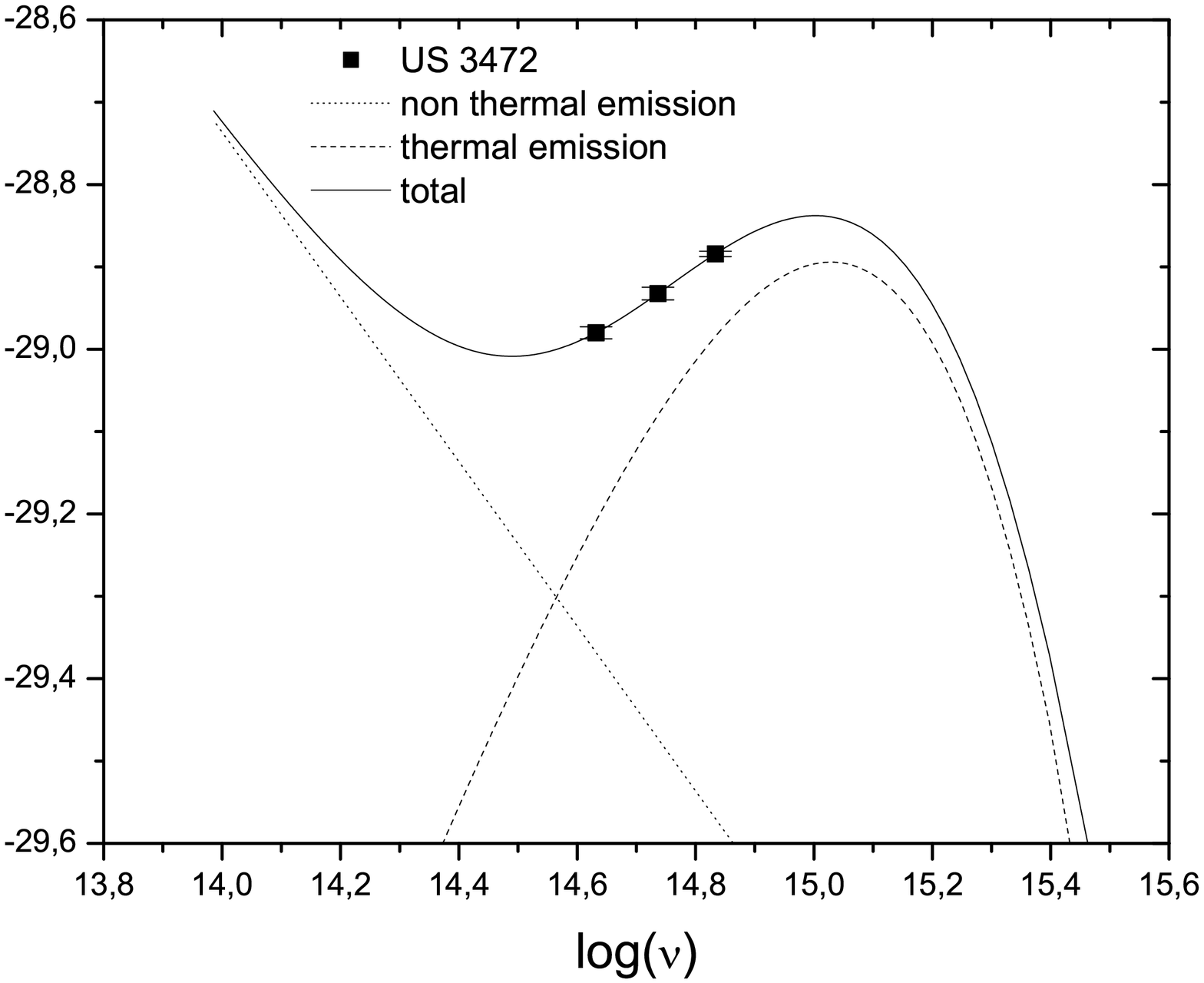}}\figurenum{7-8b}\label{fig88b}}
\subfigure[]{\scalebox{0.28}{\includegraphics{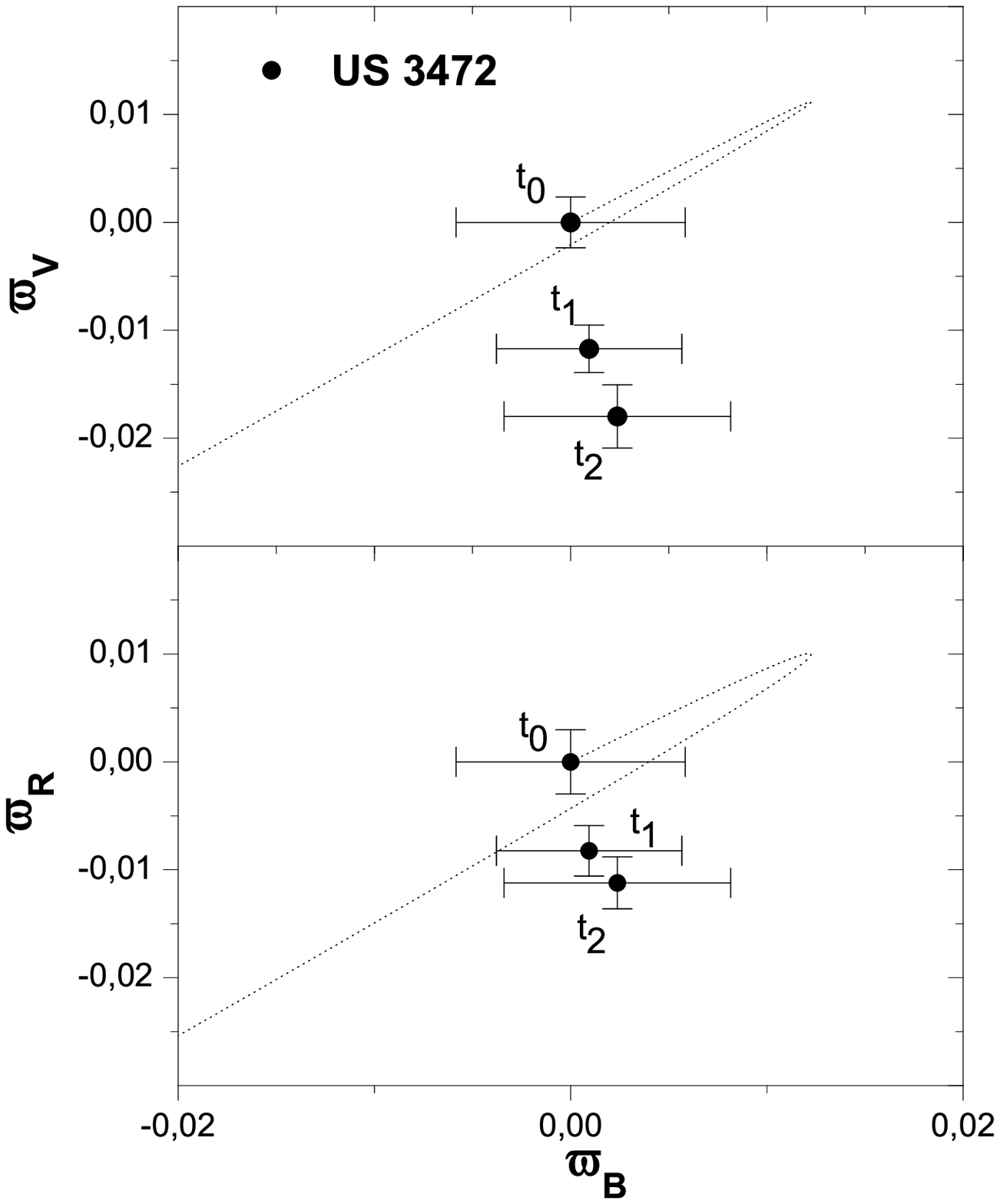}}\figurenum{7-8c}\label{fig88c}}
\subfigure[]{\scalebox{0.28}{\includegraphics{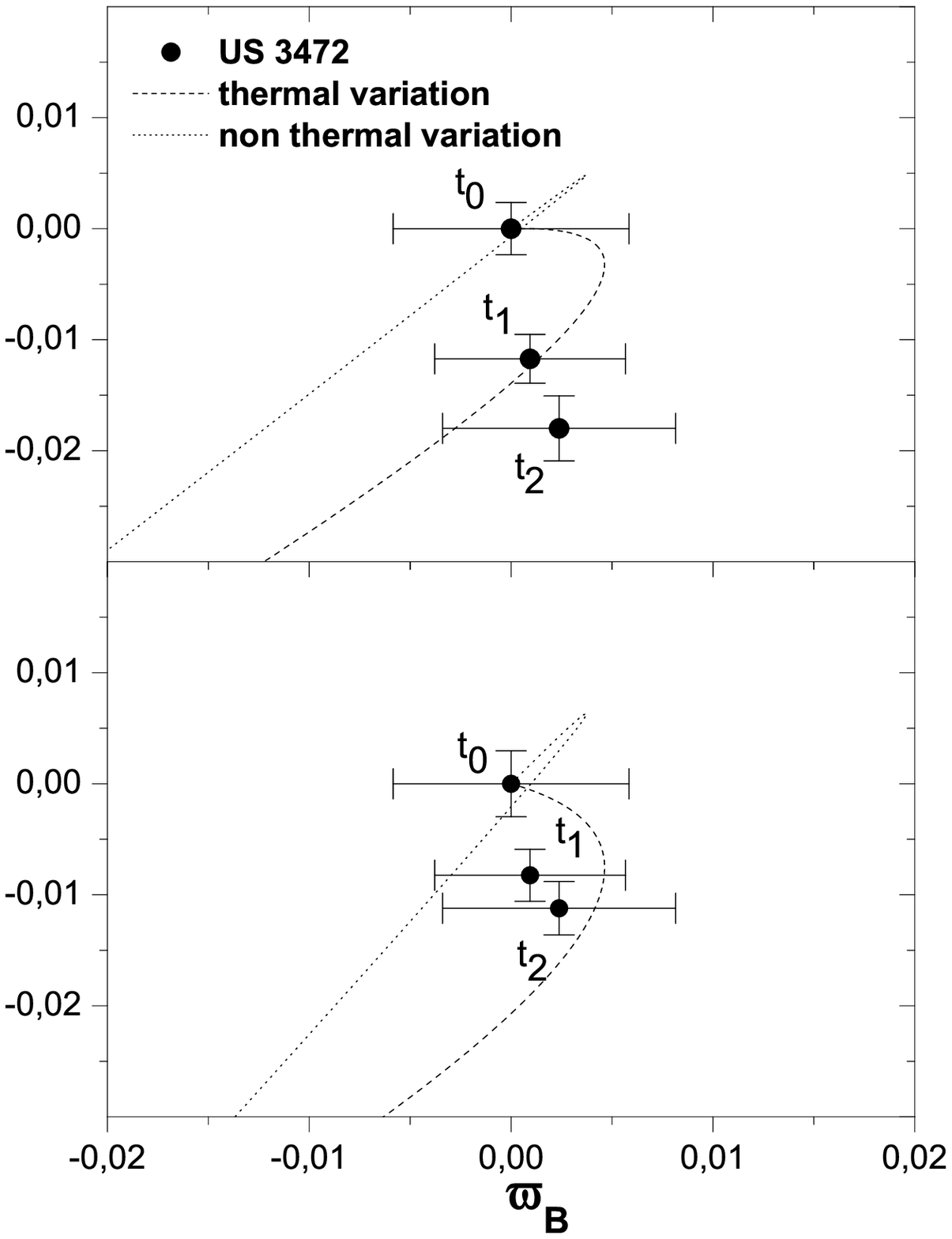}}\figurenum{7-8d}\label{fig88d}}
\caption{\sl \footnotesize Fits and variability: continuation.}
\label{fig88}
\end{center}
\end{figure*}

{\bf US~3472.} When this object was observed, its brightness was $m_V=16.25 \pm 0.01$, i.e., brighter than that reported by \citet{Lafranca92} and \citet{Veron01} ($m_V=16.99$). Initial data were fitted by a power law with  $\alpha_{t_0} = 0.48 \pm 0.01$ (Figure \ref{fig88a}). Nevertheless, with this arrangement we cannot give an explanation for the color variation shown in the Figure \ref{fig88c}. The mixture of a blackbody with $T_{t_0}\sim 27,900$ K and a non-thermal component with $a_{V t_0} \sim 29\%$ can also fit the data (Figure~\ref{fig88b}). While a change of the non-thermal component cannot explain the observed variation, a thermal variation can do it (Figure~\ref{fig88d}). However, the changes in $T$ and $n_T$ should be larger in $V$ than in $R$. In $V$, the temperature would have changed $\sim 2~000$ K with $n_{T_{t_2}} \sim 0.87$, while in $R$ the change would be of only $\sim 1~200$ K and $n_{T_{t_2}} \sim 0.92$. According to the changes of brightness in the $V$ band, the rate of the change would be of $\sim 600$ K $hr^{-1}$. The observations between $R$ and $B$ bands were obtained with a difference of $\sim 8$ minutes; then, the temperature difference would be $\sim 80$ K. Considering the first and the second data sets: $S_{Rt_0}=1.12$ and $S_{Rt_1}=1.26$. In such a case, the temperature discrepancy would be explained. The simultaneity criterion indicates that observations were not performed at the same level of brightness (considering the first and last data sets: $S_B=0.05$, $S_V=0.53$ and $S_R=0.96$), because the variation in $R$ is very quick, leading to a spurious spectral variation. In spite of this lack of simultaneity, the spectral microvariation possesses thermal characteristics. This is evident taking into account that the $B$ and $V$ bands data complies with the simultaneity criterion.

\begin{figure}[ht]
\begin{center}
\figurenum{7-9}
\subfigure[]{\scalebox{0.17}{\includegraphics{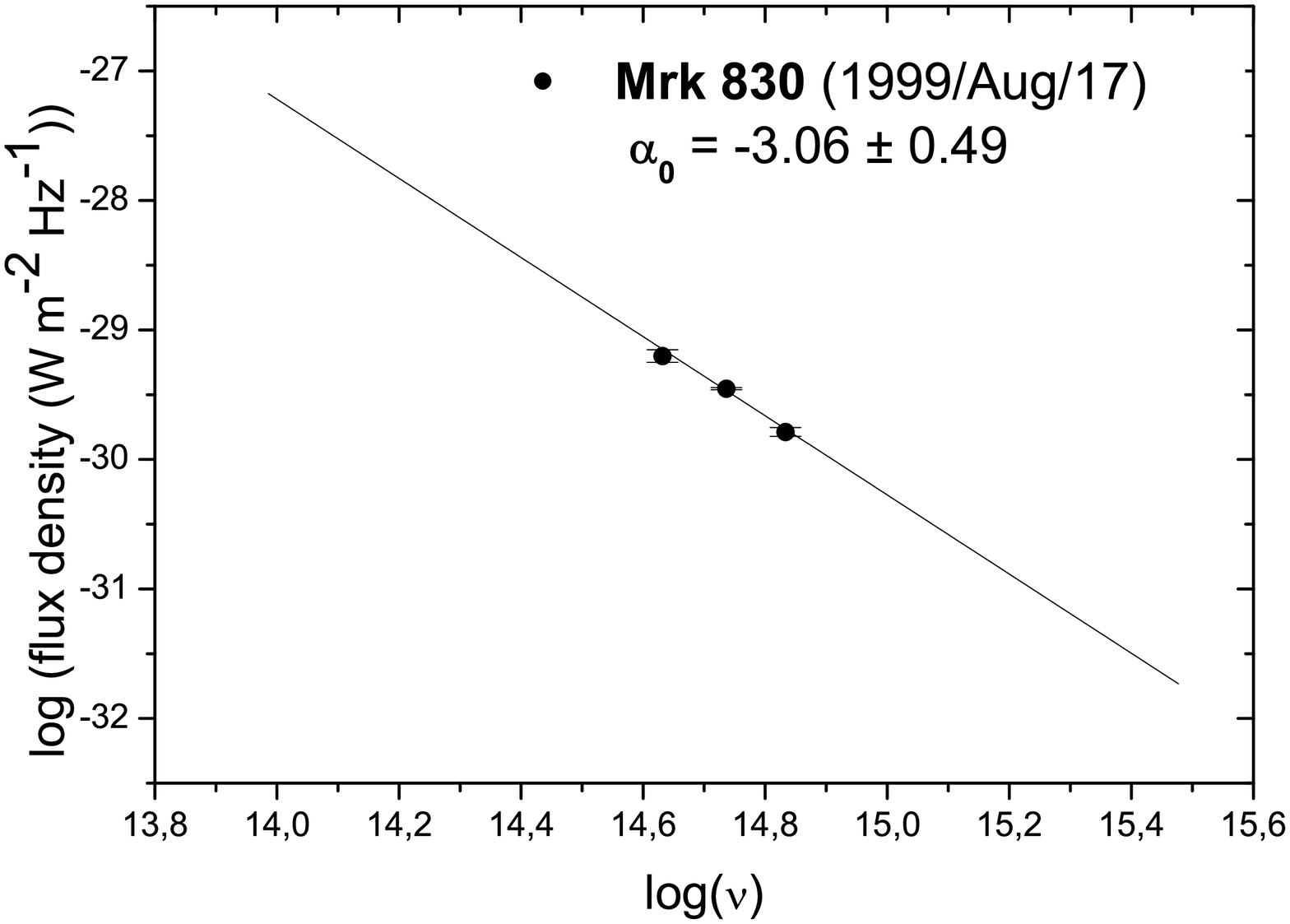}}\figurenum{7-9a}\label{fig89a}}
\subfigure{}
\subfigure[]{\scalebox{0.17}{\includegraphics{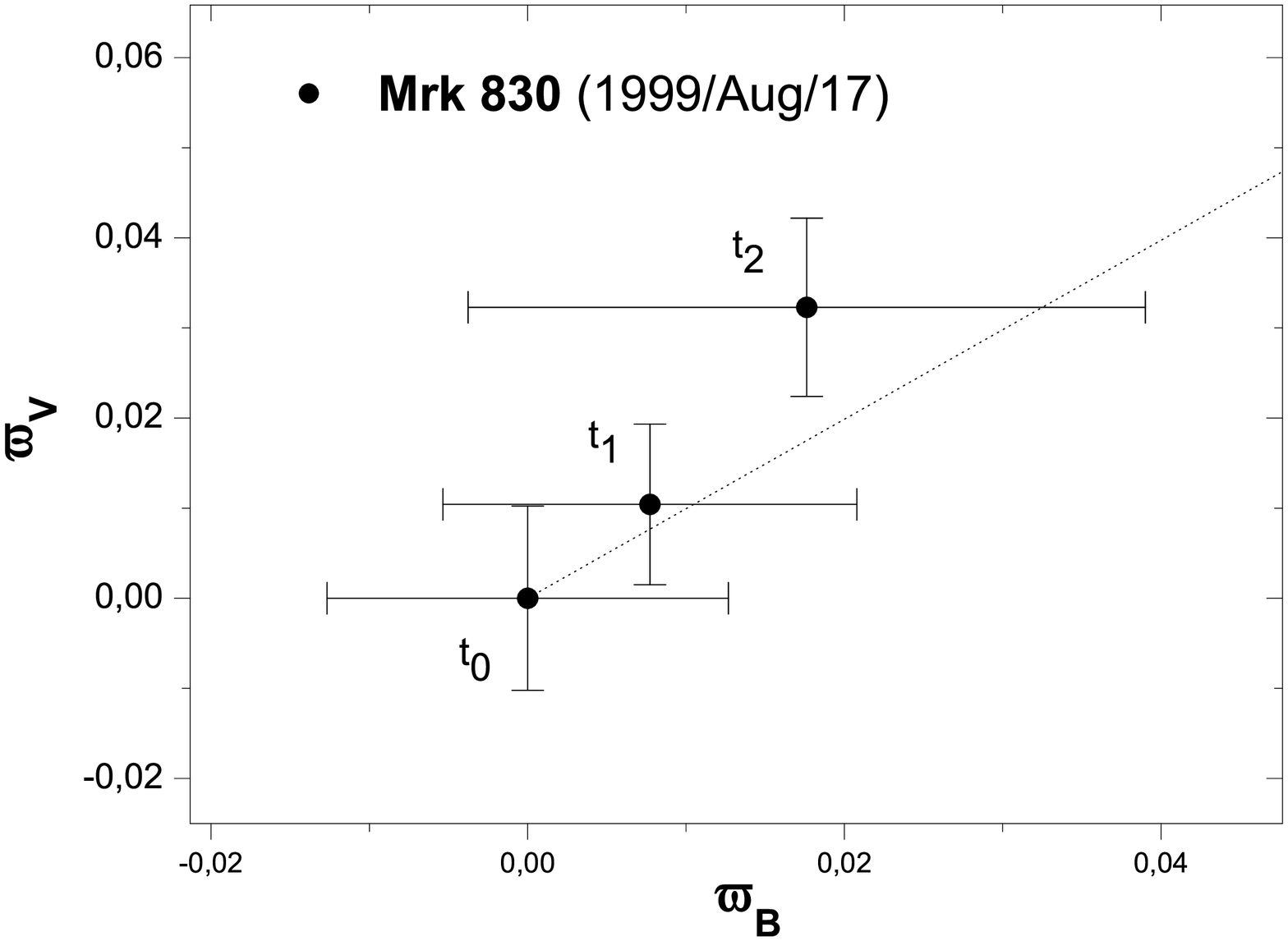}}\figurenum{7-9c}\label{fig89c}}
\caption{\sl \footnotesize Fits and variability: continuation. Since the variation detected on this object can be explained enough using a power law, we avoid to make a second fit. In this night the first set of data in the $R$ band was lost.}
\label{fig89}
\end{center}
\end{figure}

\begin{figure}[ht]
\begin{center}
\figurenum{7-10}
\subfigure[]{\scalebox{0.17}{\includegraphics{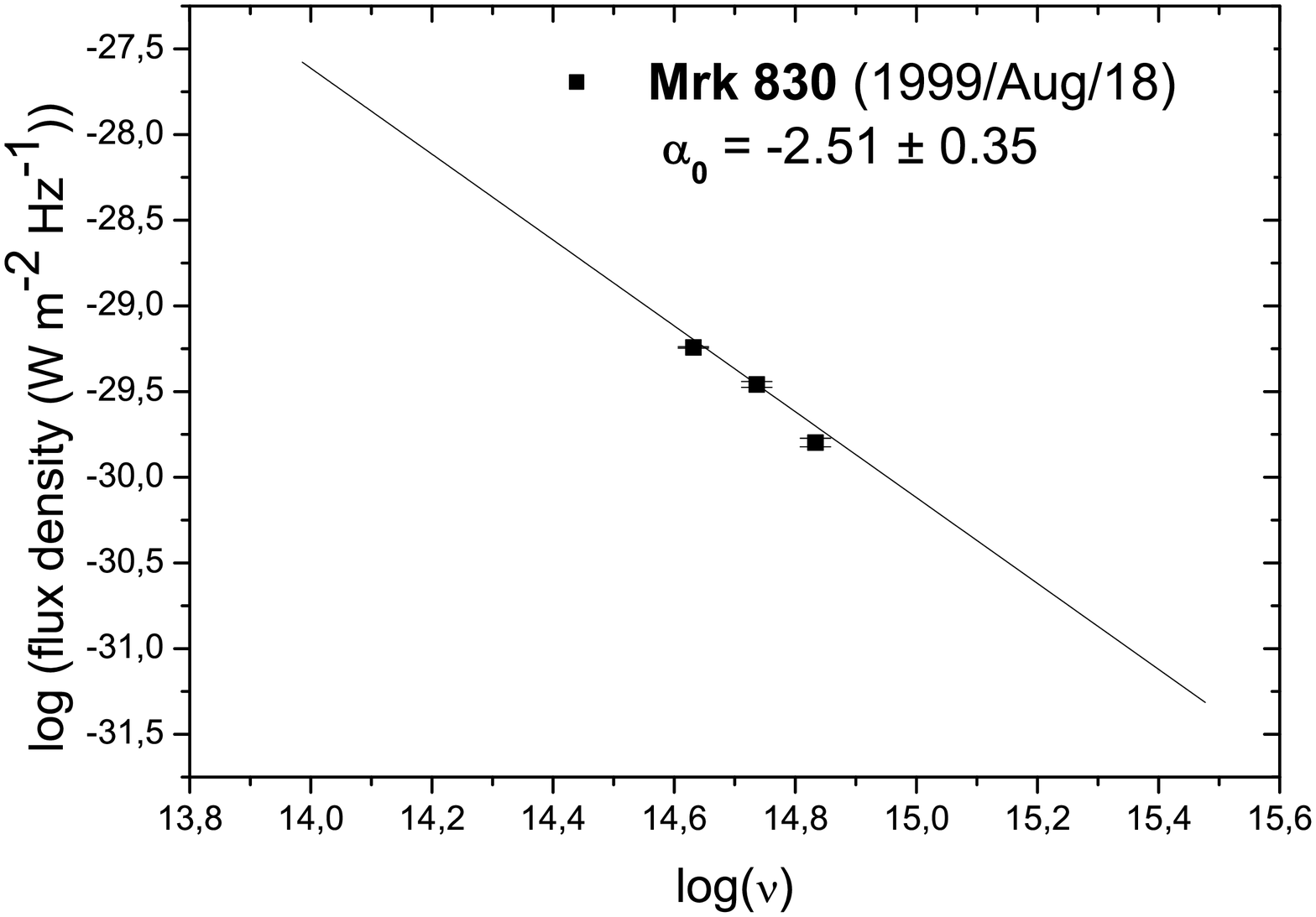}}\figurenum{7-10a}\label{fig810a}}
\subfigure{}
\subfigure[]{\scalebox{0.28}{\includegraphics{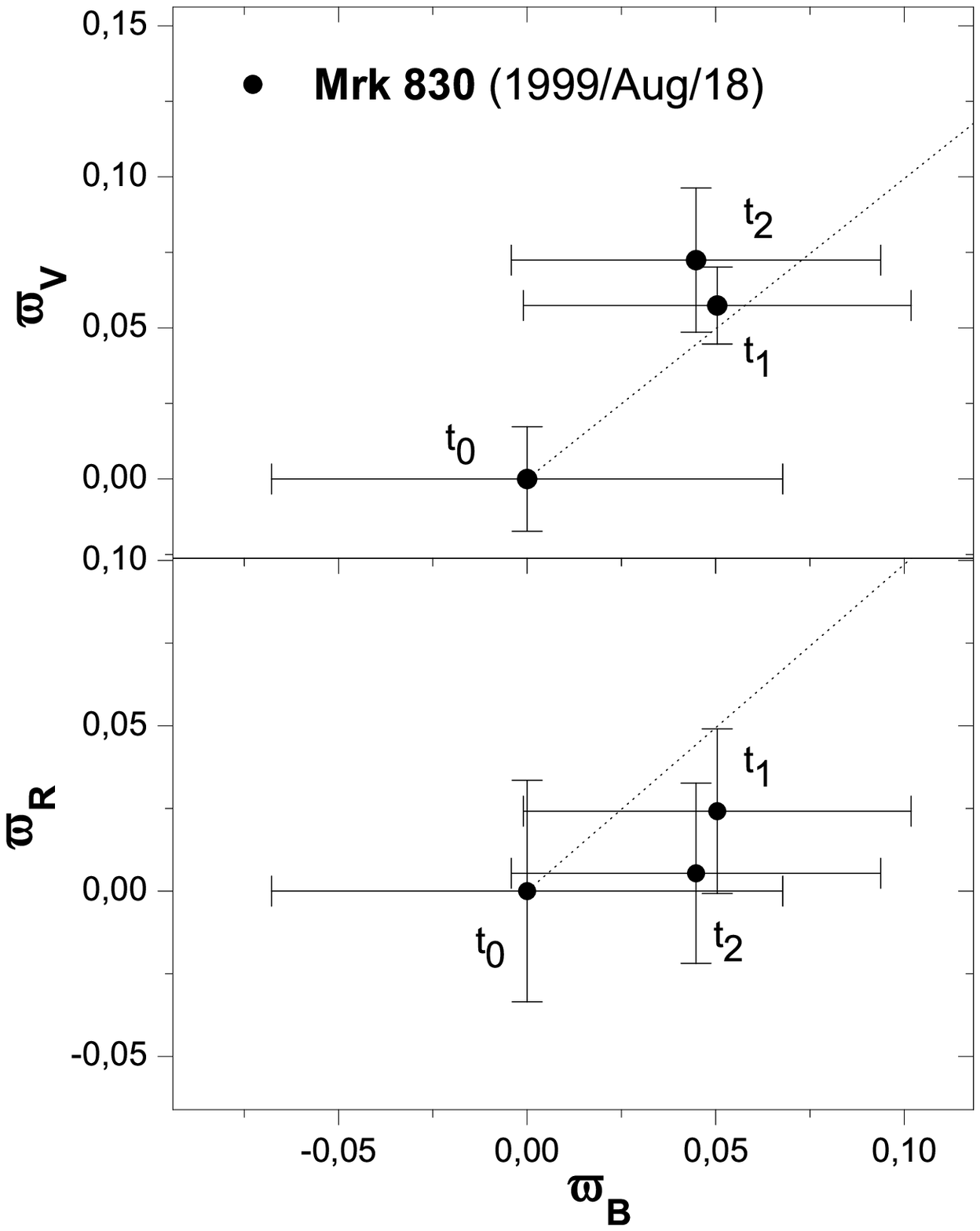}}\figurenum{7-10c}\label{fig810c}}
\caption{\sl \footnotesize Fits and variability: continuation}
\label{fig810}
\end{center}
\end{figure}

\begin{figure}[ht]
\begin{center}
\figurenum{7-11}
\subfigure[]{\scalebox{0.17}{\includegraphics{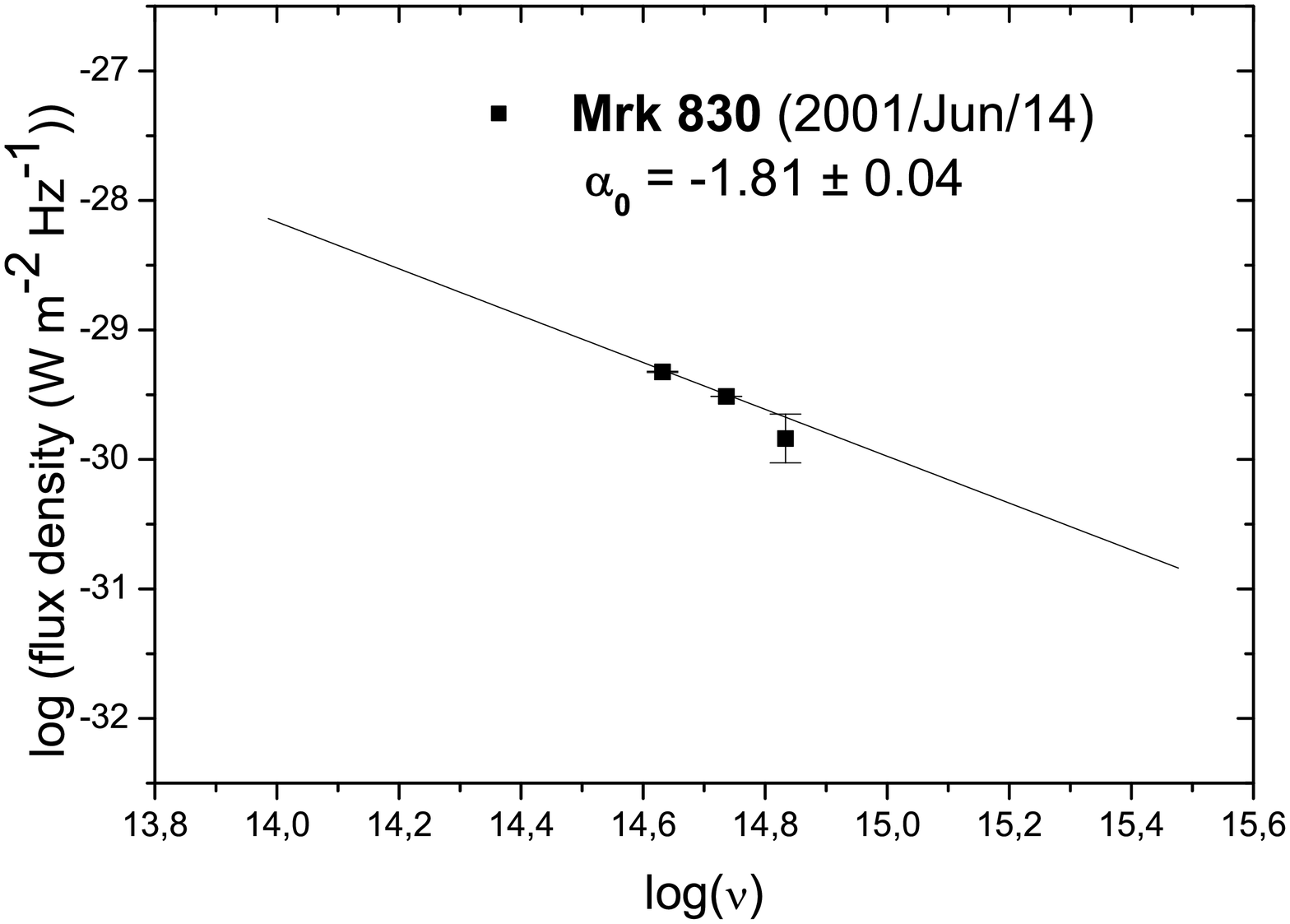}}\figurenum{7-11a}\label{fig811a}}
\subfigure{}
\subfigure[]{\scalebox{0.28}{\includegraphics{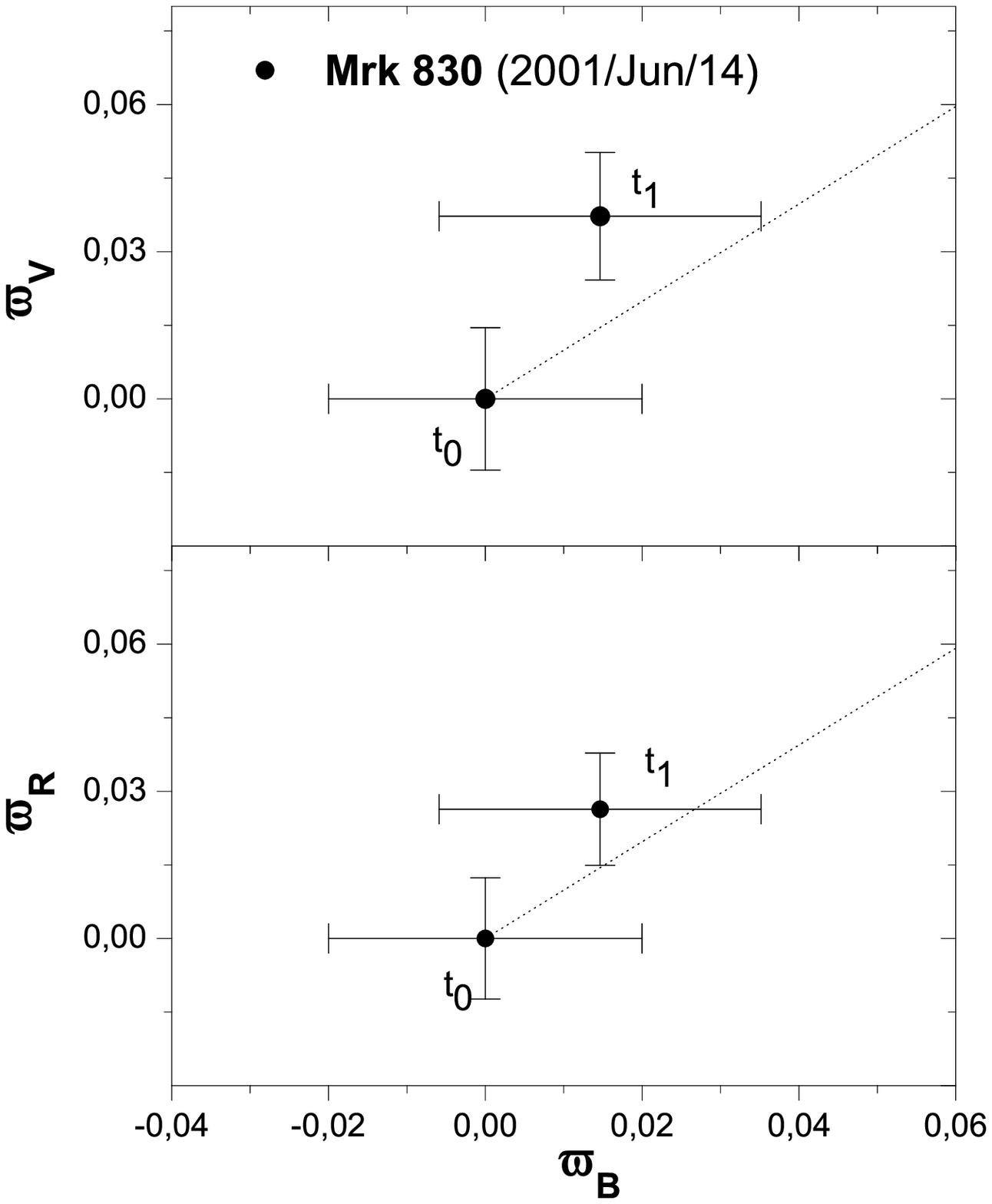}}\figurenum{7-11c}\label{fig811c}}
\caption{\sl \footnotesize Fits and variability: continuation.}
\label{fig811}
\end{center}
\end{figure}

{\bf Mrk~830.} This quasar showed variations in three different nights. Its brightness was similar during the three nights ($m_V \sim 17.6$), and near to the brightness reported by \citet{Stepanian01} ($m_V=17.29 \pm 0.03$) and by \citet{Chavu95} ($m_V=17.62$). A unique component marginally fits to the data of this object (Figures~\ref{fig89a}, \ref{fig810a}, \ref{fig811a}). Although a thermal component can be added with $T\sim 10,000$ K, contrary to the case of 1628.5+3808, the non-thermal component is sufficient to explain the spectral variation. Then, the addition of a second component is avoided.

{\it 1999 August 17.} At the beginning of the night the brightness of this quasar was $m_V=17.55 \pm 0.01$. But as the first $R$ data set was lost (see PII), then the second group was used to fit the model. A single non-thermal component fairly fits the data, with $\alpha_{t_0} = -3.06 \pm 0.49$ (Figure \ref{fig89a}). This value was used for the initial data of $B$ and $V$ bands to predict the $R$ value. It is possible to explain a microvariation event as observed with changes in this non-thermal component (Figure \ref{fig89c}). The spectral index would have increased in $9.8 \times 10^{-4}$, between the first and the third data set, while $n_n$ would have stayed constant. The simultaneity criterion indicates that observations were taken at the same level of brightness ($S_B=0.09$, $S_V=0.35$, and $S_R=0.02$).

{\it 1999 August 18.} The brightness was at the same level as for August 17, $m_V=17.56 \pm 0.03$. The first set of observations at the beginning of the session can be fitted by a non-thermal component with $\alpha_{t_0} = -2.51 \pm 0.35$ (Figure \ref{fig810a}). The variation can be described by changes in this component. In such a case, the spectral index would have decreased by a factor of $\times10^{-3}$ and the amplitude would have increased by a factor of $1\%$ ($n_{n_{t_1}} \sim 1.01$) (Figure~\ref{fig810c}). The simultaneity criterion indicates that the spectral variation is reliable ($S_B=0.27$, $S_V=0.39$, and $S_R=0.15$). However, although the model predicts that the variation could be observed in the $B$ band, it also predicts that a variations should be observed also in the $R$ band, which is not the case.

{\it 2001 June 14.} The brightness during this night, $m_V = 17.66 \pm 0.01$, was a tenth of a magnitude weaker with respect to the values measured in 1999 August. Initial data can be described by a power law, with $\alpha_{t_0} = -1.81 \pm 0.04$ (Figure \ref{fig811a}). The microvariability event can be marginally explained with a variation in the spectral index of $\sim 0.001$ while the amplitude stays constant (Figure~\ref{fig811c}). The simultaneity criterion establishes that observations have been taken without a change in brightness ($S_B=0.17$, $S_V=0.50$, and $S_R=0.37$); therefore, the color change is considered real.

\vskip0.5cm

Although the continuum emission can be fitted in all cases by a unique non-thermal component, in many of them the broad band spectral variation may require to assume the presence of a second component. Additionally, considering a unique component we have that $\alpha_{t_0} > 0$, which indicates that more spectral components must be acting.

\section{Discussion}\label{discusion}

From nine OM detections reported in PII (the two possible variations are excluded), our method can determine that three events are consistent with a thermal variation, and three with a non-thermal origin. The other three variations are undetermined, i.e., may be associated with either thermal or non-thermal processes (the cases of PKS~0003+15, MC3~1750+175, and US~3472), but with partial characteristics of thermal variations.

In agreement with \citet{de98} and \citet{Ram09}, the results presented in Section~\ref{results} show that microvariability does not depend on radio properties of quasars. Moreover, our study shows that OM related to non-thermal processes might be detectable not only in radio-loud quasars (RLQ) (e.g., 3C~281/2000/Mar/04) but also in RQQ (e.g., Mrk~830/2001/Aug/18), and that OM related to thermal processes might be detected in both kinds of objects, RQQ (e.g.,  1628.5+3808) and RLQ (e.g., PKS~1510-08).

There are several explanations to account for spectral variability originating either in the jet or the accretion disk. On the one hand, microvariability can be explained by the jet because of its physical conditions, and because time scales of OM coincide with jet dynamics (e.g., \citealt{Tingay01}; \citealt{Wiita06}). If all quasars can generate a relativistic jet, as suggested by the observational reports (\citealt{Blundell01}; \citealt{Blundell03}; \citealt{Ghisellini04}), it is not surprising to find the manifestation of non-thermal microvariability in RQQ. However, in the case of a jet, brightness temperatures exceeding the critical temperature, $T\sim 10^{12}$ K, are required (e.g., \citealt{Tingay01}; \citealt{Wiita06}, and references therein; \citealt{Fuhrmann08}). Although this problem may be overcome if large Doppler factors are considered (e.g., \citealt{Tingay01}; \citealt{Fuhrmann08}). The correlation between bands represents an additional problem for the jet thesis, because correlated variations between optic/UV and higher frequencies should be observed (e.g., \citealt{Nandra00}; \citealt{Peterson00}; \citealt{Shemmer01}; \citealt{Gaskell06}), which is not the general rule (e.g., \citealt{Edelson00}).

On the other hand, variations of diverse time scales, including OM, can be produced in accretion disks (\citealt{Webb00}; \citealt{Treve01}; \citealt{Treve02}; \citealt{Pereyra06}; see also \citealt{Mangalam93}; \citealt{Fukue03}; \citealt{Wiita06}; \citealt{Gu06}). Some of these variations may be associated with disk dynamics, or time scales associated with thermal and sound crossing time phenomena (minutes and hours; e.g., \citealt{Wiita06}). Changes in the accretion rate can modify the temperature of the disk, altering the flux level and producing short-term variations (e.g., \citealt{Webb00}; \citealt{Pereyra06}; \citealt{Wiita06}). Perturbations on the accretion disk can also produce variations with time scales of weeks, days, and even hours (e.g., \citealt{Webb00}; \citealt{Wiita06}). Recently, \citet{Chand09} have suggested that whether the OM is produced by processes related to the jet, then it is expected a smaller FWHM of the emission lines in quasars with positive detection than in those without OM detection. These author found no such behavior. Additionally, \citet{Goyal09} have reported that the microvariability detected in radio intermediate quasars has a duty cycle similar to that detected in RQQs and RLQs by \citet{Ram09} and \citet{Stalin04}. All these results suggest that variations involving the accretion disk play an important role. 

\citet{Treve02} developed a method to analyze the color changes in variability events. They determined that short-term variations in quasars may be originated by thermal processes related to changes in accretion rate, and hotspots in the accretion disk (see also \citealt{Vagne03}). The same conclusion was obtained by \citet{Webb00} for variability of quasars and Seyfert galaxies. However, although periodic or quasi-periodic variations are attributed to circling and auto-occulted hotspots in the disk (e.g., \citealt{Carrasco85}; \citealt{de90}), most variations do not display this characteristic. Nevertheless, non-periodicity might result of combination of hotspots and geometric changes, such as if the orbits in the disk are not Keplerian (e.g., \citealt{Fukue03}).

On the other hand, the contribution of each component to the total flux would play an important role to describe the spectral variations (\citealt{Vagne03} and \citealt{Gu06} arrived at a similar conclusion). We need two components to explain several spectral variations. In addition, when data are fitted exclusively with a non-thermal component, we have several cases where $\alpha_{t_0} > 0$ (3C~281, PKS~003+15, MC3~1750+175, US~3472). Since such a behavior is strange for the average SED of quasars (e.g., \citealt{Elvis86}), this result supports the idea that in general a better fit is obtained when a second component is added. Cases like those of PKS~1510-089 indicate that the largest changes of color are due to thermal variations. If a disk-jet relationship exists (e.g., \citealt{Falcke95}), more complex variations can be produced by perturbations in the disk that are propagated to the jet (e.g., \citealt{Wiita06}). Thus, our results are in agreement with \citet{Vagne03}, who reported that BL Lac objects do not show spectral variability as large as quasars, for which the thermal component contribution is larger.

Our description of the disk emission by a single blackbody may be inadequate to explain observations; however, more complicated models still produce unsatisfactory explanations even for the simplest variations. For example, the standard accretion disk model predicts time lags between optical bands (e.g., \citealt{Lawrence05}; \citealt{Pereyra06}; \citealt{Wiita06} and references therein), that are not always observed.

Because for spectral microvariability, variable and non-variable components should be considered, some spectral components that have not been considered by our model can change the value of $a_{V t_0}$. These additional components can generate spectral variations not contemplated in our model; for instance, the contribution from a line in only one of the three bands (maybe, this can give an explanation to variations as those of 3C~281 (2000 March 4), PKS~0003+15, and MC3~1750+175).

In blazars, there are evidences that the spectral variations are larger for low flux levels (\citealt{Vagne03}; \citealt{Ram04}; \citealt{Gu06}). This can be understood if at low flux level the thermal and non-thermal components have a comparable contribution to the optical flux (e.g., \citealt{Ram04}). For BL Lacertae, some OM events have been detected in which the object turned bluer when it increases its brightness (e.g., \citealt{Nesci98}; \citealt{Speziali98}; \citealt{Clements01}; \citealt{Papadakis03}; \citealt{Villata04b}; \citealt{Hu06a}). These variations should be generated by a non-thermal component (e.g., \citealt{Racine70}; \citealt{Gear86}; \citealt{Masa98}; \citealt{Vagne03}; \citealt{Villata04b}; \citealt{Hu06a}). On the other hand, the spectral changes in FSRQ seem to be larger than in BL Lac objects, because the latter have not a remarkable contribution from the thermal component (\citealt{Gu06}; \citealt{Ram04}). Then, a difference between the two blazars types may arise: the thermal component and its influence on spectral changes.

The time-lag criterion that we developed is based on the change rate of each OM event for our observational strategy. Although this criterion permits to be relatively sure that the data sets were acquired at the same flux level, it requires that the variations have constant change rates. It should be noticed that the light curves with a {\it jump}, as that of PKS~1510-089 reported in PII, can be in fact the result of continuum and gradual variations that show such jump because the observational strategy. The odd spectral variations observed in PKS~0003+15, MC3~1750+175, and US~3472 might be caused by the violation of the simultaneity criterion. The emission from additional components would also have unexpected repercussions on the color changes. 

The importance of orientation on variations originated in the jet is not discarded, but processes such as those taking place in the disk should be considered. Thermal component becomes indispensable to generate the observed variations, even in blazars (particularly at low activity), because the accretion disk in these objects must be face-on.

Although variations presented in Section~\ref{results} were classified as microvariability, we cannot rule out that they can be part of a variation of a trend with a larger characteristic time. For example, an alteration in the temperature of the accretion disk can show a variation with time scales of days, or even hours, which might presents a trend that we can observe during one night, and interpreted it as microvariability.

\section{Summary and conclusions}\label{summary}

The optical microvariations in RQQ and RLQ reported by \citet{Ram09} were analyzed to determine whether their origin is thermal or non-thermal. We suppose that while thermal variations are related to the accretion disk, non-thermal variations are related to the relativistic jet.

With such a purpose, we developed a method to discriminate between color changes by thermal and non-thermal processes in quasars during an OM event. This method consists in modeling the optical broadband continuum of quasars considering thermal and non-thermal components, and comparing the observed color changes with spectral variations derived from this model. The main result of this work is the possibility to distinguish between thermal and non-thermal origin of an optical microvariation event in quasars.

We identified that some events are consistent with a thermal origin, while some others with a non-thermal one. In other cases either component can give an explanation to the detected variation. Thus, another important contribution of this research is that, analyzing the {\it spectral microvariability}, we have found that the OM might be generated in either the disk or the jet, regardless of the radio classification of the quasars, CRLQ or RQQ. Additionally, some variations in RLQs are consistent with thermal OM and vice versa. Thus, our results indicate that microvariability in both quasar type might be originated under similar conditions.

In addition, although the continuum emission can be fitted in all cases by a unique non-thermal component, the broad band spectral variation may require to assume the presence of a second component, of thermal origin. The relative contribution of each component is an important parameter that should be taken into account for the description of spectral variability. Additional studies must be carried on to investigate the effects of these contributions on the type of quasar that we observe, and if this can be expended to the blazar class.

This result agrees with the discovery of small relativistic jets in RQQs (\citealt{Blundell01}) and explains previous results (\citealt{de98};  \citealt{Ram09}). Although this method yields interesting results applied to our data, it is desirable to have a monitoring  with a better temporal resolution,  as well as to observe in more bands. More realistic models are also necessary.

\acknowledgments{
{\footnotesize
We acknowledge the careful reading of our manuscript by an anonymous referee and the referee's comments, which helped to improve this paper. We thank Gabriel Garc\'\i a, Salvador Monrroy, Felipe Montalvo, Gustavo Melgoza, Michael Richer, Gaghik Tovmassian, Sergei Jarikov, and the staff from the OAN for assistance during the observations. A.R. is grateful to Dr. Gabriella Piccinelli for constructive comments that helped to improve this paper. D.D. is grateful for support from grant IN11161013 from PAPIIT-DGAPA, UNAM. J.A.D. and A.R. are grateful for support from CONACyT grants 50296 and 149972, respectively. A.R. is grateful for support from CONACyT grant with number of request +081535. Authors are also grateful to DGEP-DGAPA, UNAM for computational support. This research has made use of NASA/IPAC Extragalactic Database (NED), which is operated by the Jet Propulsion Laboratory, California Institute of Technology, under contract with National Aeronautics and Space Administration.}

}


\begin{thebibliography}{99}



\bibitem[Barvainis et al. (2005)]{Barvainis05} Barvainis R., Leh\'ar J., Birkinshaw M., Falcke H., Blundell K. M., 2005, ApJ, 618, 108
\bibitem[Blundell et al. (2003)]{Blundell03} Blundell K. M., Beasley A, J., Bicknell G, V., 2003, ApJ, 59, 103
\bibitem[Blundell \& Rawlings (2001)]{Blundell01} Blundell K. M., Rawlings S., 2001, ApJ, 562, L5
\bibitem[Brown et al. (1989a)]{Brown89a} Brown L. M. L., Robson E. I., Gear W. K., et al. 1989a, ApJ, 340, 129
\bibitem[Brown et al. (1989b)]{Brown89b} Brown L. M., Robson E. I. Gear W. K., Smith M. G., 1989b, ApJ, 340, 150
\bibitem[Carrasco et al. (1985)]{Carrasco85} Carrasco L., Dultzin D., Cruz-Gonz\'alez I., 1985, Nature, 314, 146
\bibitem[Chand et al. (2010)]{Chand09} Chand H., Wiita P. J., Gupta A. C., 2010, MNRAS, 402, 1059
\bibitem[Chavushyan et al. (1995)]{Chavu95} Chavushyan V. O., Stepanyan D. A., Balayan S. K., Vlasyuk V. V., 1995, AstL, 21, 804
\bibitem[Clements \& Carini (2001)]{Clements01} Clements S. D., Carini M. T., 2001,  AJ,  121, 90
\bibitem[Czerny \& Elvis (1987)]{Czerny87} Czerny B., Elvis M., 1987, ApJ, 321, 305
\bibitem[D'amicis et al. (2002)]{Damicis02} D'Amicis R., Nesci R., Massaro E., Maesano M., Montagni F., D'Alessio F., 2002, PASA, 19, 111
\bibitem[de Diego et al. (1998)]{de98} de Diego J. A., Dultzin-Hacyan, D., Ram\'\i rez, A., Ben\'\i tez E., 1998,  ApJ,  501, 69
\bibitem[de Diego \& Kidger (1990)]{de90} de Diego J. A., Kidger M., 1990, Ap\&SS, 171, 97
\bibitem[de Diego (2010)]{de10} de Diego, J. A., 2010, AJ, 139, 1269
\bibitem[Edelson et al. (2000)]{Edelson00} Edelson R., Koratkar A., Nandra K., Goad M., Peterson B. M., Collier S., Krolik J., Malkan M., Maoz D., O'Brien P., Shull J. M., Vaughan S., Warwick R., 2000, ApJ, 534, 180
\bibitem[Edelson \& Krolik (1988)]{Edelson88} Edelson R., Krolik J. H., 1988, ApJ, 333, 646
\bibitem[Elvis et al. (1986)]{Elvis86} Elvis M., Green R. F., Bechtold J., Schmidt M., Neugebauer G., Soifer B. T., Matthews K., Fabbiano G., 1986, ApJ, 310, 291
\bibitem[Falcke et al. (1995)]{Falcke95} Falcke H., Malkan M. A., Biermann P. L., 1995, A\&A, 298, 375
\bibitem[Fuhrmann et al. (2008)]{Fuhrmann08} Fuhrmann L., Krichbaum T. P., Witzel A., Kraus A., Britzen S., Bernhart S., Impellizzeri C. M. V., Agudo I., Klare J., Sohn B. W. et al. 2008, A\&A, 490, 1019
\bibitem[Fukue (2003)]{Fukue03} Fukue J., 2003, PASJ, 55, 1121
\bibitem[Gaskell (2006)]{Gaskell06} Gaskell C. M., 2006, ASPC, 360, 111
\bibitem[Gear et al. (1986)]{Gear86} Gear W. K., Robson E. I., Brown L. M. J.,, 1986, Nature, 324, 546
\bibitem[Ghisellini et al. (2004)]{Ghisellini04} Ghisellini G., Haardt F., Matt G., 2004, A\&A, 413, 535
\bibitem[Giveon et al. (1999)]{Giveon99} Giveon U., Maoz D., Kaspi S., Netzer H., Smith P. S., 1999, MNRAS, 306, 637
\bibitem[Gonz\'alez-P\'erez et al. (1996)]{GP96} Gonz\'alez-P\'erez J.-N., Kidger M. R., de Diego J. A., 1996, A\&A, 311, 57
\bibitem[Goyal et al. (2010)]{Goyal09} Goyal A., Gopal-Krishna, Joshi S., Sagar R., Wiita P. J., Anupama G. C., Sahu D. K., 2010, MNRAS, 401, 2622
\bibitem[Gu et al. (2006)]{Gu06} Gu M. F., Lee C.-U., Pak S., Yim H. S., Fletcher A. B., 2006, A\&A, 450, 39
\bibitem[Gupta \& Joshi (2005)]{Gupta05} Gupta A. C., Joshi U. C., 2005,  A\&A,  440, 855
\bibitem[Hu et al. (2007)]{Hu07} Hu S.-M., Bi X.-W., Zheng Y.-G., Mao W.-M., 2007, ASPC, 373, 205
\bibitem[Hu et al. (2006a)]{Hu06a} Hu S. M., Zhao G., Guo H. Y., Zhang X., Zheng Y. G., 2006, MNRAS, 371, 1243
\bibitem[Hu et al. (2006b)]{Hu06b} Hu S. M., Wu J. H., Zhao G., Zhou, X., 2006, MNRAS, 373, 209
\bibitem[Hufnagel \& Bregman (1992)]{Hufnagel92} Hufnagel B.R., Bregman J. N., 1992, ApJ, 386, 473
\bibitem[Kembhavi \& Narlikar (1999)]{Kembhavi99} Kembhavi A. K., Narlikar J. V., 1999, Quasars and Active Galactic Nuclei: AN INTRODUCTION, (Cambridge: Cambridge University Press)
\bibitem[Krishan \& Wiita (1994)]{Krishan94} Krishan V., Wiita P. J., 1994,  ApJ,  423, 172
\bibitem[La Franca et al. (1992)]{Lafranca92} La Franca F., Cristiani S., Barbieri C., 1992, AJ, 103, 1062
\bibitem[Landolt (1992)]{landolt92} Landolt, A.~U.\ 1992, \aj, 104, 340 
\bibitem[Lawrence (2005)]{Lawrence05} Lawrence A., 2005, MNRAS, 363, 57
\bibitem[Li \& Cao (2008)]{Li08} Li S.-H.., Cao X., 2008, MNRAS, 387, L41
\bibitem[Malkan \& Moore (1986)]{Malkan86} Malkan M. A., Moore R. L., 1986, ApJ, 300, 216
\bibitem[Malkan \& Sargent (1982)]{Malkan82} Malkan M. A., Sargent W. L. W., 1982, ApJ, 254, 22
\bibitem[Mangalam \& Wiita (1993)]{Mangalam93} Mangalam A. V., Wiita P. J., 1993, ApJ, 406, 420	
\bibitem[Massaro et al. (1998)]{Masa98} Massaro E., Nesci R., Maesano M., Montagni F., D'Alessio F., 1998, MNRAS, 299, 47
\bibitem[Nandra et al. (2000)]{Nandra00} Nandra K., Le T., George I. M., Edelson R. A., Mushotzky R. F., Peterson B. M., Turner T. J., 2000, ApJ, 2000, 544, 734
\bibitem[Nesci et al. (1998)]{Nesci98} Nesci R., Maesano M., Massaro E., Montagni F., Tosti G., Fiorucci  M., 1998, A\&A, 332, L1
\bibitem[Papadakis et al. (2003)]{Papadakis03} Papadakis I. E., Boumis P., Samaritakis V., Papamastorakis J., 2003, A\&A, 397, 565
\bibitem[Papadakis et al. (2007)]{Papadakis07} Papadakis I. E., Villata M., Raiteri C. M., 2007, A\&A, 470, 857
\bibitem[Pereyra et al. (2006)]{Pereyra06} Pereyra N. A., Vandern Berk D. E., Turnshek D. A., Hiller D. J., Wilhite B. C., Kron R. G., Schneider D. P., Brinkmann J.,  2006, ApJ, 642, 87
\bibitem[Peterson et al. (2000)]{Peterson00} Peterson B. M., McHardy I. M., Wilkes B. J., Berlind P., Bertram R., Calkins M., Collier S. J., Huchra J. P., Mathur S., Papadakis I. et al. 2000, ApJ, 542, 161
\bibitem[Racine (1970)]{Racine70} Racine R., 1970, ApJ, 159, 99
\bibitem[Ram\'\i rez et al. (2009)]{Ram09} Ram\'\i reaz A., de Diego J. A., Dultzin D., Gonz\'alez-P\'erez J.-N., 2009, AJ, 138, 991
\bibitem[Ram\'\i rez et al. (2004)]{Ram04} Ram\'\i rez A., de Diego J. A., Dultzin D., P\'erez-Gonz\'alez J. N., 2004,  A\& A,  421, 83
\bibitem[Sandage et al. (1965)]{Sandage65} Sandage A., V\'eron P., Wyndham J. D., 1965, ApJ, 142, 1307
\bibitem[Shemmer et al. (2001)]{Shemmer01} Shemmer O., Romano P., Bertram R., Brinkmann W., Collier S., Crowley K. A., Detsis E., Filippenko A. V., Gaskell C. M., George T. A., et al. 2001, ApJ, 561, 162
\bibitem[Speziali \& Natali (1998)]{Speziali98} Speziali R., Natali G., 1998, A\&A, 339, 382
\bibitem[Stalin et al. (2004)]{Stalin04} Stalin C. S., Gopal-Krishna, Sagar R., Wiita P. J., 2004,  MNRAS,  350, 175
\bibitem[Stalin et al. (2009)]{Stalin09} Stalin C. S., Kawabata K. S., Uemura M., Yoshida M., Kawai N., Yanagisawa K., Shimizu Y., Kuroda D., Nagayama S., Toda H., 2009, MNRAS, 399, 1357, DOI: 10.1111/j.1365-2966.2009.15354.x
\bibitem[Stepanian et al. (2001)]{Stepanian01} Stepanian J. A.. Green R. F., Foltz C. B., Chaffee F., Chavushyan V. H., Lipovetsky V. A., Erastova L. K., 2001, AJ, 122, 3361
\bibitem[Tingay et al. (2001)]{Tingay01} Tingay S. J., Preston R. A., Lister M. L., Piner B. G., Murphy D. W., Jones D. L., Meier D. L., Pearson T. J., Readhead A. C. S., Hirabayashi H., Murata Y., Kobayashi H., Inuoe M., 2001, ApJ, 549, 5
\bibitem[Tr\`evese \& Vagnetti (2001)]{Treve01} Tr\`evese D., Vagnetti F., 2001, MmSaI, 72, 33
\bibitem[Tr\`evese \& Vagnetti (2002)]{Treve02} Tr\`evese D., Vagnetti F., 2002, ApJ, 564, 624
\bibitem[Vagnetti et al. (2003)]{Vagne03} Vagnetti F., Trevese D, Nesci R., 2003, ApJ, 590, 123
\bibitem[V\'eron-Cetty \& V\'eron (2001)]{Veron01} V\'eron-Cetty M.-P., V\'eron P., 2001, A\&A, 374, 92
\bibitem[Villata et al. (2000)]{Villata00} Villata M., Mattox J. R., Massaro E., Nesci R., Catalano S., Frasca A., Raiteri C. M., Sobrito G., Tosti G., Nucciarelli G. et al. 2000, A\&A, 363, 108
\bibitem[Villata et al. (2004a)]{Villata04a} Villata M., Raiteri C. M., Aller H. D. , Aller M. F., H. Ter\"asranta H., Koivula P., Wiren S., Kurtanidze O. M., Nikolashvili M. G., Ibrahimov M. A., Papadakis I. E., Tosti G., Hroch F., Takalo L. O., Sillanp\"a\"a A., Hagen-Thorn V. A., Larionov V. M., Schwartz R. D., Basler J., Brown L. F., Balonek T. J., 2004, A\&A, 424, 497
\bibitem[Villata et al. (2004b)]{Villata04b} Villata M., Raiteri C. M., Kurtanidze O. M., Nikolashvili M. G., Ibrahimov M. A., Papadakis I. E., Tosti G., Hroch F., Takalo L. O., Sillanp\"a\"a A. et al. 2004, A\&A, 421, 103
\bibitem[Webb \& Malkan (2000)]{Webb00} Webb W., Malkan M., 2000, ApJ, 540, 652
\bibitem[Wiita (2006)]{Wiita06} Wiita P. J., 2006, ASPC, 350, 183
\bibitem[Wilhite et al. (2008)]{Wilh08} Wilhite B. C., Brunner R. J., Grie, C. J., Schneider D. P., Vanden Berk D. E., 2008, MNRAS, 383, 1232 
\bibitem[Xie et al. (2001)]{Xie01} Xie G.Z., Li K. H., Bai J. M., Dai B. Z., Liu W. W., Zhang X., Xing S. Y., 2001, ApJ, 548, 200



\end{thebibliography}
\end{document}